\documentclass{aa}

\usepackage[varg]{txfonts}
\usepackage{pdflscape}
\usepackage{afterpage}
\usepackage{geometry}
\usepackage{tablefootnote}
\usepackage{threeparttable}
\usepackage{microtype}
\usepackage{hyperref}
\usepackage{booktabs}
\usepackage{xcolor}
\usepackage{hyperref}
\usepackage{graphicx}
\usepackage{subcaption}
\usepackage{multirow}
\usepackage{float}

\hypersetup{%
  colorlinks=true,% hyperlinks will be black
  linkcolor=blue,% hyperlink borders will be red
  citecolor=blue
}

\usepackage{etoolbox}
\makeatletter
\patchcmd\@combinedblfloats{\box\@outputbox}{\unvbox\@outputbox}{}{%
  \errmessage{\noexpand\@combinedblfloats could not be patched}%
}%
\makeatother

\let\oldcitet=\citet

\renewcommand{\citet}[1]{\textcolor[rgb]{0,0,1}{\oldcitet{#1}}}

\begin{document}

\title{Gas phase Elemental abundances in Molecular cloudS (GEMS)}

\subtitle{II. On the quest for the sulphur reservoir in molecular clouds: the H$_{2}$S case}

\author{D. Navarro-Almaida\inst{1}
	\and
	R. Le Gal\inst{2}
	\and
	A.~Fuente\inst{1}
    \and
    P.~Rivi\`ere-Marichalar\inst{1}
    \and
    V.~Wakelam\inst{3}
    \and
    S.~Cazaux \inst{4}
    \and
    P.~Caselli\inst{5}
	\and
	Jacob C. Laas\inst{5}
	\and
	T.~ Alonso-Albi \inst{1}
	\and
	J.~C.~Loison\inst{6}
	\and
	M.~Gerin\inst{7}
    \and
    C.~Kramer\inst{8}
	\and
    E.~Roueff\inst{9}         
    \and         
    R.~ Bachiller \inst{1}
    \and            
    B.~Commerçon\inst{10}  
    \and
    R.~Friesen\inst{11}
	\and
   	S.~Garc\'{\i}a-Burillo\inst{1}
	\and
	J.~ R.~Goicoechea \inst{12}
    \and
   	B.~M.~Giuliano\inst{5} 
   	\and   
    I.~Jim\'enez-Serra\inst{13}  
    \and
    J. M. Kirk\inst{14} 
    \and
   	V.~Lattanzi\inst{5}
	\and
    J.~Malinen\inst{15,16} 
    \and
    N.~Marcelino\inst{12}     
   	\and
    R.~ Mart\'{\i}n-Dom\`enech\inst{2}
	\and
    G. M. Mu\~noz Caro\inst{13}     
    \and
    J.~Pineda\inst{5}       
    \and
    B.~Tercero\inst{1}
	\and
    S.~P. Trevi\~no-Morales\inst{17}
	\and
    O.~Roncero\inst{12}
    \and
    A.~Hacar \inst{18}
    \and
    M.~Tafalla\inst{1}
    \and
    D.~Ward-Thompson\inst{14}
    }
  %\inst{2}
  %\thanks{\emph{Present address:}
    %Department of Computer Science, Purdue University,
    %West Lafayette, IN 47907, USA}

%\offprints{R. Plemmons, \email{plemmons@...}}

\institute{ 
	Observatorio Astron\'omico Nacional (OAN), Alfonso XII, 3,  28014, Madrid, Spain
    \and
	Harvard-Smithsonian Center for Astrophysics, 60 Garden St., Cambridge, MA 02138, USA
	\and
	Laboratoire d'Astrophysique de Bordeaux, Univ. Bordeaux, CNRS, B18N, all\'ee Geoffroy Saint-Hilaire, 33615 Pessac, France
	\and
	Faculty of Aerospace Engineering, Delft University of Technology, Delft, The Netherlands; University of Leiden, P.O. Box 9513, NL, 2300 RA, Leiden, The Netherlands
	\and
	Centre for Astrochemical Studies, Max-Planck-Institute for Extraterrestrial Physics, Giessenbachstrasse 1, 85748, Garching, Germany
	\and
	Institut des Sciences Mol\'eculaires (ISM), CNRS, Univ. Bordeaux, 351 cours de la Lib\'eration, F-33400, Talence, France
	\and
	Observatoire de Paris, PSL Research University, CNRS, \'Ecole Normale Sup\'erieure, Sorbonne Universit\'es, UPMC Univ. Paris 06, 75005, Paris, France
	\and
	Instituto Radioastronom\'{\i}a Milim\'etrica (IRAM), Av. Divina Pastora 7, Nucleo Central, 18012, Granada, Spain
	\and
	Sorbonne Universit\'e, Observatoire de Paris, Universit\'e PSL, CNRS, LERMA, F-92190,  Meudon, France
	\and
	\'Ecole Normale Sup\'erieure de Lyon, CRAL, UMR CNRS 5574, Universit\'e Lyon I, 46 All\'ee d'Italie, 69364, Lyon Cedex 07, France 
	\and
	National Radio Astronomy Observatory, 520 Edgemont Rd., Charlottesville VA USA 22901
	\and
	Instituto de F\'{\i}sica Fundamental (CSIC), Calle Serrano 123, 28006, Madrid, Spain
	\and
	Centro de Astrobiolog\'{\i}a (CSIC-INTA), Ctra. de Ajalvir, km 4, Torrej\'on de Ardoz, 28850, Madrid, Spain
	\and
	Jeremiah Horrocks Institute, University of Central Lancashire, Preston PR1 2HE, UK
	\and
	Department of Physics, University of Helsinki, PO Box 64, 00014, Helsinki, Finland
	\and
	Institute of Physics I, University of Cologne, Cologne, Germany
	\and
	Chalmers University of Technology, Department of Space, Earth and Environment, SE-412 93 Gothenburg, Sweden
	\and
	Leiden Observatory, Leiden University, PO Box 9513, 2300-RA, Leiden, The Netherlands
}

\date{Received: xx 2019 / Accepted: xx 2020}

\abstract {Sulphur is one of the most abundant elements in the Universe. Surprisingly, sulphuretted molecules are not as abundant as expected in the interstellar medium and the identity of the main sulphur reservoir is still an open question.} 
{Our goal is to investigate the H$_2$S chemistry in dark clouds, as this stable molecule is a potential sulphur reservoir.}
{Using millimeter observations of CS, SO, H$_{2}$S, and their isotopologues, we determine the physical conditions and H$_{2}$S abundances along the cores TMC 1-C, TMC 1-CP, and Barnard 1b. The gas-grain model \textsc{Nautilus} is used to model the sulphur chemistry and explore the impact of photo-desorption and chemical desorption on the H$_2$S abundance.} 
{Our modeling shows that chemical desorption is the main source of gas-phase H$_2$S in dark cores. The measured H$_{2}$S abundance can only be fitted if we assume that the chemical desorption rate decreases by more than a factor of 10 when $n_{\rm H}>2\times10^{4}$. This change in the desorption rate is consistent with the formation of thick H$_2$O and CO ice mantles on grain surfaces. The observed SO and H$_2$S abundances are in good agreement with our predictions adopting an undepleted value of the sulphur abundance. However, the CS abundance is overestimated by a factor of $5-10$. Along the three cores, atomic S is predicted to be the main sulphur reservoir.}
{The gaseous H$_2$S abundance is well reproduced, assuming undepleted sulphur abundance and chemical desorption as the main source of H$_2$S. The behavior of the observed H$_{2}$S abundance suggests a changing desorption efficiency, which would probe the snowline in these cold cores. Our model, however, highly overestimates the observed gas-phase CS abundance. Given the uncertainty in the sulphur chemistry, we can only conclude that our data are consistent with a cosmic elemental S abundance with an uncertainty of a factor of 10.\\\\\\$\left.\right.$}

\keywords{astrochemistry --
  ISM: molecules -- ISM: clouds -- submillimeter: ISM}
\maketitle

\section{Introduction}

Sulphur is one of the most abundant elements in the Universe \citep[${\rm S/H}\sim 1.35\times 10^{-5}$,][]{Satoshi2017} and plays a crucial role in biological systems on Earth, so it is important to follow its chemical history in space (i.e., toward precursors of the Solar System). Surprisingly, sulphuretted molecules are not as abundant as expected in the interstellar medium (ISM). A few sulphur compounds have been detected in diffuse clouds, demonstrating that sulphur abundance in these low-density regions is close to the cosmic value \citep{Neufeld2015}. This is also the case towards the prototypical photodissociation region (PDR) in the Horsehead Nebula, where the sulfur abundance is found to be very close to the undepleted value observed in the diffuse ISM \citep{Goicoechea2006}, with an estimate of ${\rm S}/{\rm H} = (3.5\pm 1.5)\times 10^{-6}$. A wide variety of S-bearing molecules, including the doubly sulphuretted molecule S$_{2}$H, were later detected in the HCO peak in this PDR \citep{Fuente2017, Riviere2019}. However, sulphur is thought to be depleted inside molecular clouds by a factor of 1000 compared to its estimated cosmic abundance \citep{Graedel1982, Agundez2013}. The depletion of sulphur is observed not only in cold pre-stellar cores, but also in hot cores and corinos \citep{Wakelam2004}. We would expect that most of the sulphur is locked on the icy grain mantles in dense cores, but we should see almost all sulphur back to the gas phase in hot cores and strong shocks. However, even in the well-known Orion-KL hot core, where the icy grain mantles are expected to sublimate releasing the molecules to the gas phase, one needs to assume a sulphur depletion of a factor of $\sim 10$ to reproduce the observations \citep{Esplugues2014, Crockett2014}. Because of the high hydrogen abundances and the mobility of hydrogen in the ice matrix, sulphur atoms impinging in interstellar ice mantles are expected to form H$_2$S preferentially. To date, OCS is the only S-bearing molecule unambiguously detected in ice mantles because of its large band strength in the infrared \citep{Geballe1985, Palumbo1995} and, tentatively, SO$_2$ \citep{Boogert1997}. The detection of solid H$_2$S (s-H$_{2}$S hereafter) is hampered by the strong overlap between the 2558 cm$^{-1}$ band with the methanol bands at 2530 and 2610 cm$^{-1}$. Only upper limits of s-H$_2$S abundance could be derived by \citet{Jimenez-Escobar2011}, with values with respect to H$_{2}$O ice of 0.7\% and 0.13\% in W33A and IRAS183160602, respectively. 

Sulphur-bearing species have been detected in several comets. Contrary to the interstellar medium (ISM), the majority of cometary detections of sulphur-bearing molecules belongs to H$_2$S and S$_2$ \citep{Mumma2011}. Towards the bright comet  Hale Bopp, a greater diversity has been observed, including CS and SO \citep{Boissier2007}. The brighter comets C/2012 F6 (Lemmon) and C/2014 Q2 (Lovejoy) have also been shown to contain CS \citep{Biver2016}. Currently, some of the unique in-situ data are available from the Rosetta mission on comet 67P/Churyumov-Gerasimenko. With the Rosetta Orbiter Spectrometer for Ion and Neutral Analysis (ROSINA; \citealp{Balsiger2007}) onboard the orbiter, the coma has been shown to contain H$_2$S, atomic S, SO$_2$, SO, OCS, H$_2$CS, CS$_2$, S$_2$, and, tentatively, CS, as the mass spectrometer cannot distinguish the latter from CO$_2$ gases \citep{LeRoy2015}. In addition, the more complex molecules S$_3$, S$_4$, CH$_3$SH, and C$_2$H$_6$S have now been detected \citep{Calmonte2016}. Even with the large variety of S-species detected, H$_2$S remains as the most abundant, with an ice abundance of about 1.5\% relative to H$_2$O, which is similar to the upper limit measured in the interstellar medium \citep{Jimenez-Escobar2011}.

Gas phase Elemental abundances in Molecular CloudS (GEMS) is an IRAM 30m Large Program aimed at estimating the depletions of the most abundant elements (C, O, S, and N) in a selected set of prototypical star-forming filaments located in the molecular cloud complexes Taurus, Perseus, and Orion A. The first chemical study using GEMS data concluded that only 10\% of the sulphur is in gas phase in the translucent part \citep[$A_{\rm v} \sim 3-10$ mag, ][hereafter Paper I]{Fuente2019} of TMC 1. These conclusions were based on millimeter observations of CS, SO and HCS$^+$.  Here we investigate the chemistry of  H$_2$S, a key piece in the sulphur chemistry, which is still poorly understood. 

The formation of H$_2$S in the gas phase  is challenging at the low temperatures prevailing in the interstellar medium. None of the species S, SH, S$^+$, and SH$^+$ can react exothermically  with H$_2$ in a hydrogen atom abstraction reaction which greatly inhibits the formation of sulphur hydrides \citep{Gerin2016}. Alternatively, H$_2$S is thought to form on the grain surfaces where sulphur atoms impinging in interstellar ice mantles are hydrogenated, forming s-H$_2$S and potentially becoming the most important sulphur reservoir in dark cores \citep{Vidal2017}. Recent modeling and experimental studies suggest an active sulphur chemistry within the ices where  s-H$_{2}$S  might be converted in more complex compounds such as OCS or CH$_3$SH (e.g., \citealp{Laas2019}, El Akel et al. in prep). Unfortunately, both the surface chemistry reaction rates and the desorption processes are not well constrained, which makes a reliable prediction of both solid and gas phase H$_2$S abundance difficult. This paper uses a subset of the complete GEMS database to investigate the H$_2$S abundance in dark cores TMC 1 and Barnard 1b as prototypes of molecular clouds embedded in low-mass and intermediate-mass star forming regions, respectively.  
  
\section{Source sample}

\subsection{TMC 1}

\begin{figure*}
	\centering
		\begin{subfigure}[b]{0.49\textwidth}\includegraphics[width=\textwidth,keepaspectratio]{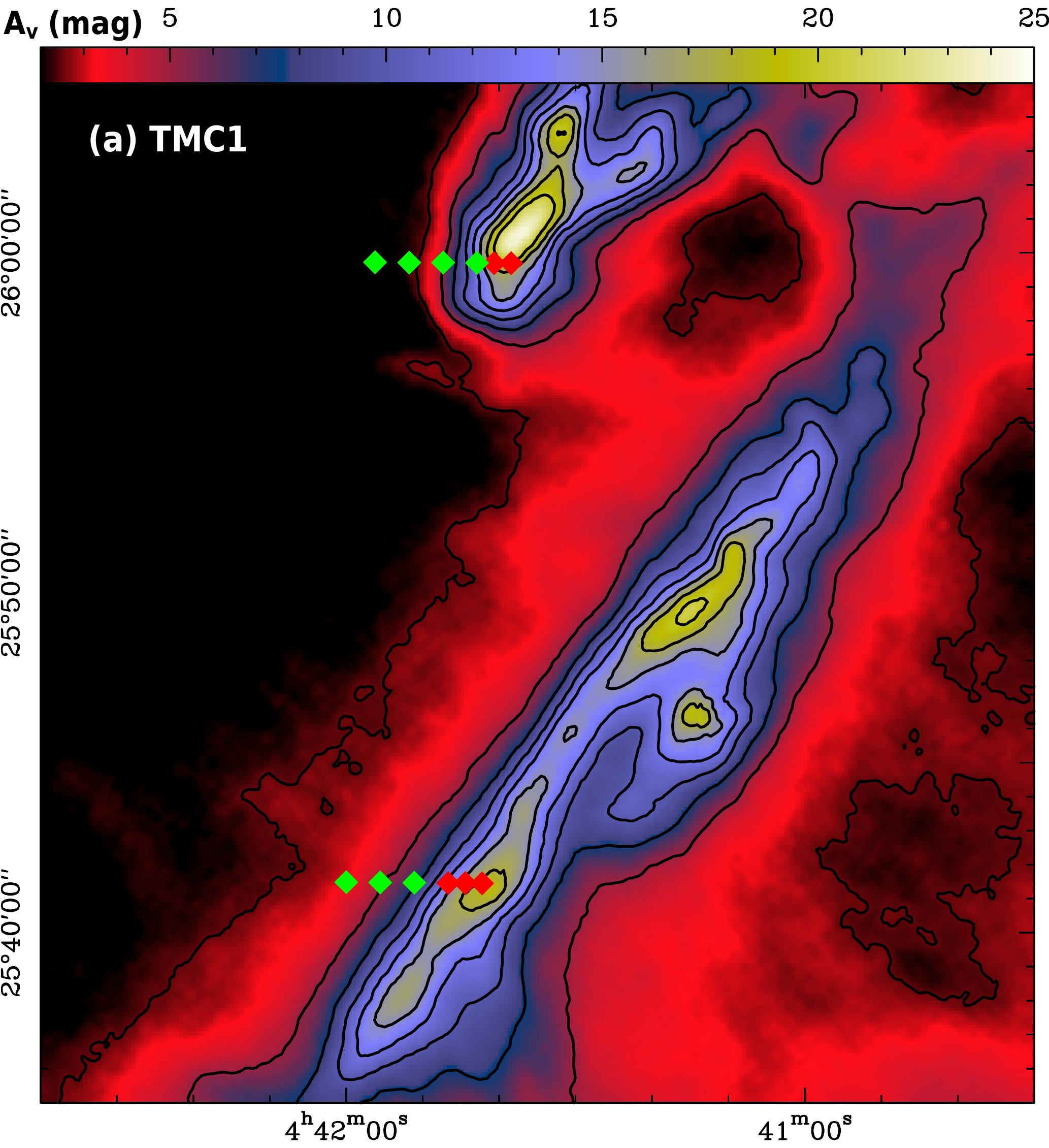}
			\label{fig:mapTMC1}
		\end{subfigure}
		~
		\begin{subfigure}[b]{0.49\textwidth}\includegraphics[width=\textwidth,keepaspectratio]{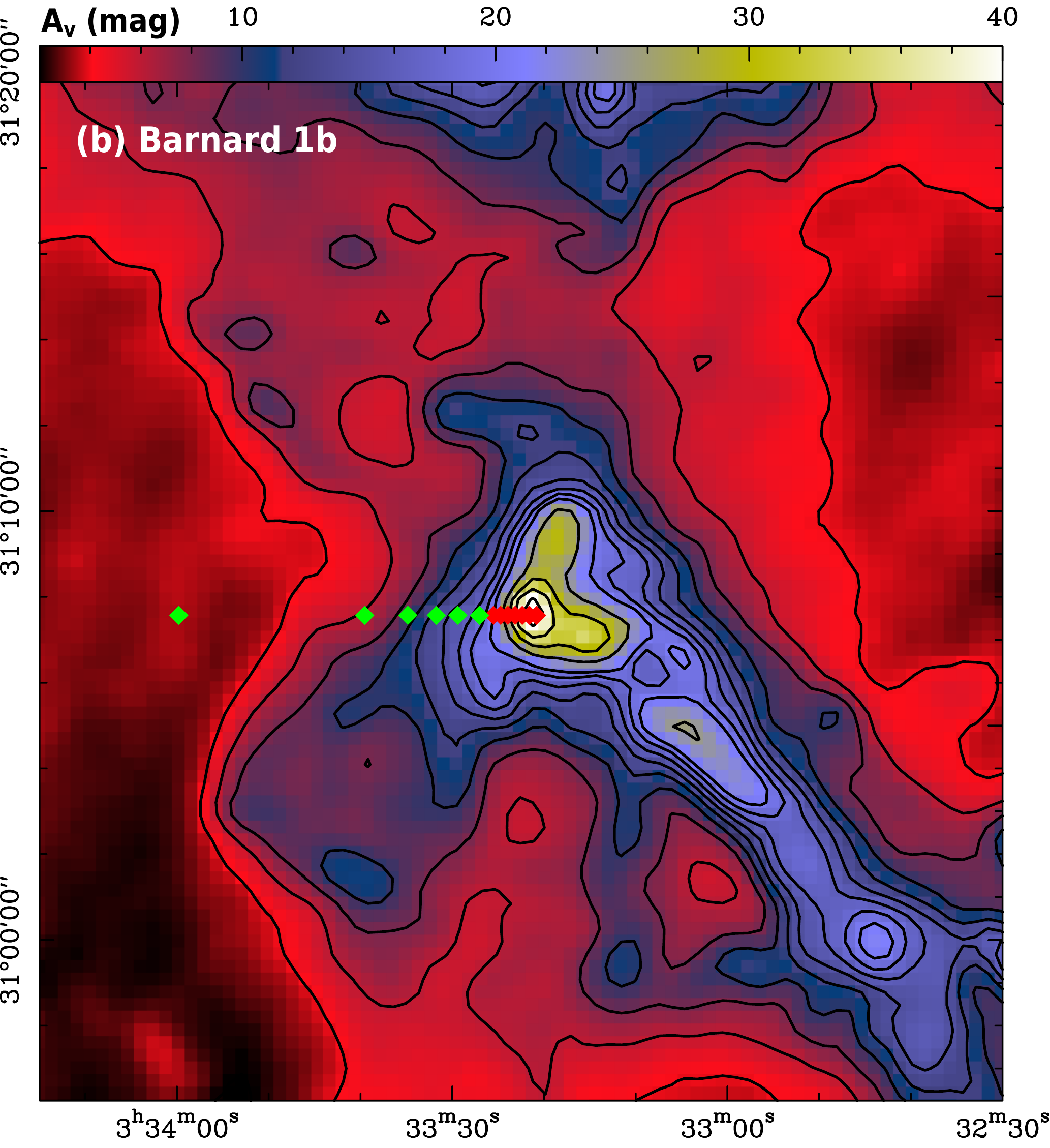}
			\label{fig:mapB1b}
		\end{subfigure}
		\caption{{\it Left panel:} TMC 1 total visual extinction (the visual extinction along the line of sight) map based on Herschel dust emission maps from Jason Kirk (private communication). {\it Right panel:} Visual extinction map of Barnard 1 from the opacity at 850 $\mu$m derived by \citet{Zari2016}. In both maps the observed positions are in lines of constant declination, or cuts,  where the red squares mark those positions in which only IRAM 30 m telescope data is obtained, and green squares mark positions with additional Yebes 40 m data.}
\end{figure*}

The Taurus molecular cloud (TMC), at a distance of 140 pc \citep{Elias1978}, is one of the closest molecular cloud complexes and is considered an archetypal low-mass star forming region. It has been the target of several cloud evolution and star formation studies (see e.g., \citet{Goldsmith2008}), being extensively mapped in CO \citep{Cernicharo1987, Narayanan2008}. The most massive molecular cloud in Taurus is the Heiles cloud 2 \citep[HCL 2, ][]{Toth2004}, which hosts the well-known region TMC 1 (\hyperref[fig:mapTMC1]{Fig. 1a}). As one of the most extensively studied molecular filaments, TMC 1 was included in the IRAM 30m Large Program GEMS. \citet{Fuente2019} modeled the chemistry of the translucent filament, deriving the gas phase elemental  abundances (C/O $\sim 0.8-1$, S/H $\sim 0.4-2.2\times10^{-6}$) and constraining  the cosmic rays molecular hydrogen ionization rate to $\sim 0.5 - 1.8\times 10^{-16}$ s$^{-1}$ in this moderate dense ($n_{\rm H_{2}}$ $< 10^4$ cm$^{-3}$) phase. Here, we will focus on the chemistry of the dense cores TMC 1-CP and  TMC 1-C (see \hyperref[fig:mapTMC1]{Fig. 1a}). TMC 1-CP has been extensively observed at millimeter wavelengths and is often adopted as template of C-rich (C/O $>1$) dense cores \citep{Feher2016, Gratier2016, Agundez2013}. Less studied, TMC 1-C was proposed as an accreting starless core with high CO depletion by \citet{Schnee2007, Schnee2010}.

\subsection{Barnard 1b}

Barnard 1 is a young, intermediate-mass star forming cloud, embedded in the western sector of the 30 pc wide molecular cloud complex Perseus (see \hyperref[fig:mapB1b]{Fig. 1b}). It hosts several dense cores in different evolutionary stages \citep{Hatchell2005}, out of which Barnard 1b is the youngest.  The Barnard 1b region was mapped in many molecular tracers, such as CS, NH$_3$, $^{13}$CO \citep{Bachiller1990}, N$_2$H$^+$ \citep{Huang2013}, H$^{13}$CO$^+$ \citep{Hirano1999}, or CH$_3$OH \citep{Hiramatsu2010,Oberg2010}. Interferometric observations of Barnard 1b revealed that this compact core hosts two young stellar objects (YSOs), B1b-N and B1b-S \citep{Huang2013, Marcelino2018}, and a third more evolved source, hereafter B1b-Spitzer, with deep absorption features from ices \citep{Jorgensen2006}. The three sources are deeply embedded in the surrounding protostellar envelope, that seems essentially unaffected by the inlaid sources,  as shown by the large column density, $N({\rm H}_2)\sim 7.6\times 10^{22}$~cm$^{-2}$ \citep{Daniel2013} and cold kinetic temperature, $T_{\rm K} = 12$ K \citep{Lis2002}. From the chemical point of view, Barnard 1b has a rich chemistry. Indeed, many molecules were observed for the first time in this object, like HCNO \citep{Marcelino2009} or CH$_3$O \citep{Cernicharo2012}. Additionally, Barnard 1b shows a high degree of deuterium fractionation and has been associated with first detections of multiple deuterated molecules, such as ND$_3$ \citep{Lis2002} or D$_2$CS \citep{Marcelino2005}, consistent with the expected chemistry in a dense and cold core. Recently, \citet{Fuente2016} proposed that this core is characterized by a low depletion of sulphur, S/H$\sim$ 10$^{-6}$. They proposed that peculiar initial conditions due to the proximity of the bipolar outflows, a rapid collapse of the parent cloud, or the imprint of the two deeply embedded protostellar objects, might be the causes. Within GEMS, we have studied the nine-point cut depicted in \hyperref[fig:mapB1b]{Fig. 1b}, with visual extinctions (extinctions along the line of sight) ranging from $\sim 3$ mag to $\sim 76$ mag.

\section{Observations}

This paper is based on a subset of the GEMS observations carried out with the IRAM 30m telescope and the 40m Yebes telescopes. The TMC 1 data used in this paper were already presented in Paper I, where we gave a detailed description of the observational procedures used within the GEMS project. The observations towards Barnard 1b were carried out using  the same observational strategy as in TMC 1 (see Paper I). We select three lines of constant declination, or cuts, through the three extinction peaks of TMC 1-C, TMC 1-CP, and Barnard 1b. We define the origin of each line as the point of highest extinction in that line. \hyperref[fig:mapB1b]{Figs. 1a-b} show the positions observed across the dense cores TMC 1-C, TMC 1-CP, and Barnard 1b. All the positions were observed in setups 1 to 4 (see Paper I) using the 30m telescope. These observations were done using frequency-switching in order to optimize the detection sensitivity. In addition to the 30m observations, the positions with A$_{v}$ < 20 mag were also observed with 40m Yebes telescope (setup 0). During the 40m observations, the observing procedure was position-switching, the OFF-position being RA(J2000)= 03:36:39.571 Dec: 29:59:53.30  for Barnard 1b. This position was checked to be empty of emission in the observed bandwidths before the observations. In the following, line intensities are in main beam temperature, $T_{\rm MB}$, for all observations. Calibration errors are estimated to be $\sim$20\% for the IRAM 30m telescope and $\sim 30\%$ for the Yebes 40m telescope. In order to constrain the gas physical conditions and the sulphur chemistry, in this paper we use the observed lines of CS and SO, in addition to H$_2$S (see the list of lines in \hyperref[table:transitions]{Table 1}, and the corresponding spectra in \hyperref[fig:TMC1_C_spectra]{Figs. A.1-2}, and \hyperref[fig:B1B_spectra]{Fig. B.1}).

\section{Multi-transition study of CS and SO}

\label{sect:AbEstim}
A precise knowledge of the gas physical conditions is required to determine accurate molecular abundances. For TMC 1, we adopted the gas kinetic temperatures and molecular hydrogen densities derived in Paper I. These values  are summarized in \hyperref[table:TMC1conditions]{Table A.1}, together with the CS column densities also derived in our previous work.

\begin{center}
	\begin{threeparttable}
	\caption{Measured transitions}
	\label{table:transitions}
	\centering
	\begin{tabular}{cccc}
		\toprule
		Species & Transition & Frequency (MHz) & Beam (arcsec) \\
		\midrule
		$^{13}$CS & $1\rightarrow 0$ & 46247.563 & 42$"$\\
		$^{13}$CS & $2\rightarrow 1$ & 92494.308 & 27$"$ \\
		$^{13}$CS & $3\rightarrow 2$ & 138739.335 & 18$"$ \\
		C$^{34}$S & $1\rightarrow 0$ & 48206.941 & 42$"$\\
		C$^{34}$S & $2\rightarrow 1$ & 96412.950 & 26$"$ \\
		C$^{34}$S & $3\rightarrow 2$ & 144617.101 & 17$"$ \\
		CS & $1\rightarrow 0$ & 48990.955 & 42$"$ \\
		CS & $2\rightarrow 1$ & 97980.953 & 25$"$ \\
		CS & $3\rightarrow 2$ & 146969.029 & 17$"$ \\
		H${_{2}}^{34}$S & $1_{1,0} \rightarrow 1_{0,1}$ & 167910.516 & 15$"$ \\
		H${_{2}}$S & $1_{1,0} \rightarrow 1_{0,1}$ & 168762.753 & 15$"$ \\
		SO & $2_{2}\rightarrow 1_{1}$ & 86093.96 & 29$"$ \\
		SO & $2_{3}\rightarrow 1_{2}$ & 99299.89 & 25$"$ \\
		SO & $3_{2}\rightarrow 2_{1}$ & 109252.18 & 23$"$ \\
		SO & $3_{4}\rightarrow 2_{3}$ & 138178.66 & 18$"$ \\
		SO & $4_{4}\rightarrow 3_{3}$ & 172181.42 & 14$"$ \\
		$^{34}$SO & $2_{3}\rightarrow 1_{2}$ & 97715.40 & 25$"$ \\
		$^{34}$SO & $3_{2}\rightarrow 2_{1}$ & 106743.37 & 23$"$ \\
		$^{34}$SO & $4_{4}\rightarrow 3_{3}$ & 168815.11 & 15$"$ \\
		\bottomrule
	\end{tabular}
%	\begin{tablenotes}{
%      		\small
%      		\item {\bf Notes:}}
%    \end{tablenotes}
\end{threeparttable}
\end{center}

We estimate the physical conditions towards Barnard 1b from a multi-transition analysis of CS and SO following the same procedure as in Paper I. We fit the intensities of the  observed CS, C$^{34}$S, and $^{13}$CS J $=1\rightarrow 0$, $2 \rightarrow 1$, and $3 \rightarrow 2$ lines using the molecular excitation and radiative transfer code RADEX \citep{vanderTak2007}. During the fitting process, the isotopic ratios are fixed to $^{12}$C/$^{13}$C = 60 and $^{32}$S/$^{34}$S = 22.5 \citep{Gratier2016}, and we assume a beam filling factor of 1 for all transitions (the emission is more extended than the beam size). Then, we let $T_{\rm k}$, $n_{{\rm H}_{2}}$, and $N({\rm CS})$ vary as free parameters. The parameter space ($T_{\rm k}$, $n_{{\rm H}_{2}}$, $N({\rm CS})$) is then explored following the Monte Carlo Markov Chain (MCMC) methodology with a Bayesian inference approach, as described in \citet{Riviere2019}. In particular, we use the \emph{emcee} \citep{Foreman2012} implementation of the Invariant MCMC Ensemble sampler methods by \citet{Goodman2010}. While $n_{{\rm H}_{2}}$ and  $N({\rm CS})$ are allowed to vary freely, we need to introduce a prior distribution to limit the gas kinetic temperatures to reasonable values in this cold region and, hence, break the temperature-density degeneracy that is usual in this kind of calculations. As estimated by \citet{Friesen2017}, the gas kinetic temperature in a wide sample of molecular clouds, based on the NH$_{3}$ (1,1) and (2,2) inversion lines, is found to be systematically 1 or 2 K lower than that obtained from Hershel maps. This is indeed corroborated in Section 5. Thus, we set a Gaussian prior distribution with mean $\mu = T_{\rm d}$ and $\sigma = 2$ K for the gas kinetic temperature. The collisional rate coefficients for the molecular excitation calculations that involve $^{13}$CS and C$^{34}$S isotopologues are those of CS. They include collisions with para and ortho-H$_{2}$ as reported by \citet{Denis2018}, with a ortho-to-para ratio of 3, and He, taken from \citet{Lique2006}. The gas temperature and the density derived for Barnard 1b from the multi-line fitting of CS and its isotopologues are shown in \hyperref[table:B1Bconditions]{Table B.1}.

A similar multi-transition analysis is carried out to derive SO column densities. In this case we have used the collisional coefficients derived by  \citet{Lique2007}. The derived molecular gas densities and SO column densities are shown in \hyperref[table:B1BconditionsSO]{Table B.2}.  In most positions, the densities derived from the SO fitting fully agree with those derived from CS. However, in the vicinity of the visual extinction peak  ($A_{\rm v}> 30$ mag), the molecular hydrogen densities derived from SO data are systematically higher than those derived with CS.  \citet{Daniel2013} derived the density structure of the core by fitting the CSO 350 $\mu$m continuum map and the IRAM 1.2mm image with a power-law radial density profile, and assuming standard values of the dust-to-gas mass ratio and dust opacity. The densities derived from the continuum images fitting are consistent with those derived from our SO observations and a factor $5-10$ higher than those derived from CS. Therefore, we adopt the densities derived from SO for the inner Barnard 1b core. The lower densities derived from CS are very likely the consequence of a high depletion of this molecule in the densest part of the core.

It is interesting to compare the SO column density obtained in this paper towards Barnard 1b with the S$^{18}$O column density derived by \citet{Fuente2016}. Comparing both values, we obtain $N({\rm SO})/N({\rm S}^{18}{\rm O}) = 78 \pm 45$, much lower than the isotopic ratio, $^{16}$O/$^{18}$O= 550. Recent chemical calculations \citep{Loison2019} using the gas-grain chemical code Nautilus \citep{Ruaud2016} in typical conditions for dark molecular clouds, showed that the isotopic fractionation might be important in the case of SO, which yields a  $^{16}$O/$^{18}$O ratio between 500 and $~80$. Following their predictions, the $N({\rm SO})/N({\rm S}^{18}{\rm O})$ ratio is expected to vary with time with changes of more than a factor of 5. The ratio measured in Barnard 1b would be consistent with a chemical age of a few 0.1 Myr.

\section{Gas kinetic temperatures from NH\texorpdfstring{$_3$}{} data}
The low densities found in the translucent phase, with $n_{\rm H}$ of the order of a few 10$^3$ cm$^{-3}$, might cast some doubts on our assumption that gas and dust are close to thermal equilibrium. To check that this is indeed the case, we have independently derived the gas kinetic temperature using the NH$_3$ (1,1) and (2,2) inversion lines as observed with the 40m Yebes telescope towards the positions with $A_{\rm v} < 20$ mag in Barnard 1b. To derive the gas kinetic temperatures, first we follow the method described by \citet{Bachiller1987} to derive the rotation temperature $T_{\rm R}$(NH$_{3}$), and then we use the prescription which relates the gas kinetic temperature $T_{\rm k}$(NH$_{3}$) with $T_{\rm R}$(NH$_{3}$) found by \citet{Swift2005}. Gas kinetic temperatures thus derived from ammonia data (\hyperref[tab:ammonia]{Table 2}) fully agree within the error with those derived from the CS multi-transition study (Table~A.1). Previous measurements of the gas kinetic temperature \citep{Lis2002} are in agreement with our values. Thus, we adopt the gas kinetic temperatures derived from CS for our further analysis. It should be noted that we have not observed ammonia lines towards the $A_{\rm v} > 20$ mag positions precluding an independent gas kinetic temperature estimate. However, taking into account the high molecular hydrogen densities measured in these regions, our assumption of gas and dust being close to thermal equilibrium is a plausible approximation. We have not detected the NH$_3$ lines towards the positions with $A_{\rm v} < 10$ mag.

\begin{table}
	\centering
		\caption{Kinetic temperatures, $T_{\rm k}$(CS), in Kelvin, of several $A_{\rm v} < 20$ mag positions in Barnard 1b, compared to the estimated rotation temperature $T_{\rm R}$(NH$_{3}$) and the corresponding kinetic temperature $T_{\rm k}$(NH$_{3}$) from the ammonia (1,1) and (2,2) inversion lines.}
			\begin{tabular}{c|ccc}
				\toprule
				Offset (",")\  & \ T$_{\rm k}$(CS) (K) & T$_{\rm R}$(NH$_{3}$) (K) & T$_{k}$(NH$_{3}$) (K) \\ \midrule 
					(+80", 0") & $13.2\pm 1.8$ & $11.5\pm 1.5$ & $12.0\pm 2.1$\\ 
					(+110", 0") & $14.4\pm 1.9$ & $11.6\pm 1.8$ & $12.1\pm 2.4$\\ 
					(+140", 0") & $14.2\pm 1.0$ &	$12.7\pm 1.9$ & $13.4\pm 2.6$\\ 
    			\bottomrule
			\end{tabular}
			\label{tab:ammonia}
\end{table}

\section{H\texorpdfstring{$_2$}{}S abundance}

\begin{figure}
  	\includegraphics[width=0.49\textwidth,keepaspectratio]{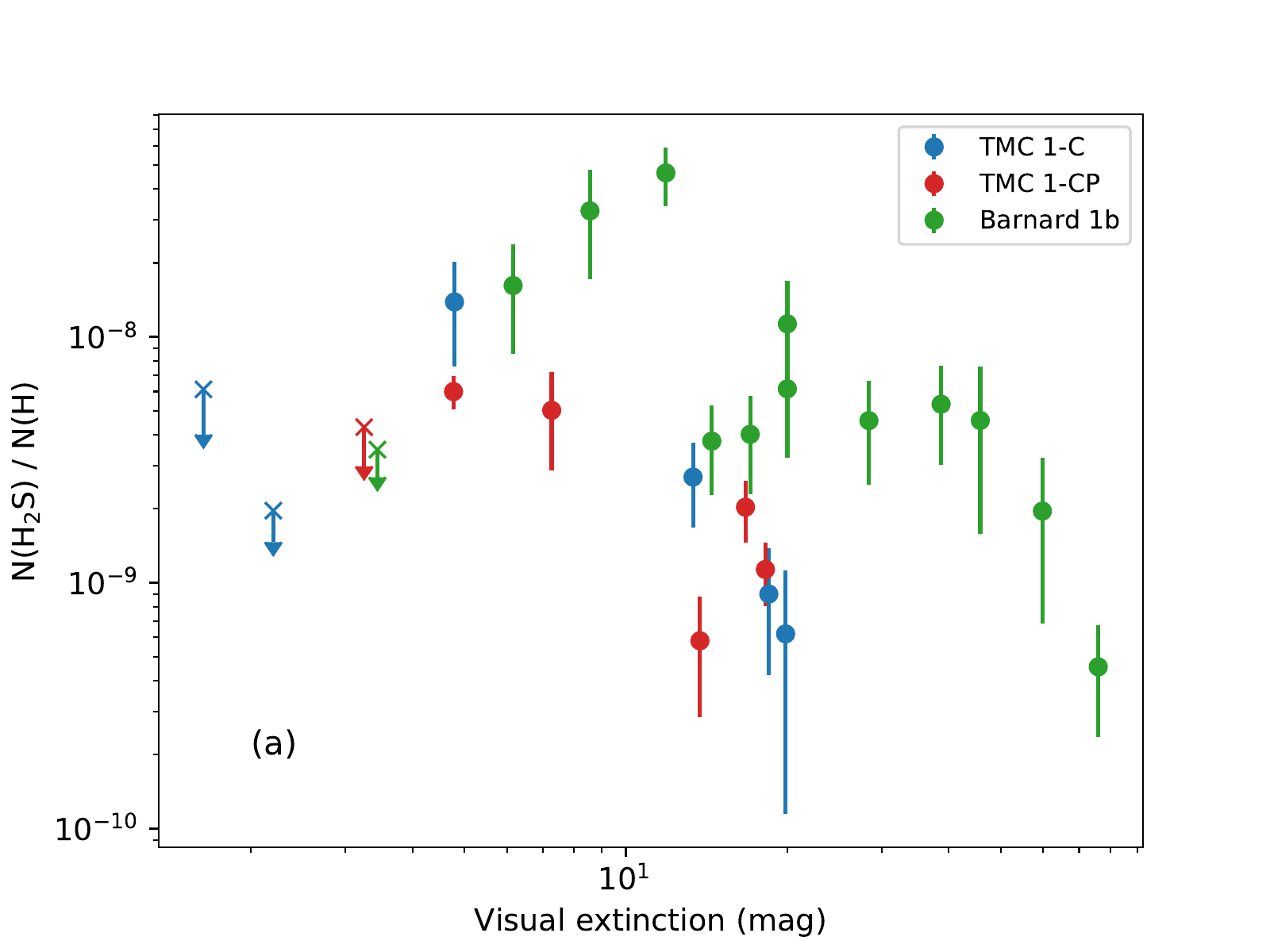}
%  	\label{fig:H2S-Dens}
 % 	\caption{Observed H$_{2}$S abundance as a function of the H$_{2}$ density in TMC1 cores TMC1-C (blue), TMC1-CP (red), and B1b (green).}
%\end{figure}
 
%\begin{figure}
 % 	\includegraphics[width=0.5\textwidth,keepaspectratio]{SqrtBehavior}
 % 	\label{fig:nSqrtT}
 % 	\caption{Observed H$_{2}$S gas-phase abundance, plotted against $n_{\rm gr}\sqrt{T}$ in log scale, showing the sticking behavior described in Eq. \eqref{eq:timescale}.}
%\end{figure}

%\begin{figure}
  	\includegraphics[width=0.49\textwidth,keepaspectratio]{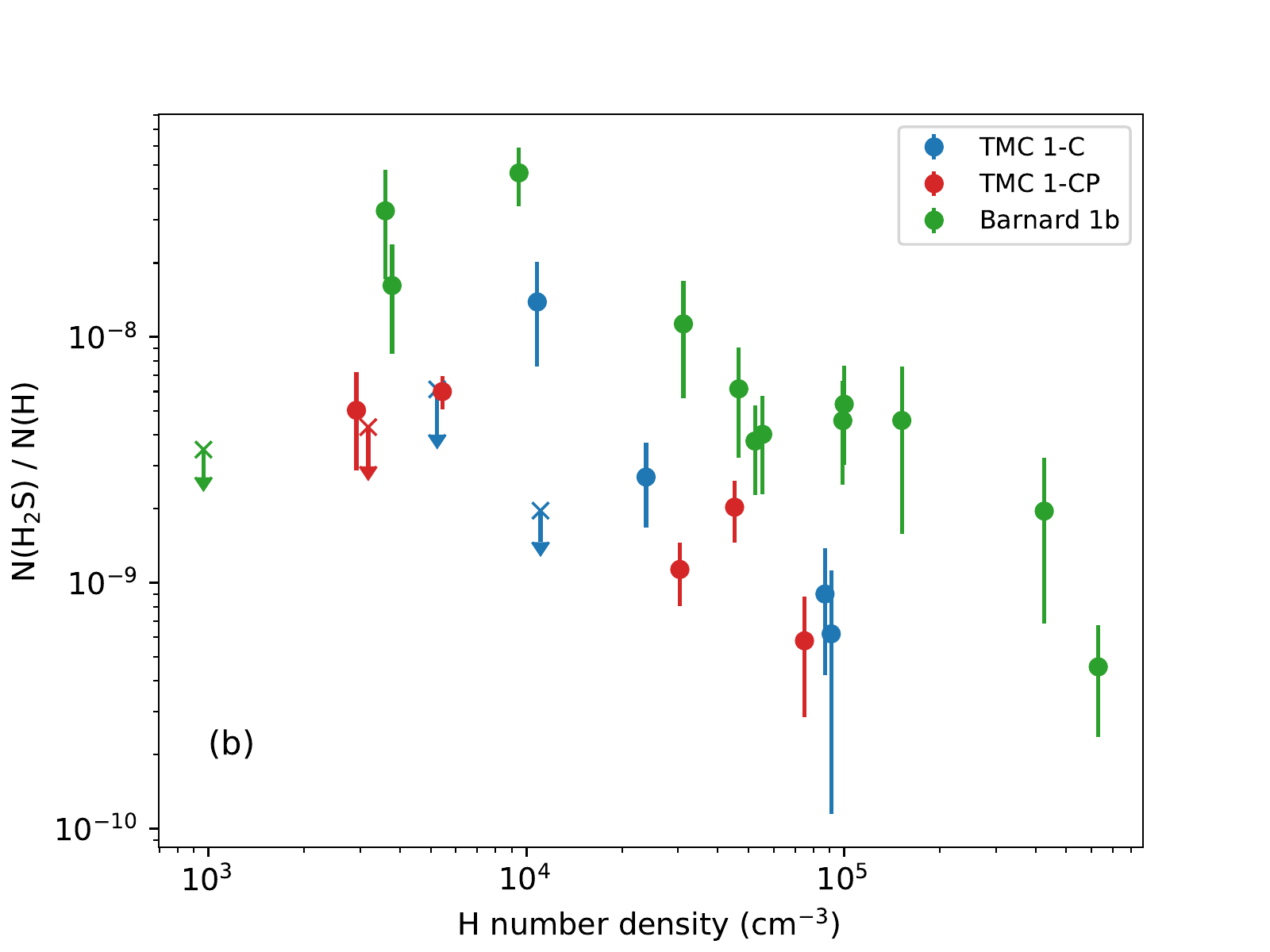}
%  	\label{fig:H2Smag}
 % 	\caption{Observed H$_{2}$S gas-phase abundance, plotted against the visual magnitude in TMC1 cores TMC1-C (blue), TMC1-CP (red), and B1b (green).}
%\end{figure}

%\begin{figure}
  	\includegraphics[width=0.49\textwidth,keepaspectratio]{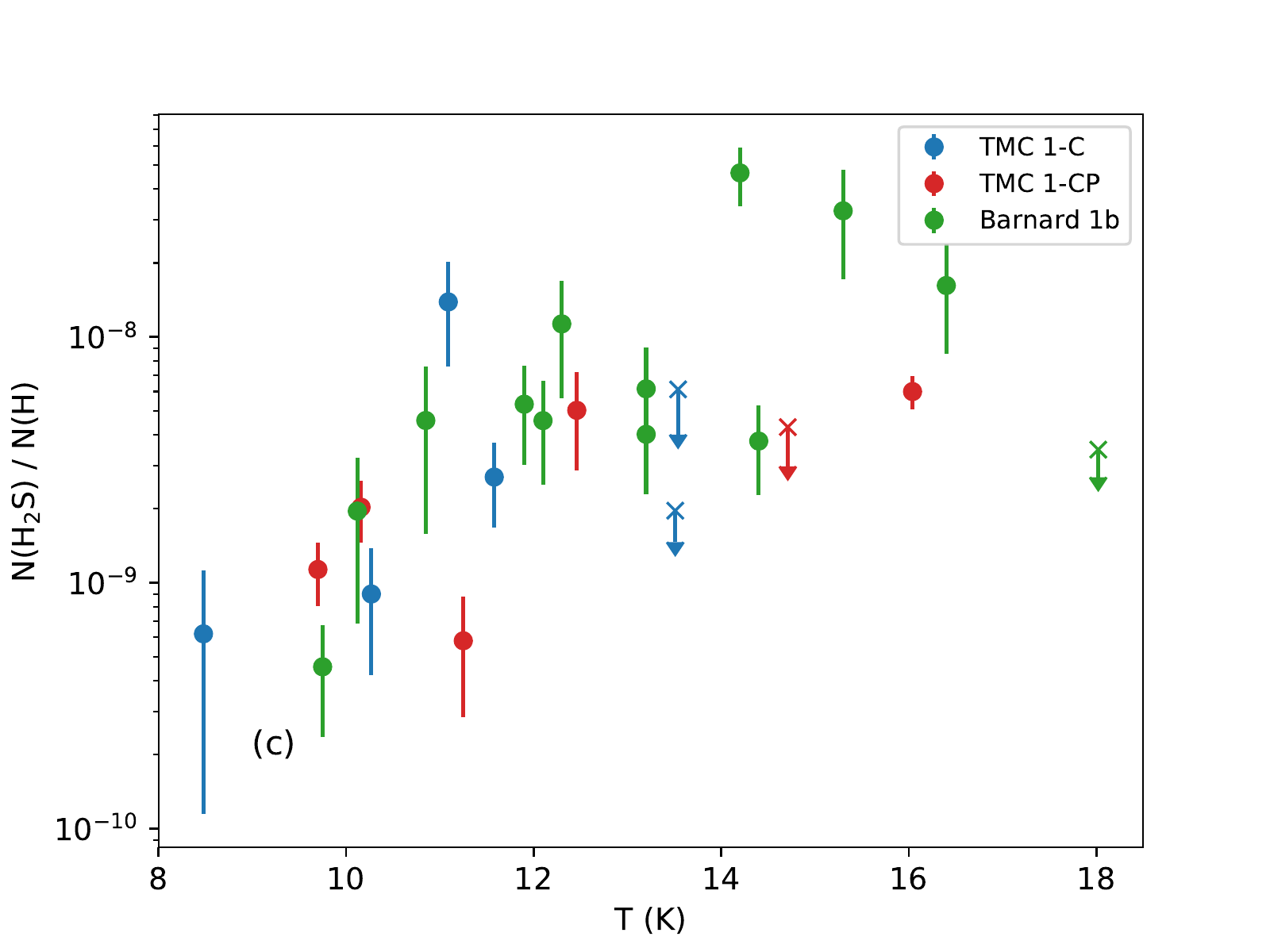}
 	\label{fig:H2S}
  	\caption{In (a), the observed H$_{2}$S abundance is plotted as a function of visual extinction in TMC 1-C (blue), TMC 1-CP (red), and Barnard 1b (green). In a similar fashion, in (b) the observed H$_{2}$S gas-phase abundance against the H number density in TMC 1 and Barnard 1b cores is shown. Crosses represent upper bound values. Finally, (c) displays the H$_{2}$S abundance plotted against the gas kinetic temperature.}
\end{figure}

The gas physical conditions derived in previous sections are used to compute the H$_2$S column densities. First, we estimate the ortho-H$_{2}$S column density using  RADEX with the ortho-H$_2$O collisional coefficients calculated by \citet{Dubernet2009}, scaled to ortho-H$_2$S. The H$_2$S abundance is then calculated adopting an ortho-to-para ratio of 3. The resulting molecular abundances are listed in \hyperref[table:TMC1conditions]{Table A.1} and \hyperref[table:B1Bconditions]{Table B.1}, and shown in \hyperref[fig:H2S]{Figs. 2a-c}. 

In these plots, we try to find correlations between the H$_{2}$S abundance and physical quantities like visual extinction, gas temperature, and H density. Here, hydrogen nuclei number density is twice the $n_{\rm H_{2}}$ density obtained with the MCMC approach described earlier. As seen in such plots, the relationship between the H$_{2}$S abundance and the physical parameters is mostly monotonic. To assess the degree of correlation, we compute the Kendall's tau coefficient \citep{Kendall1938} in each case: 1 for perfect correlation, 0 means no correlation, and -1 stands for perfect anti-correlation. \hyperref[fig:H2S]{Fig. 2a} shows the estimated H$_2$S abundances as a function of the visual extinction towards the observed positions. They are loosely anti-correlated (Kendall test results: $\tau_{\rm K}=-0.3$, p-value $=0.06$), as the H$_2$S abundance is, in general, at its minimum towards the visual extinction peak, with values of the order of 10$^{-10}$, and increases towards lower visual extinctions (\hyperref[fig:H2S]{Fig. 2a}). In TMC 1, the maximum H$_2$S abundance is measured at $A_{\rm v}\sim 5$ mag, with values of $\sim 2\times 10^{-8}$. In Barnard 1b, the H$_2$S abundance peaks at  $A_{\rm v} \sim 10$ mag with values of $\sim 5\times 10^{-8}$ and then, decreases again, reaching values of $\sim 2\times10^{-8}$ at $A_{\rm v} \sim 5$ mag. We do not detect H$_2$S when $A_{\rm v} < 5$ mag (\hyperref[fig:H2S]{Fig. 2a}).

\begin{figure}
  	\includegraphics[width=0.49\textwidth,keepaspectratio]{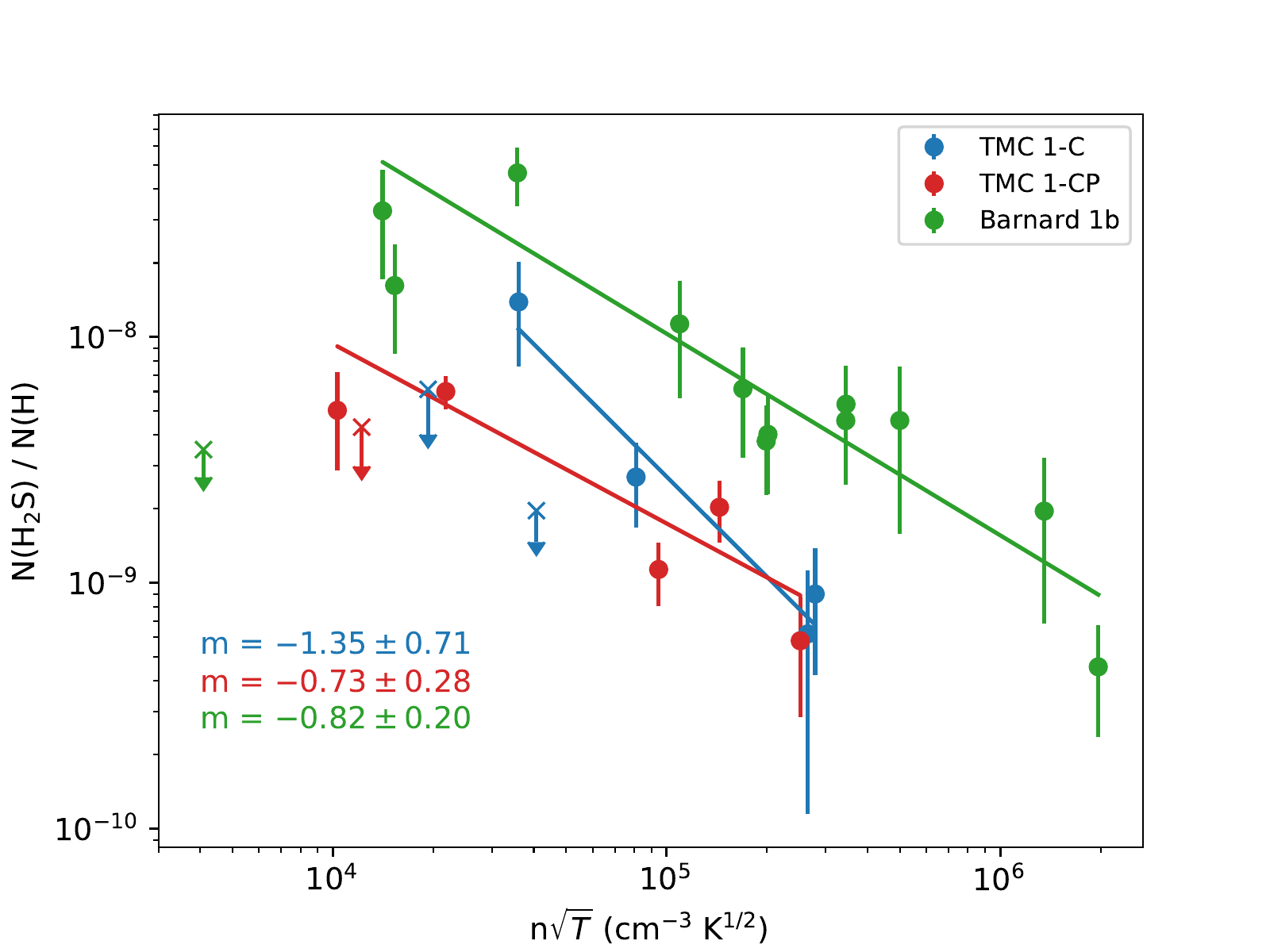}
  	\label{fig:nSqrtT}
  	\caption{Observed H$_{2}$S gas-phase abundance plotted against $n_{\rm gas}\sqrt{T}$ (cm$^{-3}$ K$^{1/2}$) in log scale, showing the sticking behavior described by Eq. \eqref{eq:timescale}. Crosses mark upper bound values, which were not taken into account in this analysis.}
\end{figure}

The H$_{2}$S gas-phase abundance is moderately anti-correlated with the hydrogen nuclei density ($\tau_{\rm K}=-0.4$, p-value $<10^{-2}$). This kind of anti-correlation is expected when the freeze-out of molecules on grain surfaces is determining the molecular abundance. The probability of collisions between gas and grains is proportional to the gas density and the thermal velocity. Assuming that the sticking efficiency is $\sim 1$, that is, the H$_{2}$S molecules stick on grains in every collision, we can define a depletion timescale given by the inverse of this probability such as

\begin{equation}
	\label{eq:timescale}
	X({\rm H}_{2}{\rm S}) \propto t_{\rm dep} \propto \frac{1}{n_{\rm gas}\sigma_{\rm gr}v} \propto \frac{1}{n_{\rm gas}\sqrt{T_{\rm kin}}},
\end{equation}
where $X({\rm H}_{2}{\rm S})$ is the abundance of ${\rm H}_{2}{\rm S}$ respect to H nuclei, $n_{\rm gas}$ is the density of the gas, $\sigma_{\rm gr}$ is the grain cross-section, and $v$ is the thermal velocity of the species, related to the kinetic temperature $T_{\rm kin}$. In \hyperref[fig:nSqrtT]{Fig. 3} we plot the H$_{2}$S gas-phase abundance against $n_{\rm gas}\sqrt{T_{\rm kin}}$ . The H$_2$S abundance decreases with $n_{\rm gas}\sqrt{T_{\rm kin}}$ following a power-law with index $m \approx -1$ in the three studied cores, which confirms this behavior (we do not include upper bound values in this analysis). Interestingly, the abundance of H$_2$S in Barnard 1b is more than a factor of $\sim 3$ higher than towards TMC 1 for each density, suggesting a higher formation rate or lower destruction rate in Barnard 1b. In \hyperref[fig:H2S]{Fig. 2c}, we plot the H$_{2}$S gas-phase abundance as a function of the kinetic temperature. The H$_2$S abundance is strongly correlated with the gas kinetic temperature (Kendall test gave $\tau_{\rm K} = 0.6$ and p-value $p<10^{-3}$) as long as $A_{\rm v} > 5$ mag, which translates into a lower abundance towards the cooler quiescent region TMC 1 compared with slightly warmer Barnard 1b core that is located in an active star forming region. In the following we determine the physical structure and model the chemistry of these two regions.

\section{Physical structure: core density profiles}
\label{sec:phystruct}

\begin{figure*}
	\centering
		\begin{subfigure}[b]{0.49\textwidth}\includegraphics[width=\textwidth,keepaspectratio]{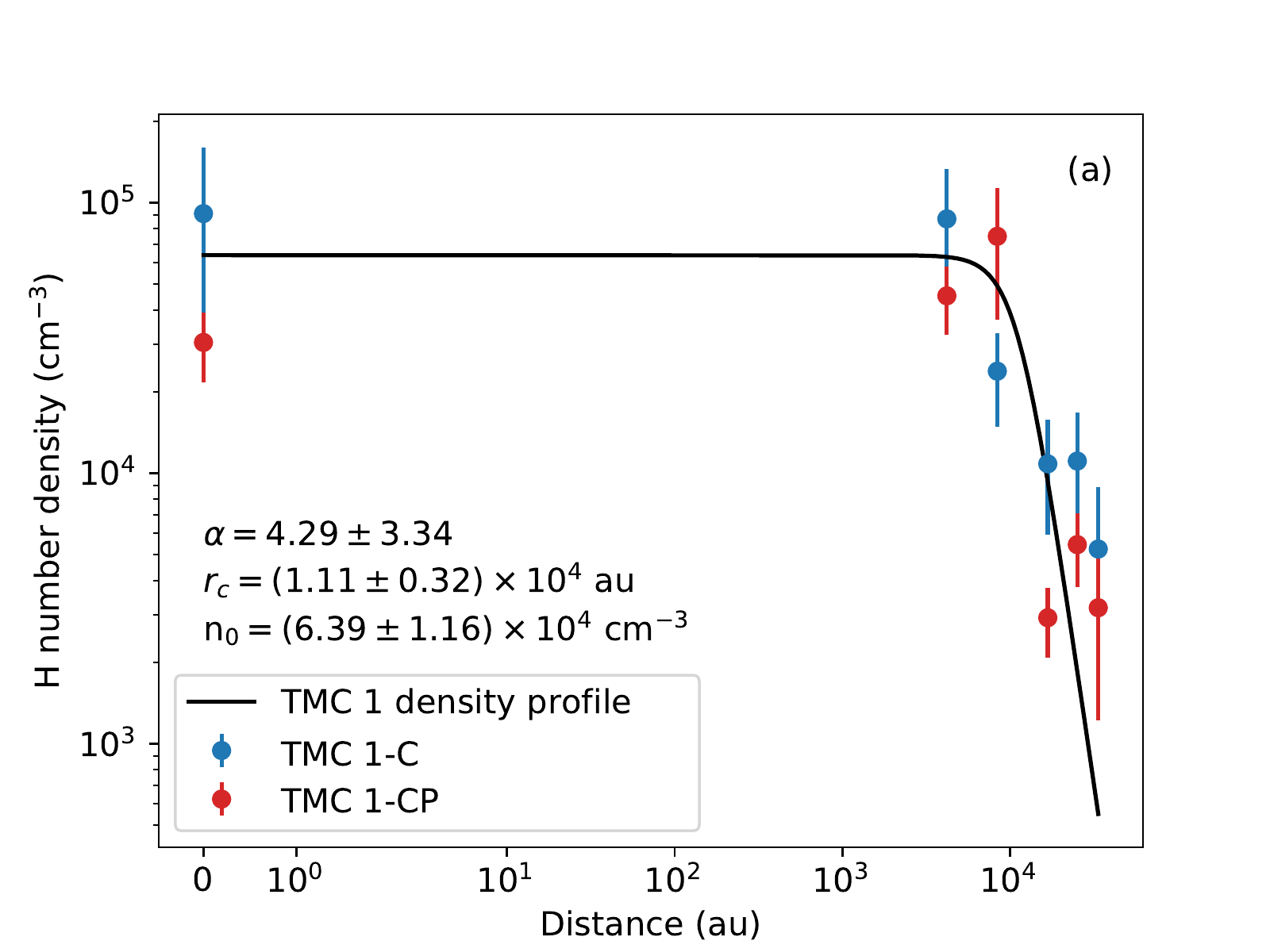}
			\label{fig:TMC1BE}
		\end{subfigure}
		~
		\begin{subfigure}[b]{0.49\textwidth}\includegraphics[width=\textwidth,keepaspectratio]{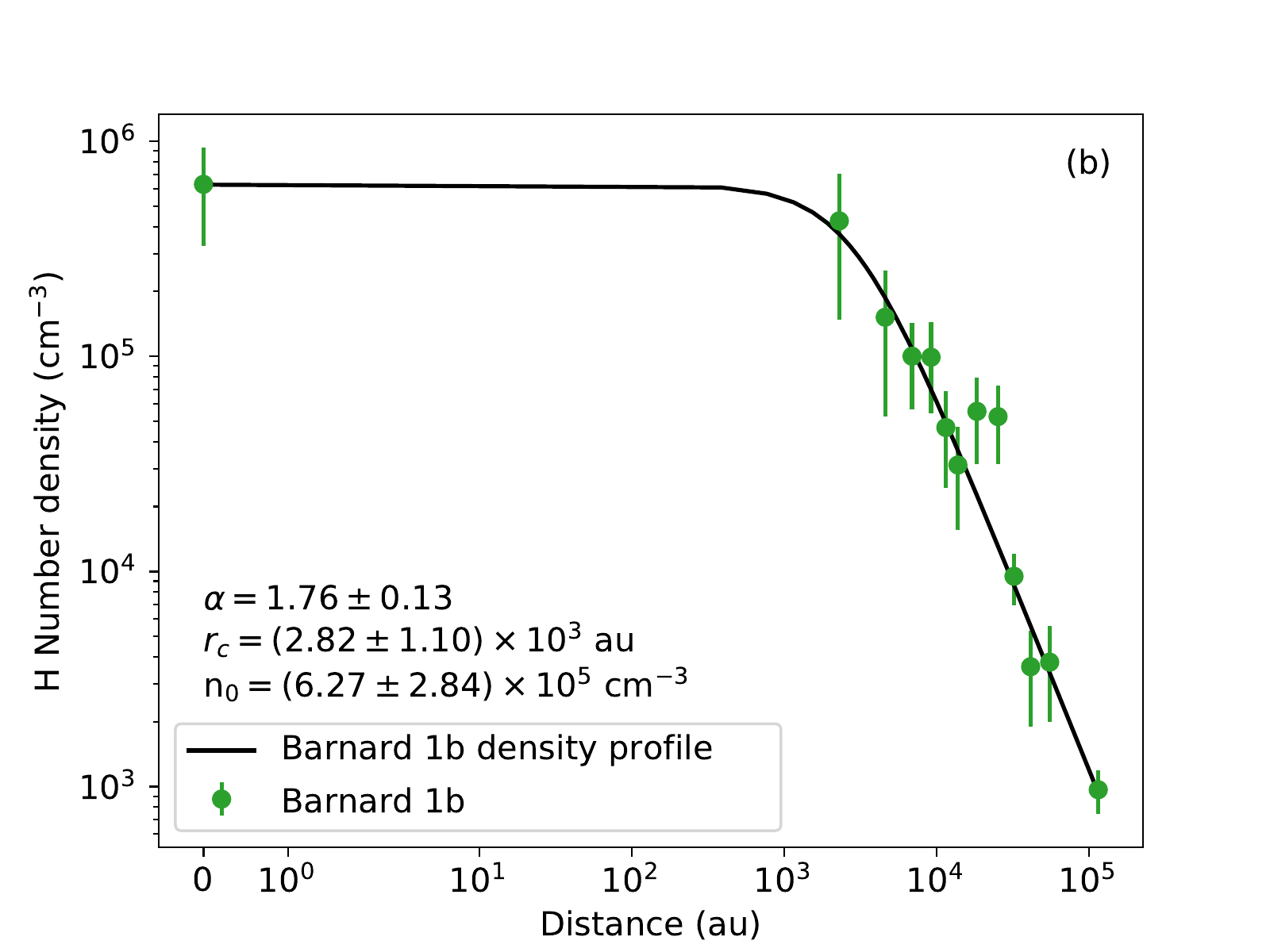}
			\label{fig:B1BBE}
		\end{subfigure}
		\caption{H number density profiles of TMC 1-C and TMC 1-CP (a) and Barnard 1b (b) as the result of fitting the data in \hyperref[table:TMC1conditions]{Table A.1.} and \hyperref[table:B1Bconditions]{Table B.1.}, respectively, to the Equation \eqref{eq:BEprofile}.}
\end{figure*}

The densities at the observed positions in TMC 1 and Barnard 1b (those marked in \hyperref[fig:mapTMC1]{Fig. 1a} and \hyperref[fig:mapB1b]{Fig. 1b}) were derived using the procedure described in Section 4. To fully characterize the density structure of the cores along the line of sight, we consider the simplest case of spherically symmetric, thermally supported, and gravitationally bound clouds. The solution for such a cloud is the well-known Bonnor-Ebert sphere (BE, hereafter). We assume that the radial density structure of the filament is that of a BE, which has been widely parameterized by the approximate analytical profile \citep[see e.g.,][]{PriestleyViti2018, Tafalla_2002}:

\begin{equation}
	\label{eq:BEprofile}
	n_{\rm H}(r) = \frac{n_{0}}{1+\left(\frac{r}{r_{c}}\right)^{\alpha}},
\end{equation}
where $r$ is the distance from the origin, that is , either TMC 1-C, TMC 1-CP, or Barnard 1b, ${\rm n}_{0}$ is the hydrogen nuclei number density at the origin $r = 0$, $r_{c}$ is the flat radius of the BE, and $\alpha$ is the asymptotic power index. At high distances, density falls as $\propto r^{-\alpha}$. As \hyperref[fig:TMC1BE]{Fig. 4a} and \hyperref[fig:B1BBE]{Fig. 4b} show, the family of curves defined by \eqref{eq:BEprofile} is in good agreement with the observed values (to include the origin in a logarithmic scale, we use the symmetrical logarithmic scale from Matplotlib Python library, which by default uses a linear scale between 0 and 1 and the logarithmic scale for the rest of the data range). The central density, $n_{0}$, ranges between $(5.23-7.55)\times 10^{4}$ cm$^{-3}$ in TMC 1 and $(3.43-9.11)\times 10^{5}$ cm$^{-3}$ in Barnard 1b. The value of $r_{c}$, the flat radius of the BE sphere, is found to be $(0.28\pm 0.11)\times 10^{4}$ au and $(1.11\pm 0.32)\times 10^{4}$ au in Barnard 1b and TMC 1, respectively. Finally, the asymptotic power index $\alpha$ ranges from 1.76 in the case of Barnard 1b to 4.29 in TMC 1. In a simple parametrization of a BE collapse, a larger flat radius corresponds to a younger, less evolved BE \citep{Aikawa2005, PriestleyViti2018}. This is in agreement with our parameterizations, since TMC 1-C and TMC 1-CP, starless cores, have larger flat radii than Barnard 1b, which hosts a first hydrostatic core (FHSC for short) and is, therefore, a more evolved core than TMC 1-C and TMC 1-CP. Similarly, the higher asymptotic power index $\alpha$ of TMC 1 is indeed a feature of younger BEs, according to the parameterizations derived by \citet{PriestleyViti2018}.\\

\begin{figure*}%
	\centering
		\begin{subfigure}[b]{0.49\textwidth}\includegraphics[width=\textwidth,keepaspectratio]{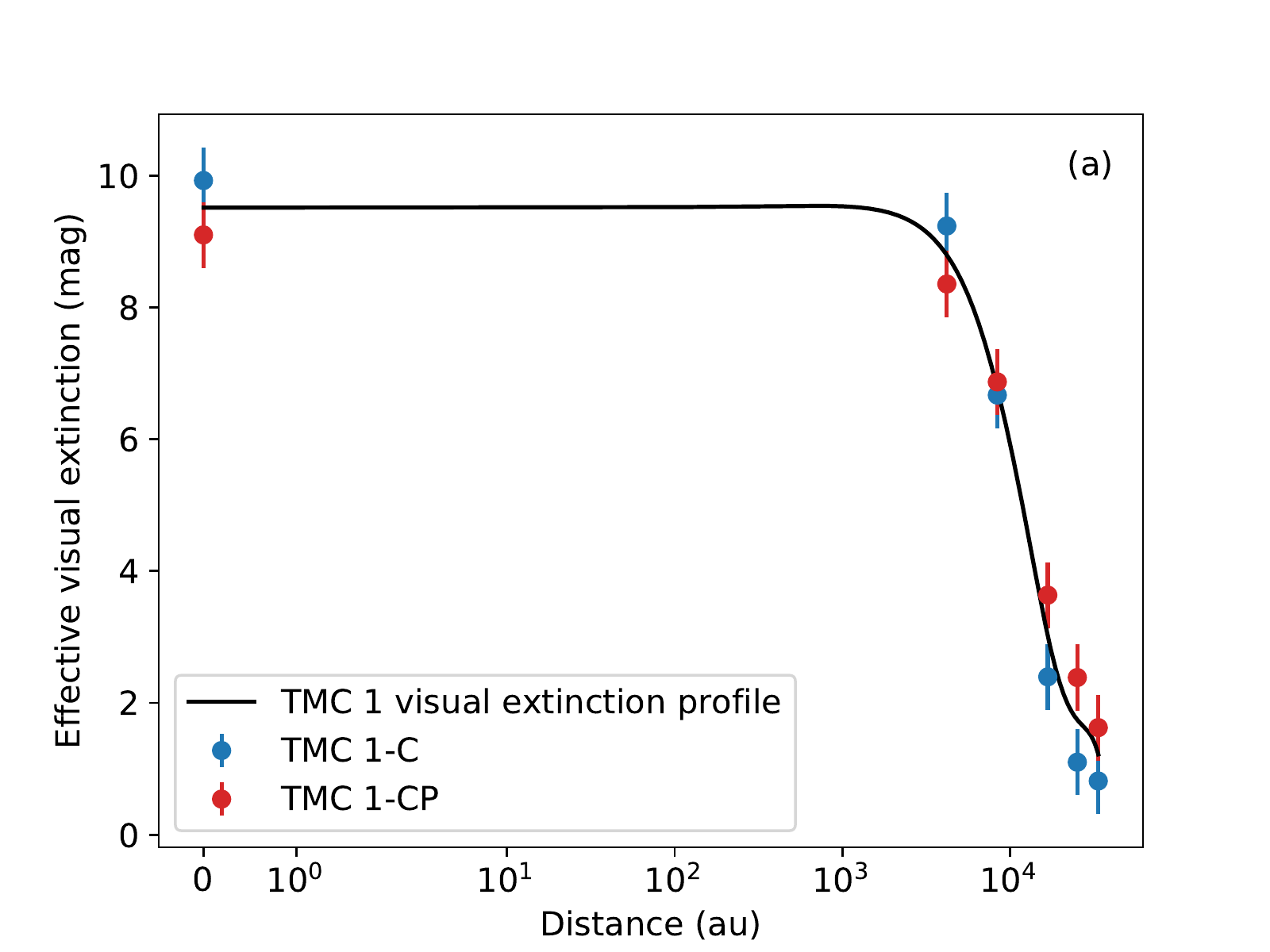}
			\label{fig:TMC1Av}
		\end{subfigure}
		~
		\begin{subfigure}[b]{0.49\textwidth}\includegraphics[width=\textwidth,keepaspectratio]{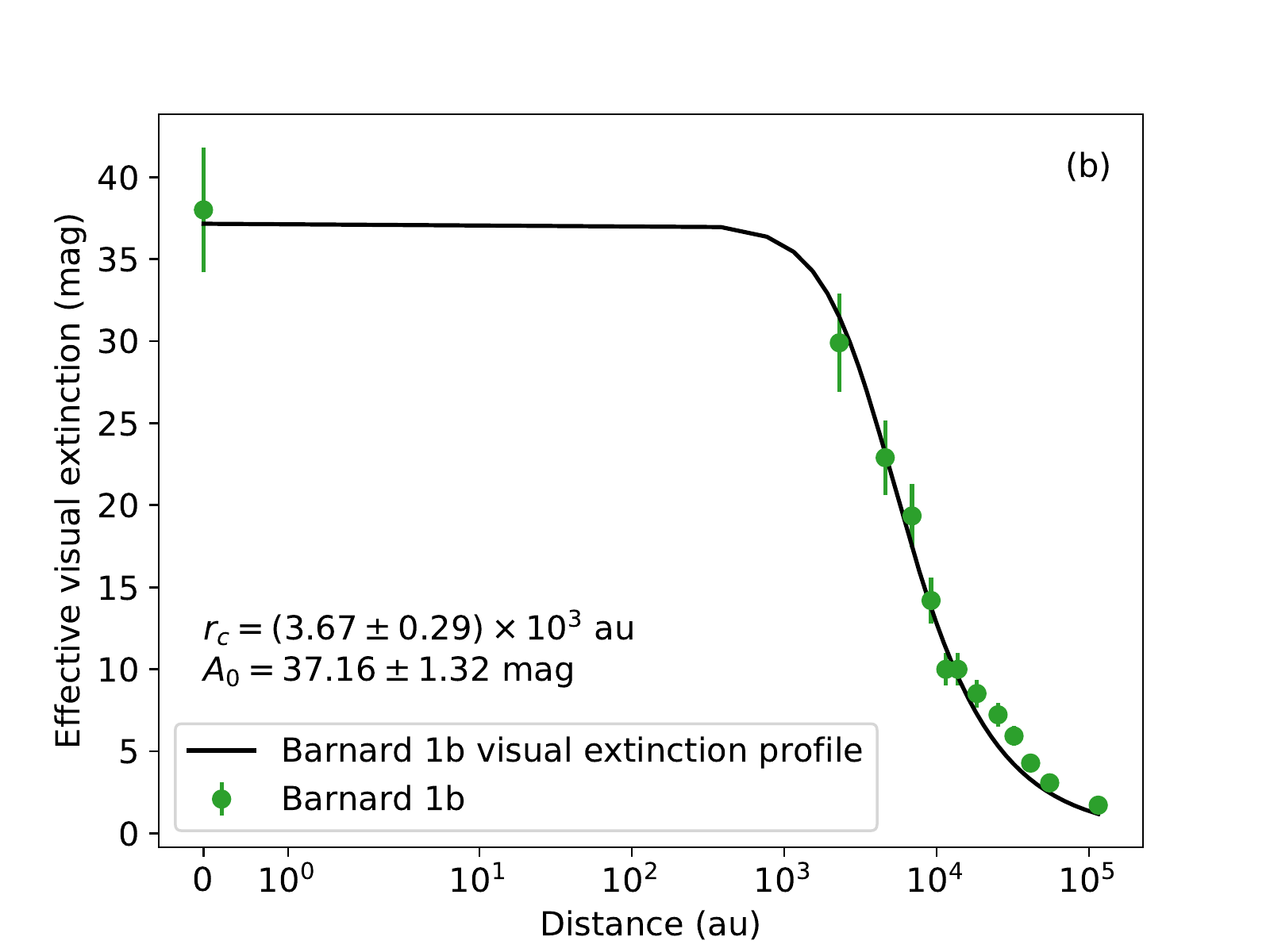}
			\label{fig:B1BAv}
		\end{subfigure}
		\caption{Effective visual extinction in magnitudes of TMC 1-C and TMC 1-CP (a) and Barnard 1b (b) as the result of fitting the data in \hyperref[table:TMC1conditions]{Table A.1.} and \hyperref[table:B1Bconditions]{Table B.1.}, respectively, to the Equation \eqref{eq:av_profile}.}
\end{figure*}

\begin{figure*}
	\centering
		\begin{subfigure}[b]{0.49\textwidth}\includegraphics[width=\textwidth,keepaspectratio]{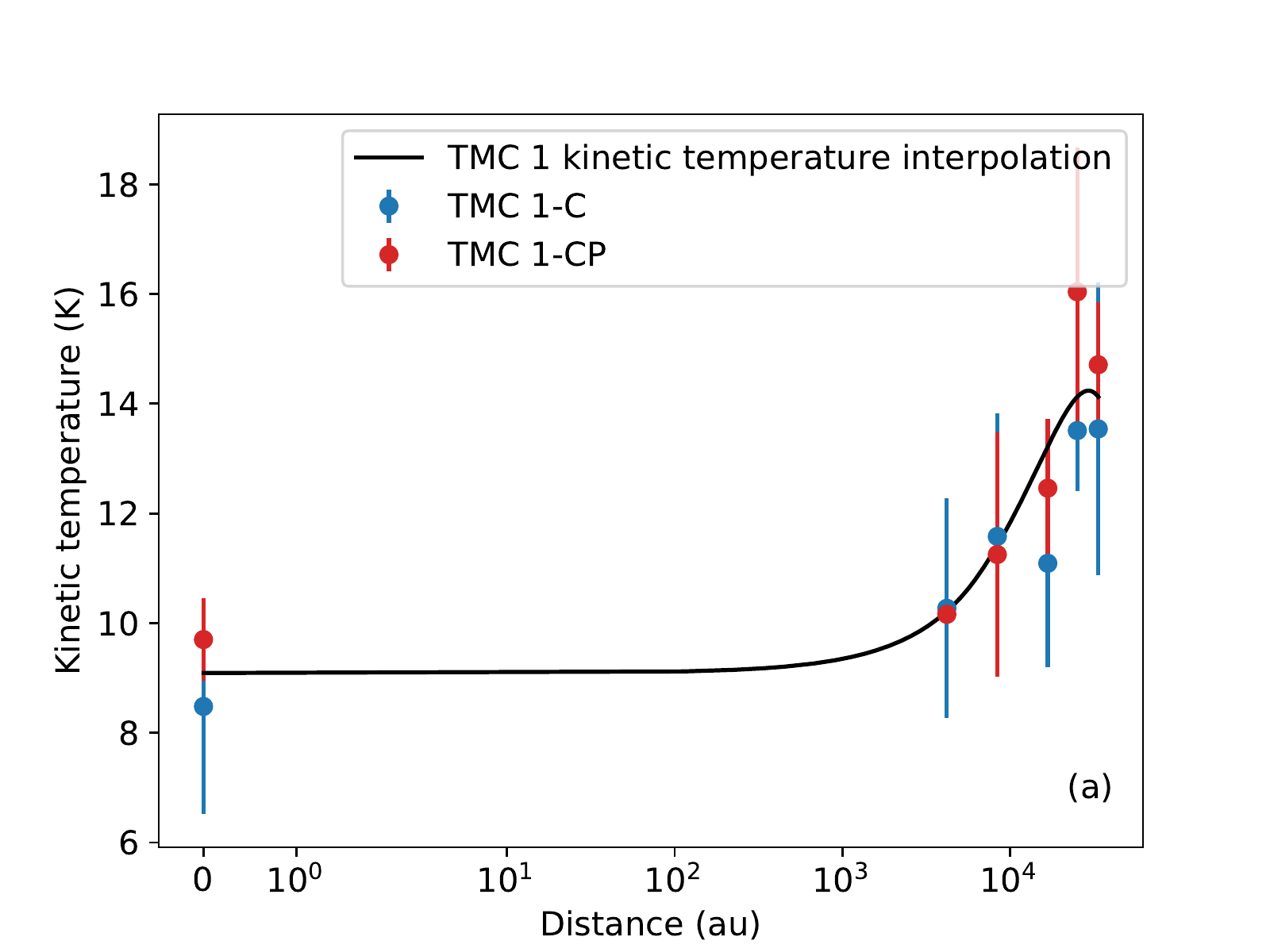}
			\label{fig:TMC1Tkin}
			\end{subfigure}
		~
		\begin{subfigure}[b]{0.49\textwidth}\includegraphics[width=\textwidth,keepaspectratio]{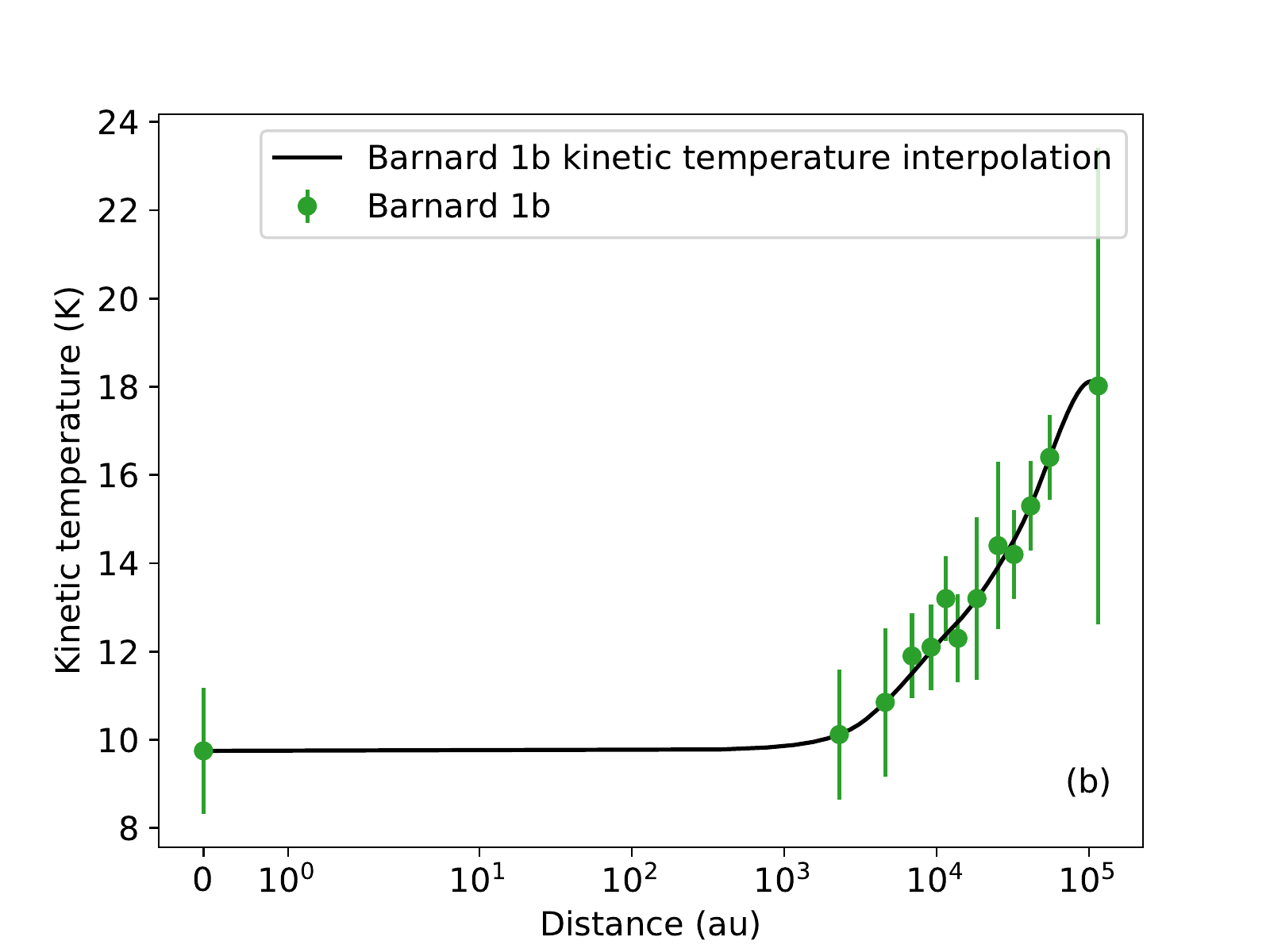}
			\label{fig:B1BTkin}
		\end{subfigure}
		\caption{Kinetic temperature in the observed positions of TMC 1-C and TMC 1-CP (a) and Barnard 1b (b) and the quadratic spline interpolation of the error-weighted average at each position (solid black line).}
\end{figure*}

The visual extinction along the line of sight of the selected positions in TMC 1 and Barnard 1b cores is taken from the visual extinction maps in \hyperref[fig:mapTMC1]{Fig. 1a} and \hyperref[fig:mapB1b]{Fig. 1b}. A parameter that will be needed in the chemical modeling of the cores (Section 9) is the shielding from the external UV field at each point. Assuming spherical symmetry and isotropic UV illumination, each point inside the cloud is shielded from the external UV field by an extinction that is, approximately, half of that measured in the extinction maps. We thus define an effective visual extinction $A_{{\rm v}\,\rm eff}$ as half of the visual extinction measured in the extinction maps: $A_{{\rm v}\,\rm eff}(r) = A_{{\rm v}}(r)/2$. In \hyperref[fig:TMC1Av]{Figs. 5a-b}, the effective visual extinction is plotted as a function of the distance towards the origin. The effective visual extinction across TMC 1 is interpolated using a cubic spline passing through the average of the effective extinction at each position (\hyperref[fig:TMC1Av]{Fig. 5a}). In Barnard 1b, we fit the effective visual extinction with the following family of functions: 

\begin{equation}
	\label{eq:av_profile}
	A_{{\rm v}\,\rm eff}(r) = \frac{A_{0\,\rm eff}}{\sqrt{1+r/r_{\rm c}}},
\end{equation}
where $r$ is the distance from the origin, $A_{0\,\rm eff}$ is the effective visual extinction at the origin, and $r_{\rm c}$ is the flat radius of the profile, the distance at which effective visual extinction starts falling significantly. This kind of profile is well-suited for BEs with an asymptotic power index $\alpha\sim 2$ \citep[see][]{DappBasu2018}. The flat radius of Barnard 1b is found to be in agreement with the one previously estimated via BE fitting. Finally, as shown in \hyperref[fig:TMC1Tkin]{Fig. 6a} and \hyperref[fig:B1BTkin]{Fig. 6b}, we parameterize the gas temperature of the observed positions.

\subsection*{Incident UV field in TMC 1 and Barnard 1b}

\label{sect:uvfield}

The incident UV field is a key parameter in the photodesorption of H$_2$S. The increase of the grain temperature at the cloud borders is only understood as the consequence of dust heating by the interstellar radiation field (ISRF). However, to our knowledge, it is not straightforward to derive the incident UV field from the dust temperature. On the one hand, the exact relation between dust temperature and the ISRF is dependent on the poorly known grain composition and its detailed variation across the cloud. On the other hand, in our part of the Galaxy, the dust heating is dominated by the visible part of the ISRF, and the visible and IR parts of the ISRF do not scale in a simple way with the UV part. Remaining aware of all these problems, in Paper I we obtain a first guess of the local UV flux in TMC 1 using the analytical expression by \citet{Hocuk2017}. This expression relates the dust temperature of a region with a given visual extinction and the incident UV field in units of the Draine field \citep{Draine1978}. In this paper, we use the same procedure to estimate the incident UV radiation in Barnard 1b.

\begin{figure*}
	\centering
		\begin{subfigure}[b]{0.48\textwidth}\includegraphics[width=\textwidth,keepaspectratio]{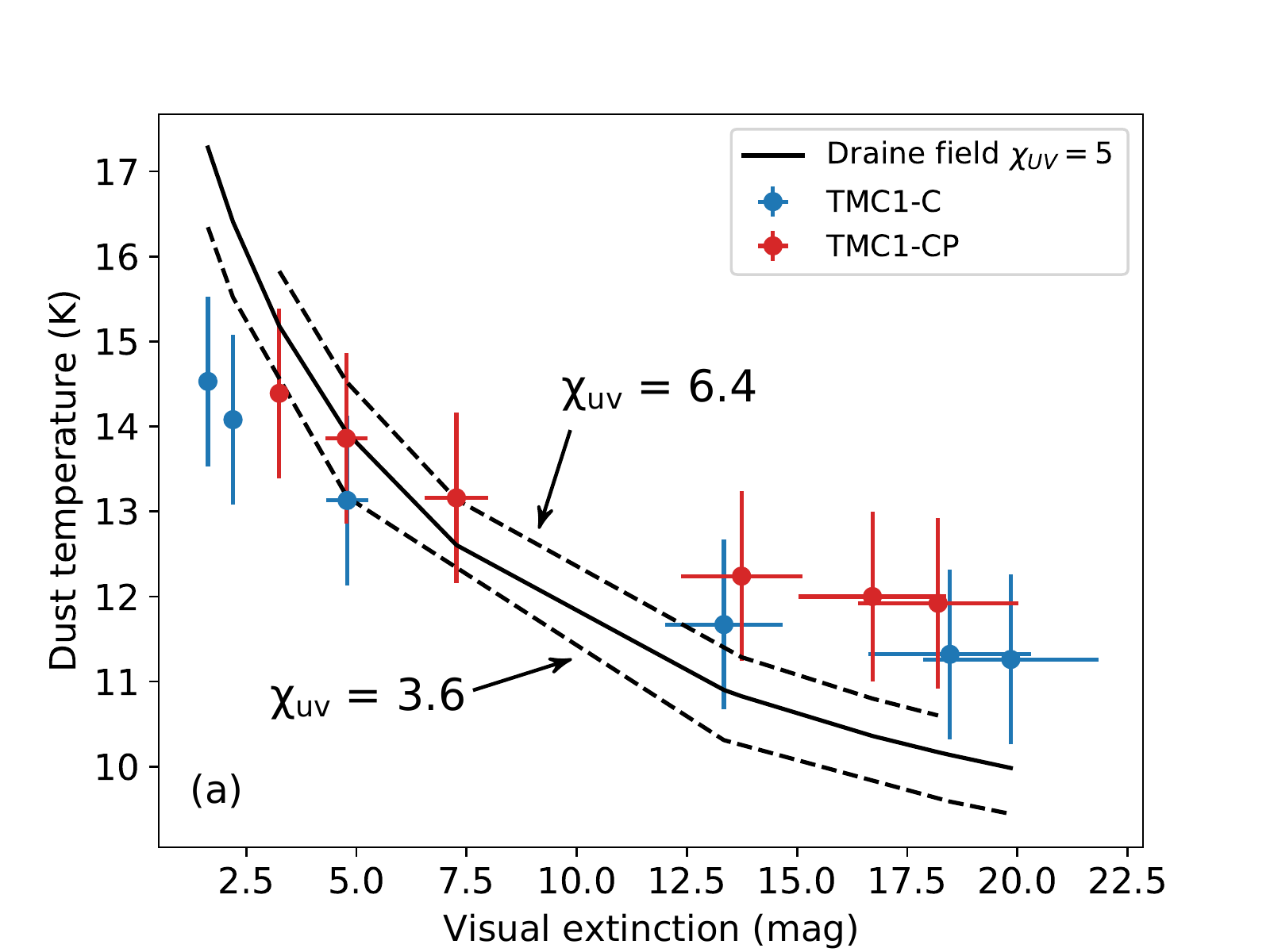}
			\label{fig:UVTMC1}
		\end{subfigure}
		~
		\begin{subfigure}[b]{0.47\textwidth}\includegraphics[width=\textwidth,keepaspectratio]{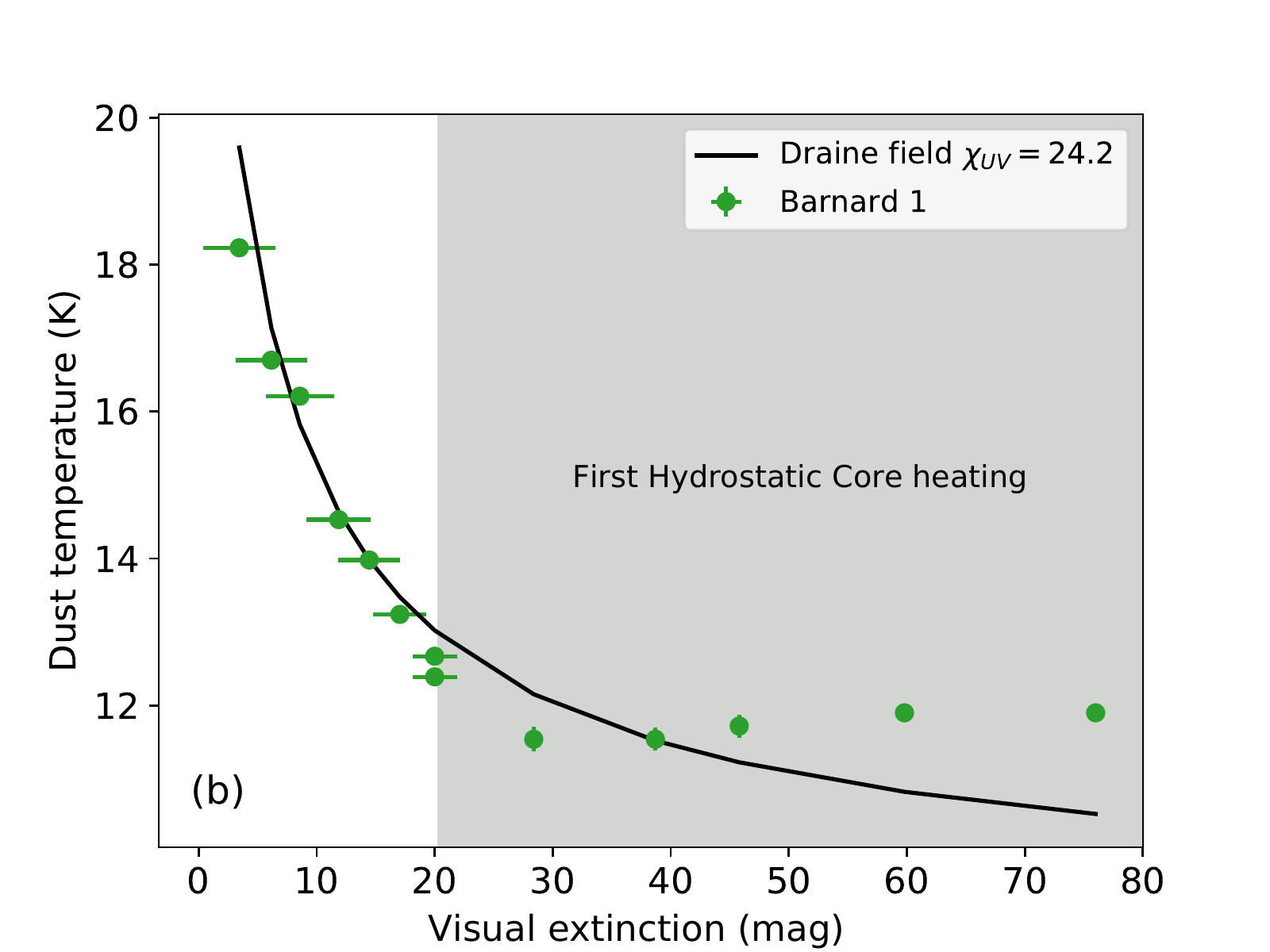}
			\label{fig:UVB1b}
		\end{subfigure}
		\caption{(a) shows the values of the UV field following the \citet{Hocuk2017} parameterization for TMC 1-C (blue) and TMC 1-CP (red) cuts (dashed lines). We also show the best fit (solid line) for the whole TMC 1 region, with $\chi_{\rm UV}({\rm TMC 1}) = 5$. Similarly, in (b), the parameterization yields a Draine field of $\chi_{\rm UV}({\rm B1b})\sim 24$ for Barnard 1b. It is worth noticing the raise of dust temperature in the $A_{\rm v} > 30$ mag region, probably due to the FHSC heating of dust or the presence of a thick ice covering.}
\end{figure*}

\hyperref[fig:UVTMC1]{Fig. 7a} and \hyperref[fig:UVB1b]{Fig. 7b} show the $T_{\rm dust}-{A}_{\rm v}$ plots for the three cuts considered in this paper. None of the cuts can be fitted with a single value of the UV field. In TMC 1, the dust temperature of the dense core ($A_{\rm v} > 7.5$ mag) is better fitted with an illuminating UV field such that $\chi_{\rm UV} \sim 6.4$, while the translucent region ($A_{\rm v} < 7.5$ mag) is better fitted with $\chi_{\rm UV} \sim 3.6$. In Barnard 1b, dust temperatures are best fitted with $\chi_{\rm UV}=24$ in the region $A_{\rm v} < 20$ mag. However, this value of the incident UV field underestimates the dust temperature in the region $A_{\rm v} > 30$ mag. The reason for this underestimation is unclear. A thick layer of ice would allow the dust to be warmer by up to 15\% at high visual extinctions \citep{Hocuk2017}. It is interesting to note that Barnard 1b is located in an active star forming region, and hosts the two very young protostellar cores B1b-N and B1b-S. In fact, as seen with ALMA, dust temperatures above 60 K have been detected in scales of $0.6"$ in the Barnard 1b core. Therefore, we cannot exclude an additional dust heating because of the presence of these young protostars. Due to the large column density $N({\rm H}_{2})\sim 7.6\times 10^{22}$ cm$^{-2}$ in Barnard 1b, this heating would only affect high extinction areas. Since we are interested in the effects of photo-desorption in the low-extinction layers of the cloud, we will adopt  $\chi_{\rm UV}=5$ in TMC 1 and $\chi_{\rm UV}=24$ in Barnard 1b in our calculations.

\section{Photodesorption of H\texorpdfstring{$_2$}{}S: a simple accretion-photodesorption model}

To study the role of photodesorption in the formation of gas-phase H$_2$S, we adapted the simplified PDR model proposed by \citet{Hollenbach2009} to the case of H$_2$S. In this model, the grains are supposed to be covered by an ice layer, from where H$_2$S is released into the gas phase via photodesorption. Only two processes, freezing onto grain mantles and photodissociation, are considered for gas-phase H$_2$S destruction. The gas-phase H$_2$S formation rate is then proportional to the attenuated local UV interstellar flux $1.7\chi_{\rm UV}F_{0}e^{-0.9A_{\rm v}}+\Phi_{\rm SP}$, where $F_{0}$ is the local interstellar flux, $A_{\rm v}$ the visual extinction in magnitudes, $\Phi_{\rm SP}$ is the flux of secondary photons, and $\chi_{\rm UV}$ is the incident UV field in Draine units. This rate is also proportional to the photodesorption efficiency  $Y_{\rm H_{2}S}$, which is set to 1.2 $\times$ 10$^{-3}$ molecules per incident photon, as calculated by \citet{Fuente2017}, the fraction of desorption sites occupied by H$_{2}$S ice $(f_{s,{\rm H_{2}S}})$, and the probability of collision $(n_{\rm gr}\sigma_{\rm gr})$. The destruction rate however must account for dissociation and freezing of H$_{2}$S molecules. The probability of a UV photon to dissociate a H$_{2}$S molecule is given by the attenuated UV field flux times the density of H$_{2}$S, $1.7\chi_{\rm UV}R_{\rm H_{2}S}e^{-0.9A_{\rm v}}n({\rm H_{2}S})$, where $R_{H_2S}$ is the H$_{2}$S photodissociation rate, and $n({\rm H_{2}S})$ is the H$_{2}$S number density. As discussed previously, the freezing probability is given by the probability of collision between grains and H$_{2}$S molecules $(n({\rm H_{2}S})v_{H_{2}S}n_{\rm gr}\sigma_{\rm gr})$. In the stationary state, both creation and destruction rates are equal, and therefore: 

\begin{multline}\label{eq:rates}
	(1.7\chi_{\rm UV}F_{0}e^{-0.9A_{\rm v}}+\Phi_{\rm SP})Y_{\rm H_{2}S}f_{s,{\rm H_{2}S}}n_{\rm gr}\sigma_{\rm gr} = \\ = 1.7\chi_{\rm UV}R_{\rm H_{2}S}e^{-0.9A_{\rm v}}n({\rm H_{2}S})
	+ n({\rm H_{2}S})v_{H_{2}S}n_{\rm gr}\sigma_{\rm gr}.
\end{multline}

\begin{figure}
  	\includegraphics[width=0.49\textwidth,keepaspectratio]{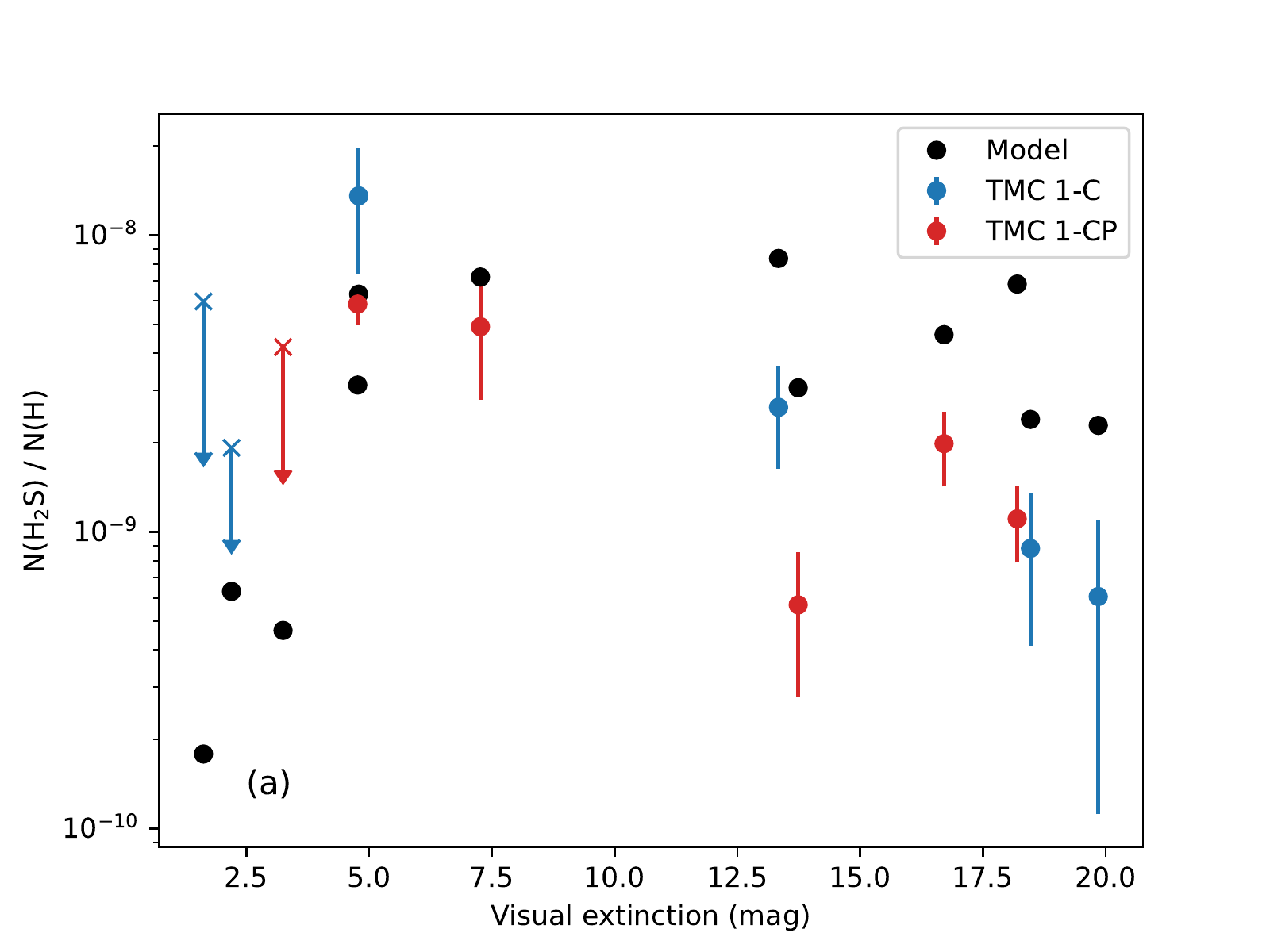}
  	\label{fig:modelTMC1}
%  	\caption{H$_{2}$S gas-phase abundance from our observations in TMC1 (blue), and the model prediction (red) presented as a function of the visual magnitude.}
%\end{figure}
%\begin{figure}
  	\includegraphics[width=0.49\textwidth,keepaspectratio]{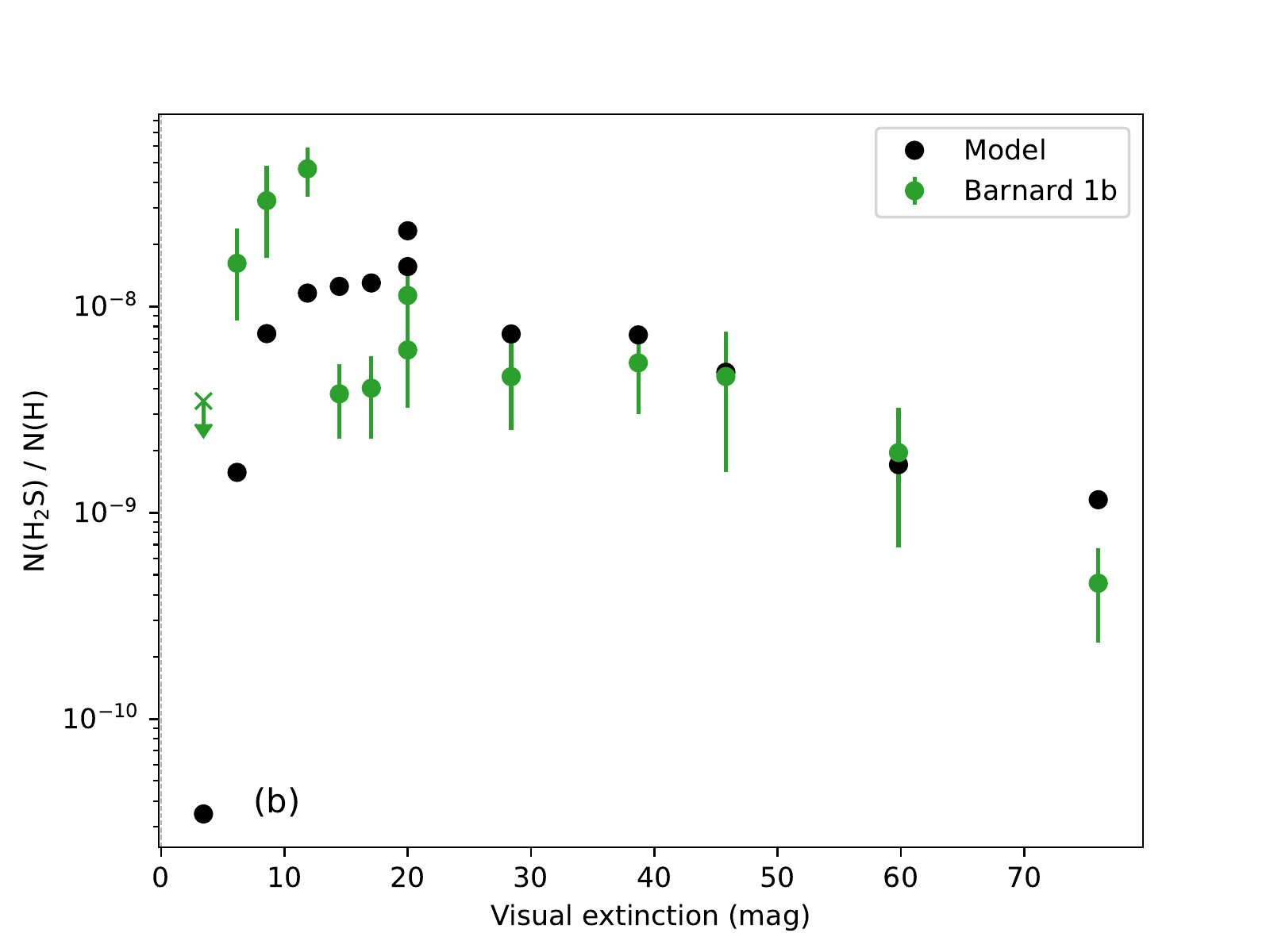}
%  	\label{fig:modelB1B}
  	\caption{Observed H$_{2}$S gas-phase abundance in TMC 1-C, TMC 1-CP, and Barnard 1b (blue, red, and green, respectively), compared to the PDR model prediction (black). The abundances are calculated using expression \eqref{eq:abparam} and assuming $\Phi_{\rm SP}$ = $2\times 10^4$ photons cm$^{-2}$ s$^{-1}$ in TMC 1 and  $\Phi_{\rm SP}$ = $7 \times 10^4$ photons cm$^{-2}$ s$^{-1}$ in Barnard 1b. The value of  $\Phi_{\rm SP}$ has been selected to match the observations.}
\end{figure}

It should be noted that secondary photons are not considered to contribute to the photodissociation rate, $R_{H_2S}$, and are assumed to follow the same extinction law as the FUV radiation. Defining $\sigma_{\rm H}\equiv\frac{n_{\rm gr}(A_{\rm v})\sigma_{\rm gr}}{n_{\rm H}(A_{\rm v})}$, and assuming that the ratio $\frac{n_{\rm gr}(A_{\rm v})}{n_{\rm H}(A_{\rm v})}$ is constant for the range of $A_{\rm v}$ considered,

\begin{equation}\label{eq:abparam}
	X({\rm H_{2}S})=\frac{(1.7\chi_{\rm UV}F_{0}\ e^{-0.9A_{\rm v}}+\Phi_{\rm SP})\,Y_{\rm H_{2}S}\,f_{s,{\rm H_{2}S}}\,\sigma_{\rm H}}{1.7\chi_{\rm UV}R_{\rm H_{2}S}\ e^{-0.9A_{\rm v}}+v_{H_{2}S}\,n_{\rm H}(A_{\rm v})\, \sigma_{\rm H}}.	
\end{equation}
The fractional coverage of the surface by H$_{2}$S ice, $f_{s,{\rm H_{2}S}}$, is given by equating the sticking of S atoms to grain surfaces to the photodesorption rate of H$_{2}$S
\begin{equation*}
	(1.7\chi_{\rm UV}F_{0}e^{-0.9A_{\rm v}}+\Phi_{\rm SP})Y_{\rm H_{2}S}f_{s,{\rm H_{2}S}}n_{\rm gr}\sigma_{\rm gr} = n({\rm S})v_{S}n_{\rm gr}\sigma_{\rm gr},
\end{equation*}
and therefore,

\begin{equation}\label{eq:fraction}
	f_{s,{\rm H_{2}S}}=\frac{n({\rm S})v_{\rm S}}{Y_{\rm H_{2}S}(1.7\chi_{\rm UV}F_{0}e^{-0.9A_{\rm v}}+\Phi_{\rm SP})}.	
\end{equation}
Combining Equations \eqref{eq:abparam} and \eqref{eq:fraction}, we are able to predict the H$_2$S abundance for given physical conditions. The abundance relative to water found in comets is of the order of 2\% \citep{Bockelee2000}, thus we take as saturation value, $f_{s,{\rm H_{2}S}}$ = 0.02.  We  obtain the best fitting of the observed H$_2$S abundances in TMC 1 with $\Phi_{\rm SP}$ = $2\times 10^4$ photons cm$^{-2}$ s$^{-1}$ (\hyperref[fig:modelTMC1]{Fig. 8a}). However, we do not reproduce the H$_2$S abundances in Barnard 1b with the same flux of secondary photons, specially at higher extinction regions (see \hyperref[fig:modelTMC1]{Fig. 8b}). To explain the abundances observed in Barnard 1b we should adopt a higher value, $\Phi_{\rm SP}$ = $7 \times 10^4$ photons cm$^{-2}$ s$^{-1}$. Secondary photons are generated by the interaction of H$_2$ with cosmic rays \citep{Prasad1983}. While the UV photons are absorbed by dust, cosmic rays penetrate deeper into the cloud maintaining a flux of UV photons even in the densest part of starless cores. The flux of UV secondary photons in dense clouds has a typical value of $10^4$ photons cm$^{-2}$ s$^{-1}$, with a cosmic ray molecular hydrogen ionization rate of $3.1\times 10^{-17}$ s$^{-1}$, and a factor of 3 uncertainty \citep{Shen2004}. Recent works point to a lower cosmic ionization molecular hydrogen rate in Barnard 1b than in TMC 1  \citep{Fuente2016, Fuente2019} which is in contradiction with assuming higher  $\Phi_{\rm SP}$ in Barnard 1b, rendering our assumptions very unlikely. It is true, however, that the uncertainties of these estimates are large, of around a factor of $\sim 5$, not allowing us to fully discard this explanation.

As we have seen, this simple model presents several difficulties. It neglects the chemically active nature of the icy grains, as it does not include any chemical reaction on grain surfaces. Furthermore, recent studies have shown that chemical desorption might be an important source of gas-phase H$_{2}$S \citep{Oba2018}, which is not included here. Finally, due to the many ad-hoc parameters introduced, it lacks of any predicting power. It is therefore necessary to consider a more complete chemical modeling of these cores. In the next section we carry out a 1D modeling of the cores using a full gas-grain chemical network to have a deeper insight into the H$_2$S chemistry.

\section{Complete modeling of the TMC 1 and Barnard 1b chemistry}\label{sect:CompleteModel}

\textsc{Nautilus 1.1} \citep{Ruaud2016} is a numerical model suited to study the chemistry in astrophysical environments. It solves the kinetic equations for the gas-phase and the heterogeneous chemistry at the surface of interstellar dust grains. \textsc{Nautilus} is now a three-phase model, in which gas, grain surface and grain mantle phases, and their interactions, are considered. Given a set of physical and chemical parameters, \textsc{Nautilus} computes the evolution of chemical abundances. In the following, we adopt the physical properties of TMC 1 and Barnard 1b derived in Section 7. Since both TMC 1-C and TMC 1-CP are described with identical physical parameters, we consider a common chemical model for them. 

\subsection{Chemical network}

We have developed an up-to-date sulfur chemical network, based on the KInetic Database for Astrochemistry (KIDA), including recent updates \citep{Fuente2017, LeGal2017, Vidal2017, LeGal2019}. Our chemical network is composed of 1126 species (588 in the gas phase and 538 in solid phase) linked together via 13155 reactions, with 8526 reactions in gas phase and 4629 reactions in solid phase. The gas-phase reactions are composed of (i) bi-molecular reactions, such as radiative associations, ion-neutral and neutral-neutral reactions), (ii) recombinations with electrons, (iii) ionization and dissociation reactions by direct by UV-photons, cosmic rays, and secondary photons (i.e., photons induced by cosmic rays). The solid phase reactions are composed of both surface and bulk iced grain mantle reactions that occurs for most of them through the diffusive Langmuir-Hinshelwood mechanism. The bulk and surface of the grain mantle interact via swapping processes \citep{Garrod2013, Ruaud2016}. Desorption into the gas phase is only allowed for the surface species, considering both thermal and non-thermal mechanisms. The latter include desorption induced by cosmic-rays \citep{Hasegawa1993}, photodesorption, and chemical desorption \citep{Garrod2007}. We further describe this last process in Section 9.2. The binding energies considered in this work can be found in Table 2 of \citet{Wakelam2017} and on the KIDA database website\footnote{\url{http://kida.obs.u-bordeaux1.fr/}}.

\begin{table}
	\centering
		\caption{Initial abundances}
			\begin{tabular}{cc}
				\toprule
%				Element &  \multicolumn{1}{c}{Abundance wrt hydrogen} \\  \hline
%				 & Low sulphur depletion & High sulphur depletion \\ \midrule
					He & \multicolumn{1}{c}{$9.00\times 10^{-2}$} \\
					N & \multicolumn{1}{c}{$6.20\times 10^{-5}$} \\
					O & \multicolumn{1}{c}{$2.40\times 10^{-4}$} \\
					C$^+$ &  \multicolumn{1}{c}{$1.70\times 10^{-4}$} \\
					S$^+$ &  \multicolumn{1}{c}{$1.50\times 10^{-5}$}  \\
					Si$^+$ & \multicolumn{1}{c}{$8.00\times 10^{-9}$} \\
					Fe$^+$ & \multicolumn{1}{c}{$3.00\times 10^{-9}$} \\
					Na$^+$ & \multicolumn{1}{c}{$2.00\times 10^{-9}$} \\
					Mg$^+$ & \multicolumn{1}{c}{$7.00\times 10^{-9}$} \\
					P$^+$ & \multicolumn{1}{c}{$2.00\times 10^{-10}$} \\
					Cl$^+$ & \multicolumn{1}{c}{$1.00\times 10^{-9}$} \\
					F & \multicolumn{1}{c}{$6.68\times 10^{-9}$} \\
    			\bottomrule
			\end{tabular}
			\label{tab:initab}
\end{table}

\subsection{Chemical desorption: 1D modeling}

\begin{figure*}
	\centering
	\includegraphics[width=\textwidth,keepaspectratio]{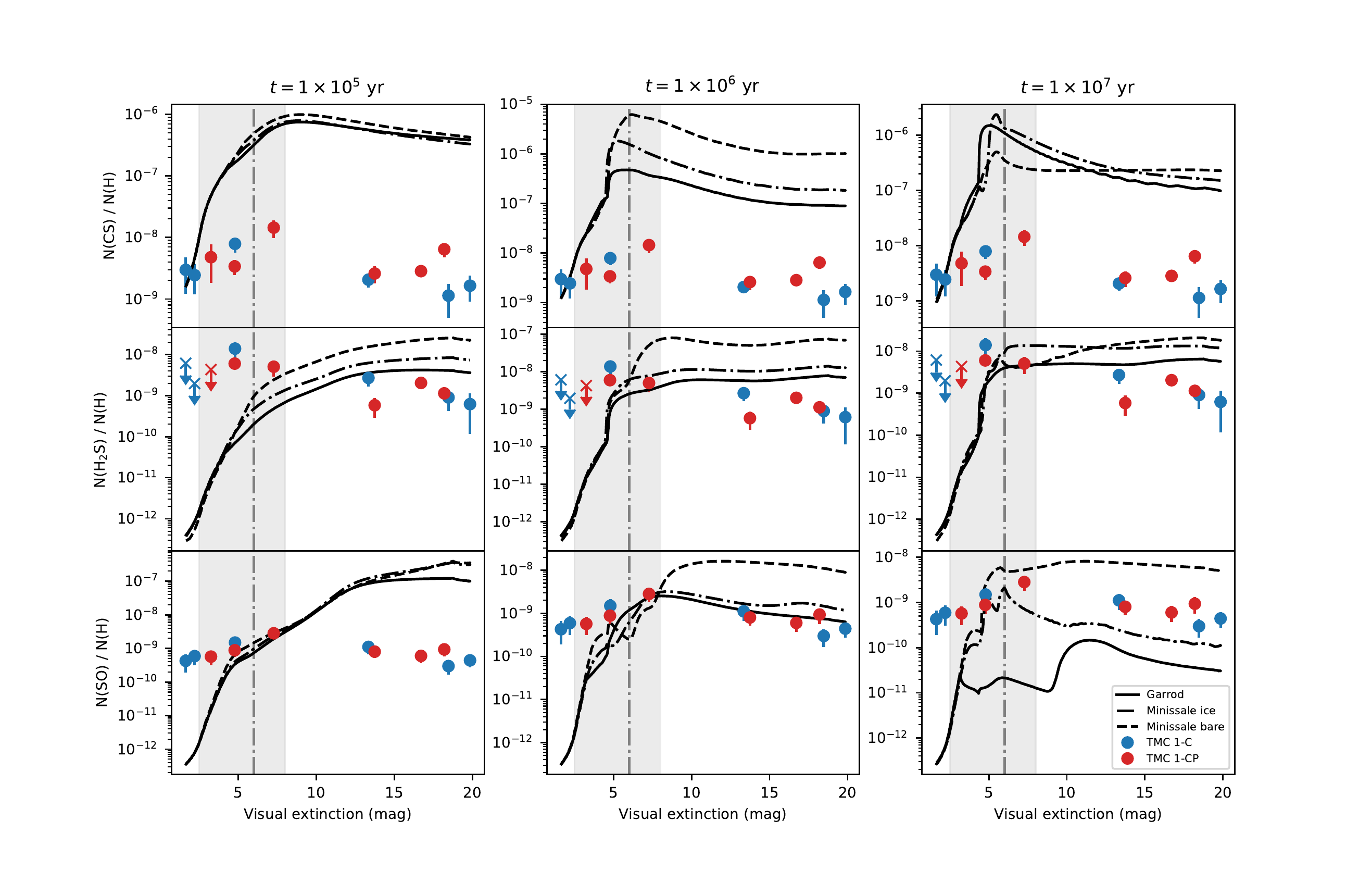}
	\caption{Predicted abundances (solid lines) of the CS (top row), H$_{2}$S (middle row), and SO (bottom row) by models with different chemical desorption schemes, together with the observed abundances in TMC 1, at times $0.1-1-10$ Myrs. Notice that there is an extinction interval in which the observed H$_{2}$S fit to different chemical desorption schemes. This can be interpreted as a change in the grain surface composition. The vertical dashed line corresponds to $A_{\rm v} = 6$ mag. The interval $2.5-8$ mag around this value is shaded in all subplots.}
	\label{fig:TMC1Comparison}
\end{figure*}

\begin{figure*}
	\centering
	\includegraphics[width=\textwidth,keepaspectratio]{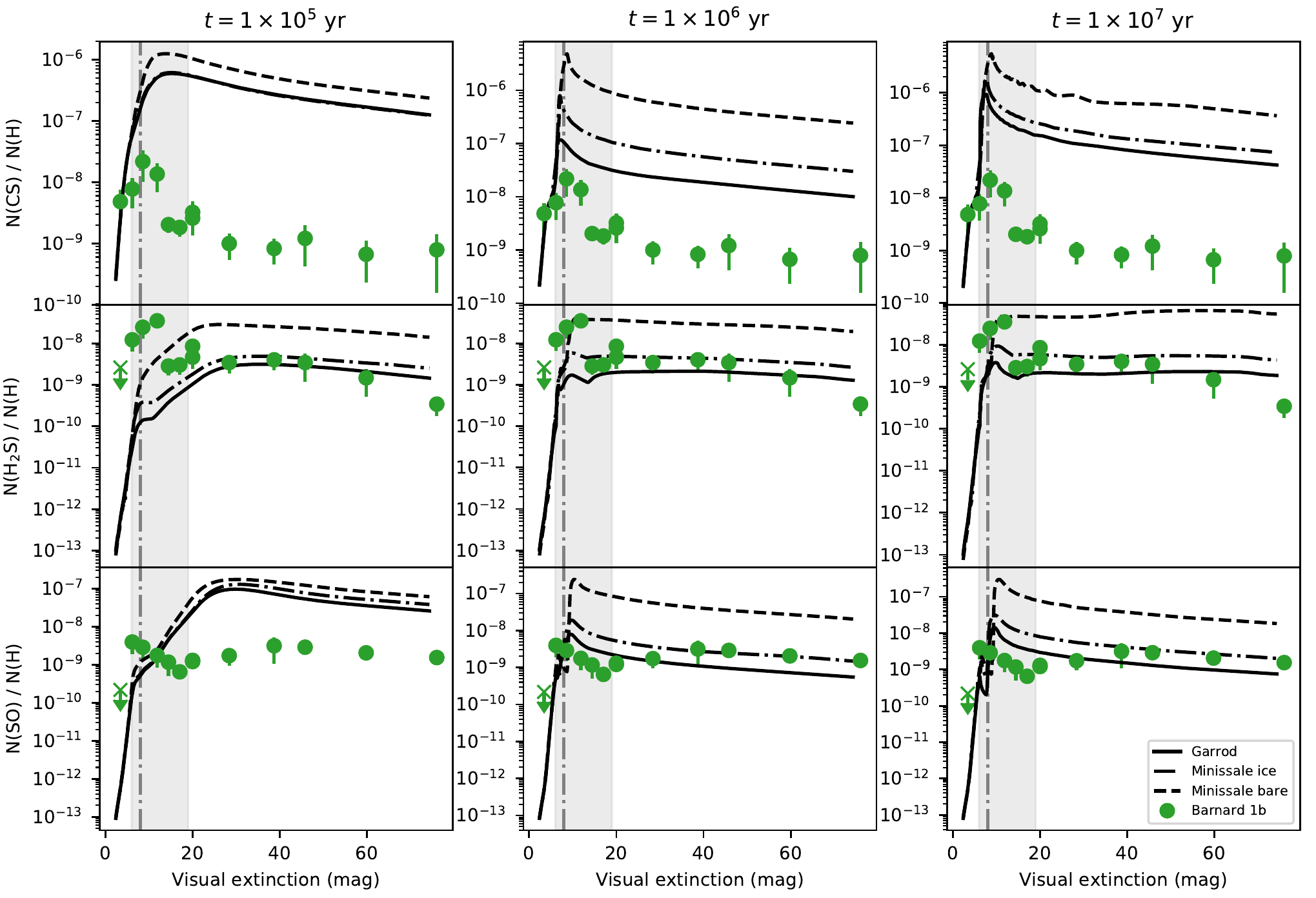}
	\caption{Plots that show the predicted abundances (solid lines) of the CS (top row), H$_{2}$S (middle row), and SO (bottom row) by models with different chemical desorption schemes, together with the observed abundances in Barnard 1b, at times $0.1-1-10$ Myrs. The vertical dashed line represents $A_{\rm v} = 8$ mag, and the shaded area encloses the range of extinctions $6-19$ mag.}
	\label{fig:B1BComparison}
\end{figure*}

In Fig.~\ref{fig:TMC1Comparison} and Fig.~\ref{fig:B1BComparison}, we show the comparison of our model with the observations of TMC 1 and Barnard 1b, respectively. We have computed the chemical abundances of different sulphuretted molecules in TMC 1 and Barnard 1b using the physical structures derived in Section 7. As initial abundances, we adopt undepleted sulphur abundance (see \hyperref[tab:initab]{Table 3}). Dust and gas temperatures are assumed to be equal. Other relevant parameters are $\chi_{\rm UV} = 5$ and $\zeta_{\rm CR} = 1.15\times 10^{-16}$ s$^{-1}$ for TMC 1 \citep{Fuente2019}, and $\chi_{\rm UV} = 24$ and $\zeta_{\rm CR} = 6.5\times 10^{-17}$ for Barnard 1b \citep{Fuente2016}. To compare the output of the chemical model with the observations, we compute a weight-averaged abundance over the line of sight, weighted with the density at each position. Assuming the spherically symmetric density profiles $n_{\rm H}$ given in \hyperref[fig:TMC1BE]{Fig. 4a} and \hyperref[fig:B1BBE]{Fig. 4b}, the weight-averaged abundance ${\rm [X]_{\rm ac}}$ of an element X along the line of sight with offset $r$ is calculated here as: 

\begin{equation}\label{eq:tielens}
	{\rm [X]_{\rm ac}}(r) = \frac{\sum_{i}\left(l_{i+1}-l_{i}\right)\left(n_{\rm H}(s_{i}){\rm [X]}(s_{i})+n_{\rm H}(s_{i+1}){\rm [X]}(s_{i+1})\right)}{\sum_{j}\left(l_{j+1}-l_{j}\right)\left(n_{\rm H}(s_{j})+n_{\rm H}(s_{j+1})\right)},
\end{equation}
where $s_{i}=\sqrt{r^2+l_{i}^{2}}$, $l_{i}$ is a discretization of the segment along the line of sight $l_{\rm max} > \dots > l_{i+1} > l_{i} > \dots > 0$, with $l_{\rm max} = \sqrt{r^2+r_{\rm max}^{2}}$ and $r_{\rm max}$ being the radius of the density profile $n_{\rm H}(r)$. 

In dark clouds, where the temperature of grain particles is below the sublimation temperature of most species, non-thermal desorption processes are needed to maintain significative abundances of molecules in gas phase. This is especially important in the case of H$_2$S that is thought to be mainly formed on the grain surfaces. Differently from the other processes, chemical desorption links the solid and gas phases without the intervention of any external agents such as photons, electrons, or other energetic particles  \citep{Garrod2007}. In other words, it could be efficient in UV-shielded and low-CR environments where photodesorption or sputtering cannot be efficient, becoming more likely the most efficient mechanism in dark cores. This mechanism seems also to be responsible for the abundance of complex molecules such as methanol and formaldehyde in dense cores and PDRs, becoming dominant against phototodesorption \citep{Esplugues2019, LeGal2017, Vasyunin2017, Esplugues2016}. New laboratory experiments proved that chemical desorption might be important for H$_2$S formation on water ice \citep{Oba2018}. 

Achieving the correct inclusion of chemical desorption in chemical models is difficult. The efficiency of this process depends not only on the specific molecule involved, but also on the detailed chemical composition of the grain surface on which the reaction occurs, making a correct estimate of its value very difficult to obtain. The chemical desorption process starts from the energy excess of some reactions. In bare grains, as described in \citet{Minissale2014}, its efficiency essentially depends on four parameters: enthalpy of formation, degrees of freedom, binding energy, and mass of newly formed molecules. However, this efficiency becomes lower when dust grains are covered with an icy mantle because part of the energy released in the process is absorbed by the ice matrix. \citet{Minissale2016}, based on experimental results, suggested that when taking into account the icy surface, the efficiency of the chemical desorption should be reduced to 10\% of the value calculated for a bare grain, when dealing with the water-dominated icy coated mantles of dense regions. We run our chemical model to compare with our observations considering the three chemical desorption scenarios, i) bare grains following Minissale prescription, ii) icy coated grains following Minissale prescription, and iii) ice coated grains assuming Garrod prescription, in which the ratio of the surface–molecule bond-frequency to the frequency at which energy is lost to the grain surface is set to 0.01. The calculations following these three different scenarios are shown in Fig.~\ref{fig:TMC1Comparison} and Fig.~\ref{fig:B1BComparison}. 

\begin{figure*}
	\centering
	\includegraphics[width=0.495\textwidth,keepaspectratio]{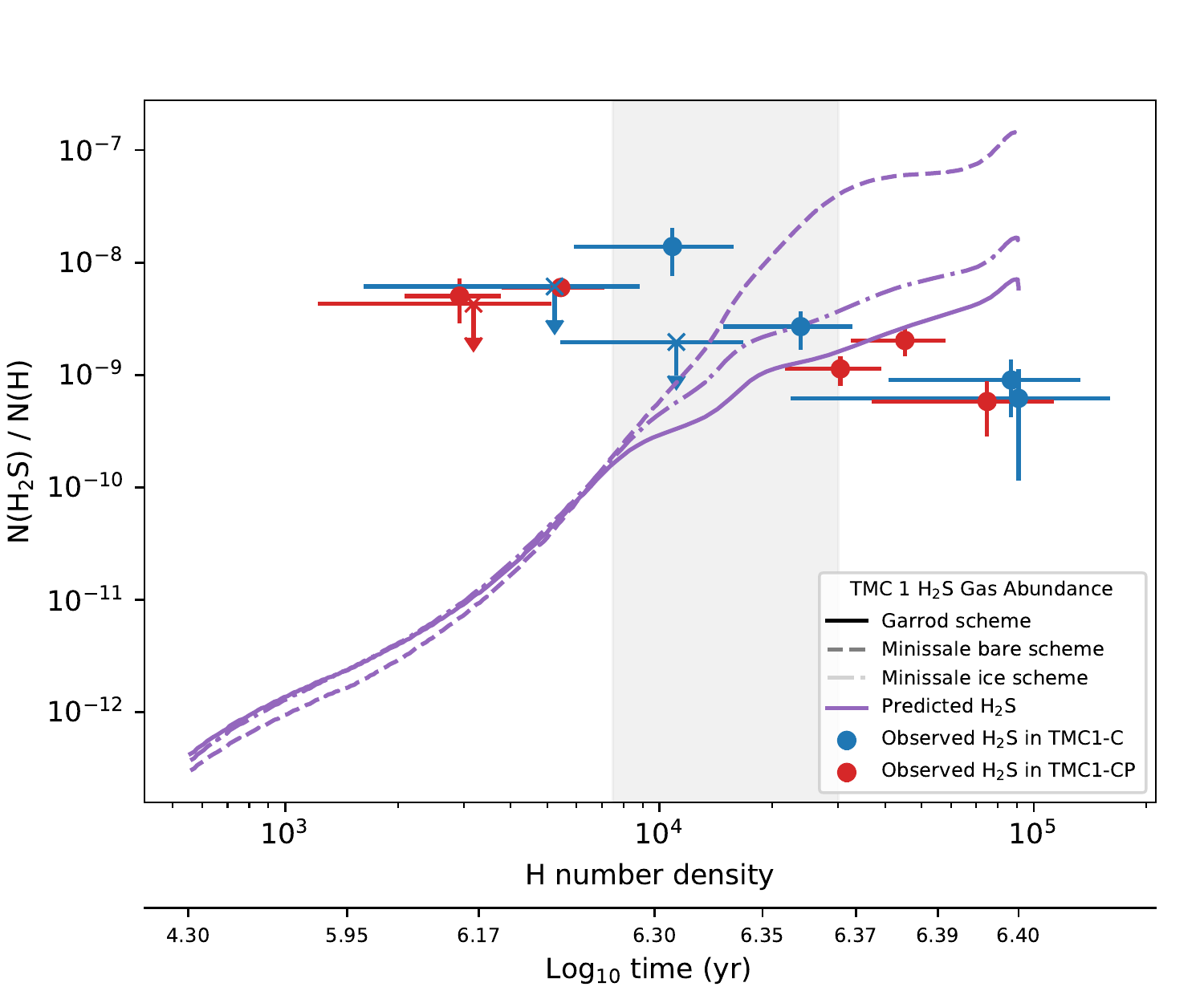}
    \includegraphics[width=0.495\textwidth,keepaspectratio]{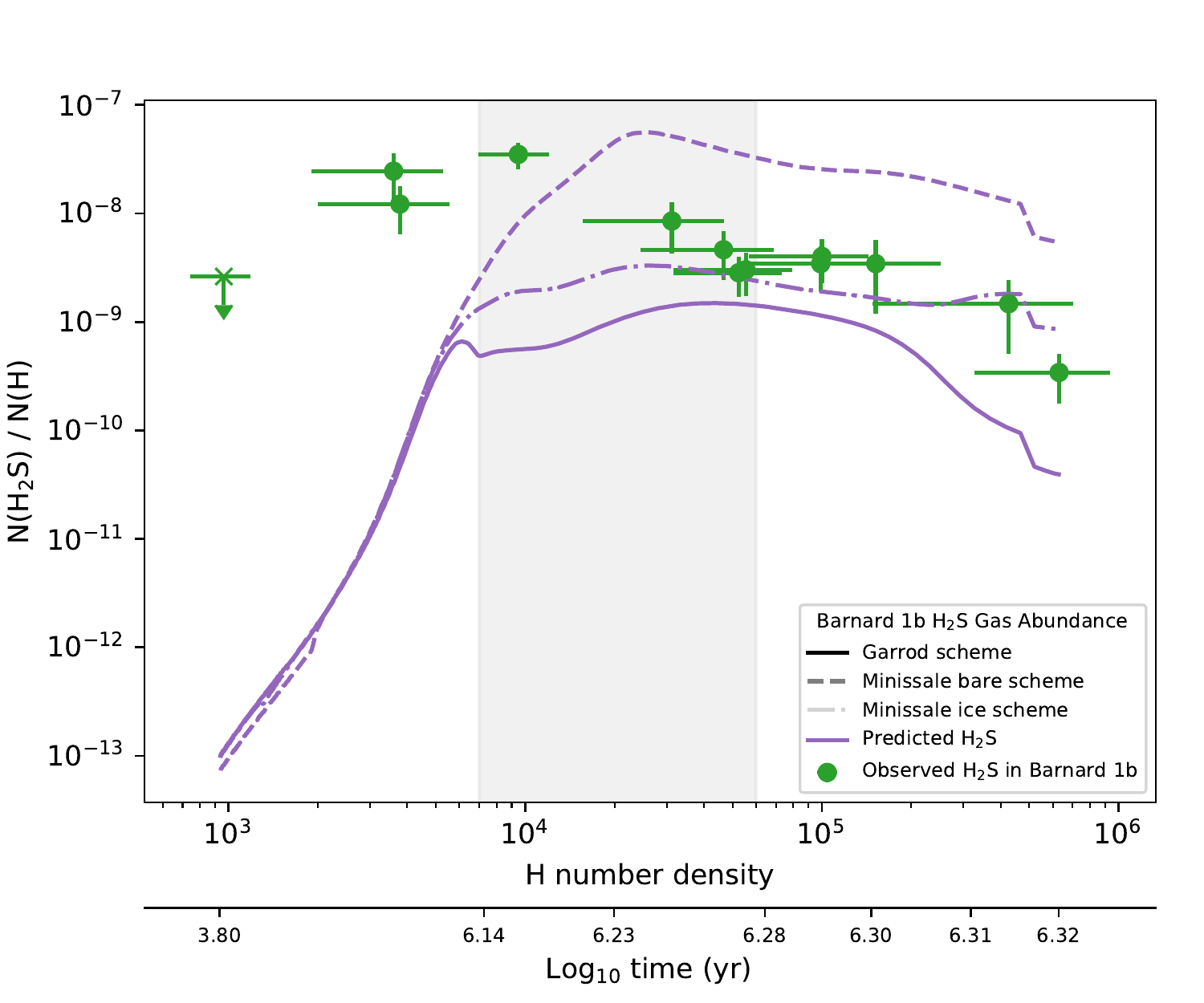}
	\caption{Detail of the predicted H$_{2}$S gas-phase abundance in TMC 1 (left) and Barnard 1b (right) according to the different desorption schemes in a 0-dimensional simulation, and the observed H$_{2}$S gas-phase abundances, plotted as a function of density and the time according to a freefall collapse. The shaded band encloses the density interval where the change in desorption scheme occur.}
	\label{fig:TMC1H2Sfreefall}
\end{figure*}

The observed H$_2$S abundances are in reasonable agreement with those predicted by \textsc{Nautilus} at $1$ Myr. At this time, the main reactions that lead to the creation and destruction of gas-phase H$_{2}$S at low and high extinctions in TMC 1 and Barnard 1b are found in \hyperref[tab:reactions]{Table 4}. In all scenarios, chemical desorption is the main formation mechanism of gas-phase H$_{2}$S, and its prevalence diminishes at high extinctions, where density increases and gas-phase reactions become more important to form gas-phase H$_{2}$S.
%It is worth noticing that the physical structure in Barnard 1b corresponds to a more evolved stage in the collapse of a BE sphere, also consistent with the presence of two very young protostellar objects in its interior. The fact that their chemical age is similar in the two clouds might suggest either a faster collapse in Barnard 1b, as proposed by \citet{Fuente2016}, or a higher radiation field. 
Interestingly, the H$_2$S observations cannot be fitted using the same chemical desorption scheme at every position. In fact, we see that the observational data are in better agreement with the bare grain chemical desorption scheme as described by \citet{Minissale2016} towards the edge of the clouds and to an ice-covered grain chemical desorption scheme towards the more shielded regions. This suggests a change in the chemical composition of grain surfaces, which become covered by a thick ice mantle as the density increases towards the core center, and opens the possibility to use the H$_2$S abundance to estimate the transition from bare to ice coated grains. Based on our observations, grains become coated with a thick ice mantle at ${A}_{\rm v} \sim 8 \ {\rm mag}$ $({A}_{\rm v\ eff} = {A}_{\rm v}/2 \sim 4 \ {\rm mag})$ in TMC 1 and  ${A}_{\rm v} \sim 12 \ {\rm mag}$ $({A}_{\rm v\ eff} \sim 6 \ {\rm mag})$ in Barnard 1b. As a sanity check for our interpretation, we compute a rough estimation of the upper bound of the visual extinction required to photodesorb the ice covering the grains with the expression \citep{Tielens2010Book}:

\begin{equation}
	A_{{\rm v}\,\rm eff}^{\rm ice} < 4.1 + \ln\left[1.7\chi_{\rm UV}\left(\frac{\rm Y_{\rm pd}}{10^{-2}}\right)\frac{10^{4}\ {\rm cm}^{-3}}{n}\right],
\end{equation}
with $\chi_{\rm UV}$ the incident UV field in Draine units, $Y_{\rm pd}$ the photodesorption yield, and $n$ the number density. Assuming  ${Y_{\rm pd}} = 10^{-3}\ {\rm photon}^{-1}$ \citep{Hollenbach2009}, we obtain that grains would mainly remain bared for ${A}_{v\,\rm eff}^{\rm ice} < 6 \ {\rm mag}$ in TMC 1 and  for ${A}_{v\,\rm eff}^{\rm ice} < 8 \ {\rm mag}$ in Barnard 1b. These values are in qualitative agreement with our results, within a factor of $\sim 2$, as grains are expected to remain bare deeper in  Barnard 1b than in TMC 1 because of the higher incident UV field. However, this simple expression does not quantitatively reproduce our values of ${A}_{v\,\rm eff}^{\rm ice}$. This is not surprising taking into account that our estimate of the local ISRF is uncertain.  Furthermore, the above expression is an approximation based on the equilibrium between photo-desorption and freeze-out of water molecules on the grain surfaces. To test the consistency of our result,  we will use our chemical model to investigate the ice composition in Section 10.

It is also interesting to compare the output of our models with the observations of the other two most abundant S-bearing molecules, CS and SO. One first result is that our models overestimate by more than one order of magnitude the CS abundance  in TMC 1 and Barnard 1b. The problem of the overestimation of the CS abundance was already pointed out by \citet{Vidal2017} and \cite{Laas2019}.  \citet{Vidal2017} suggested that it could come from wrong branching ratios of the dissociative recombination of HCS$^+$ or a potentially new sink for CS that might be the O + CS reaction whose reaction rate has not been measured at low temperatures. The predicted SO abundances are in reasonable agreement with the observed values if we assume the Garrod prescription to describe chemical desorption even at low visual extinctions. The difficult match between chemical model predictions for S-bearing species and observations make it challenging to determine the sulphur elemental abundance with an accuracy better than a factor of 10.

\begin{figure*}
	\centering
	\includegraphics[width=0.49\textwidth,keepaspectratio]{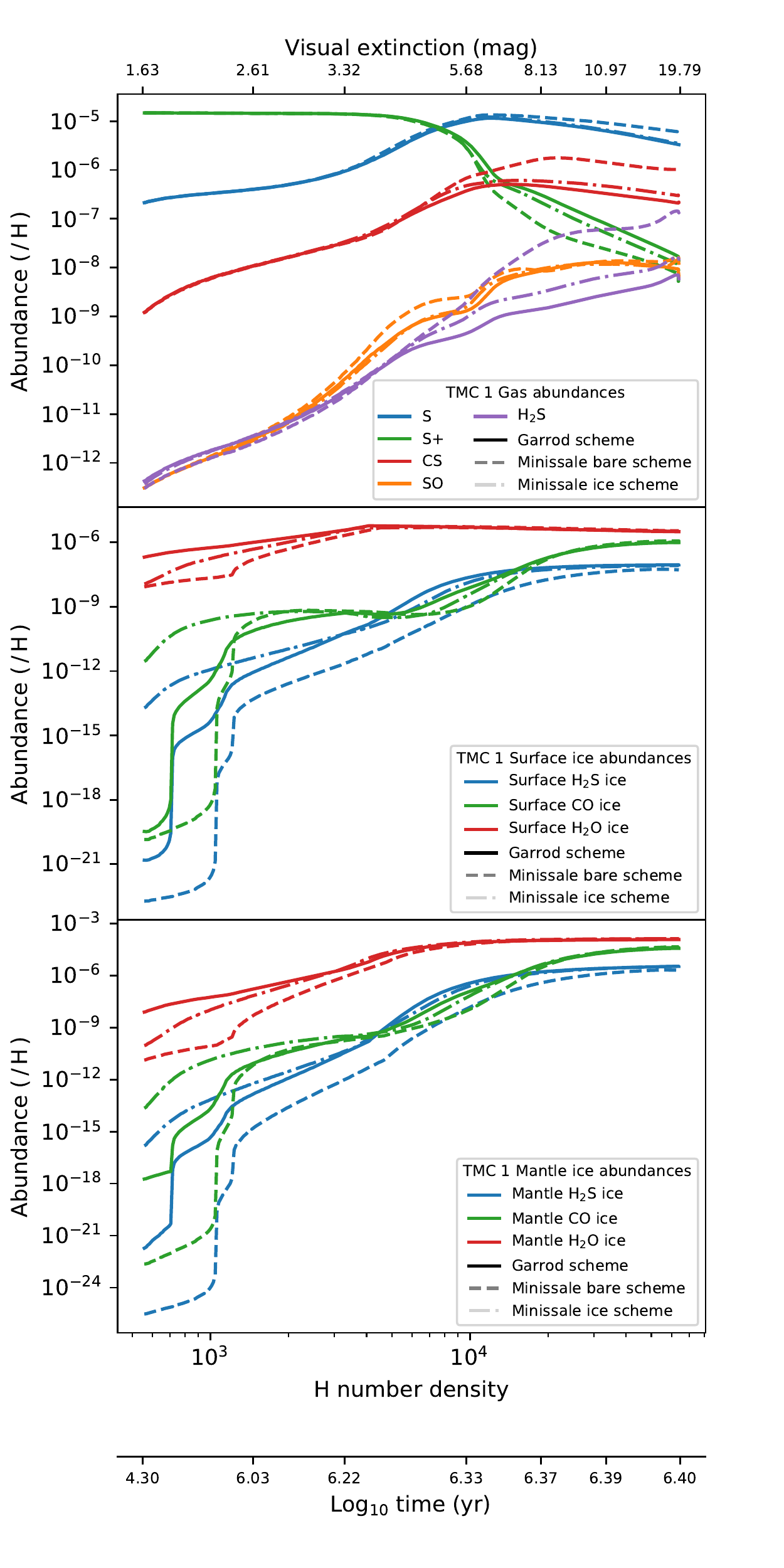}
	\includegraphics[width=0.49\textwidth,keepaspectratio]{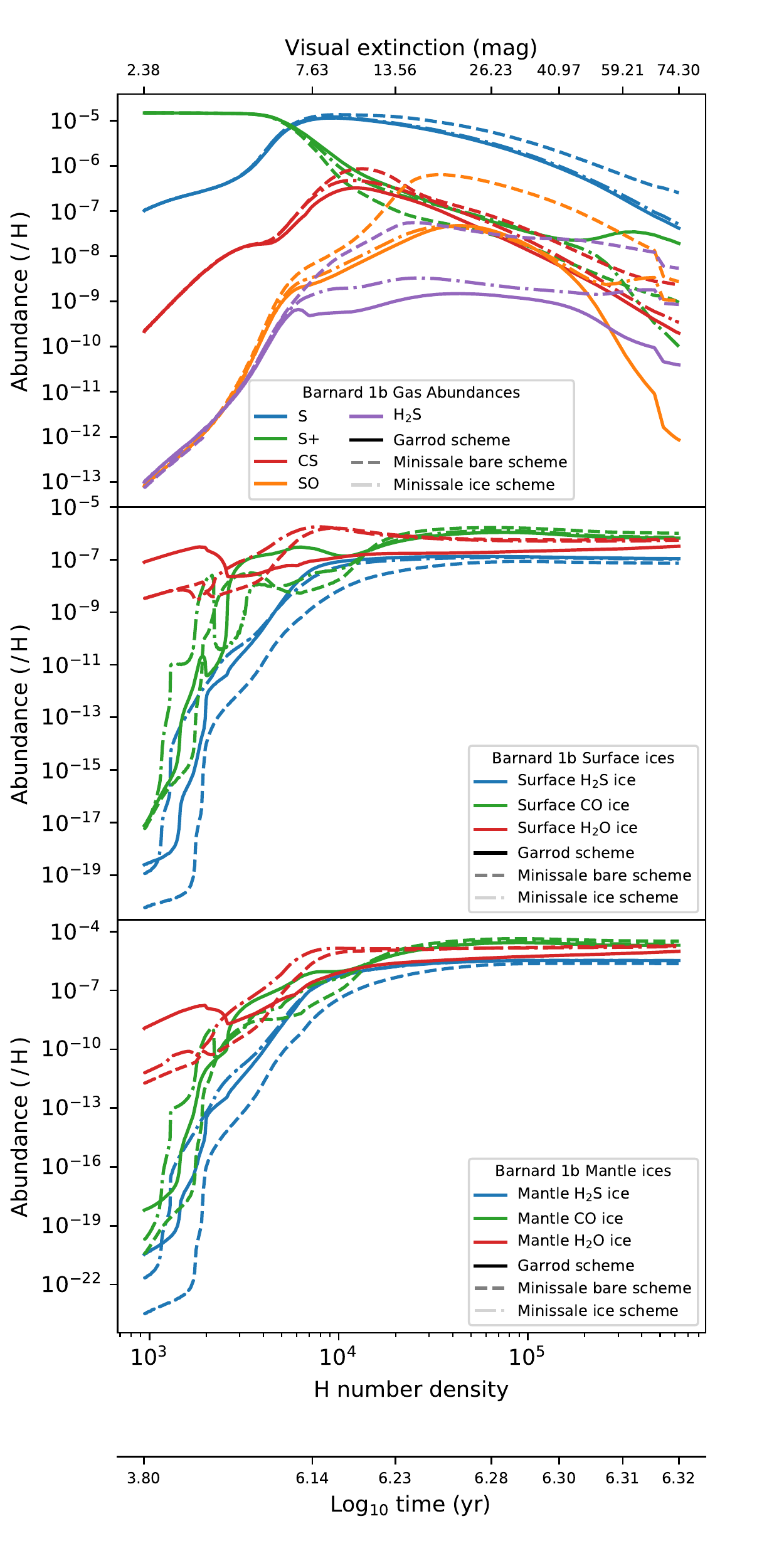}
	\caption{Predicted abundances of different molecules in gas phase (top row), ice surfaces (middle row), and ice mantles (bottom row) of TMC 1 (left column), and Barnard 1b (right column), according to the different desorption schemes. They are plotted as a function of the number density, together with the visual extinction that corresponds to such densities (see \hyperref[fig:TMC1BE]{Fig. 4a-b}, \hyperref[fig:TMC1Av]{Fig. 5a-b}), and the time according to a freefall collapse.}
	\label{fig:TMC1freefall}
\end{figure*}

\section{Discussion: Is  H$_2$S tracing the snow line in dark cores?}

As commented above, the H$_2$S chemical desorption efficiency seems to decay at a visual extinction of $A_{\rm v}\sim 8$ mag in TMC 1 and  $A_{\rm v}\sim 12$ mag in Barnard 1b. We propose that this jump in the chemical desorption efficiency might be caused by a change in the chemical composition on the surface of grains, in particular the formation of a thick ice mantle. In this section, we use our chemical model to explore the link between the efficiency of the H$_2$S chemical desorption as traced by our observations and the ice chemical composition in TMC 1 and Barnard 1b.

In \textsc{Nautilus}, two ice phases are considered, the mantle and the surface. The mantle is assumed to grow by incorporating material from the surface. The mantle is hence a fossil record of the grain history. To determine the chemical composition of the ice mantle we should reproduce the trajectory of a cell of gas and dust during the core contraction. As a reasonable approximation, we have converted our 1D profile in a cell trajectory as follows. First, the density, temperature and visual extinction profiles shown in \hyperref[fig:TMC1BE]{Fig. 4a-b}, \hyperref[fig:TMC1Av]{Fig. 5a-b} and \hyperref[fig:TMC1Tkin]{Fig. 6a-b} are divided in 300 bins that we use to define the cell trajectory. Second, the time elapsed by the material to reach a bin $i$ is equal to the difference between the free-fall time $t_{\rm ff}(\rho) \approx \sqrt{3/2\pi G\rho}$ of the initial bin and the current bin, that is, $t(i)=t_{\rm ff}(\rho_{0}) - t_{\rm ff}(\rho_{i})$, and consequently, the duration of the whole trajectory is equal to the free-fall time assuming the density of the first bin. This time, for the lowest density in our cuts $n_{\rm H}\sim 1\times 10^{3}$ cm$^{-3}$ (see \hyperref[table:B1Bconditions]{Table B.1}), is $t_{\rm ff}\sim 10^{6}$ yr. It is remarkable that this is the time that best fit the observations in Section 9, thus suggesting this choice to be a good guess of the collapse time. Third, in each time-step, the chemical abundances of the previous bin are taken as initial abundances for the next step. These trajectories describe the changes in density, temperature and local UV radiation experienced by a given cell once the collapse has started. Finally, we need to establish the chemical composition of the initial molecular cloud. To set the initial abundances of the cloud, we run the chemical model in a pre-phase, using the physical conditions of the first bin and letting chemistry evolve during 1 Myr, the typical time for this stage. The abundances at the beginning of the cloud collapse are then those that result from the pre-phase.

Following this procedure, we predict the chemical composition of the gas and dust as a function of time. In \hyperref[fig:TMC1H2Sfreefall]{Fig. 11} we show the H$_2$S abundance as a function of H number density in TMC 1 and Barnard 1b. The density is continuously increasing with time and provides an easier comparison with observations. We recall, however, that in this comparison we are not taking into account the different physical and chemical conditions along the line-of-sight. According to our results in Section 9, for the low densities prevailing in the cloud border, we need to assume the chemical desorption prescription for bare grains to account for the high observed H$_2$S abundances, but at higher densities, our model overestimates the H$_2$S abundance. In both sources we find the same behavior with a break point at densities around $n_{\rm H}=2\times10^4$ cm$^{-3}$. In the case of Barnard 1b, the abundances are well accounted using the Minissale prescription for ice coated grains for  $n_{\rm H}$ $>$ 2 $\times$ 10$^4$ cm$^{-3}$. In TMC 1, the observed abundances are 5$-$10 times lower than our predictions in any of the scenarios. To relate the H$_2$S abundance with the ice composition, we show the surface and mantle chemical composition as a function of density in TMC 1 and Barnard 1b in \hyperref[fig:TMC1freefall]{Fig. 12}. It is worth noting that the density $n_{\rm H}=2\times 10^4$ cm$^{-3}$ corresponds to the phase in which basically the whole ice mantle is formed, achieving the highest values for the s-H$_2$O and s-CO abundances. Moreover, solid H$_{2}$S (s-H$_{2}$S) becomes the most abundant S-bearing molecule locked in the ices, with an abundance of $\sim 3\times 10^{-6}$ in both targets, one fifth of the cosmic value. Furthermore, according to our predictions, the main sulphur reservoir is gaseous atomic sulphur all along the cores. 

%One additional result is the different ice composition predicted in TMC 1 and Barnard 1b. While the predicted abundance of water ice is as high as $\sim 1.3 \times 10^{-4}$ in TMC 1, its abundance is $1.7 \times 10^{-5}$ in Barnard 1b. In Barnard 1b, the oxygen is found mainly in the form of CO$_{2}$ ice, with an abundance of X(s-CO$_{2}$)$\sim 7.4\times 10^{-5}$. The ratio s-CO$_{2}$/s-H$_{2}$O depends on the diffusivity of the surface and the grain temperature since the efficiency of surface processes involving O, S, CO and SO is enhanced when surface diffusivities and ice mobilities are increased \citep{Semenov2018}. The higher predicted s-OCS abundance in Barnard 1b is also a consequence of this high mobility of heavy atoms. The Spitzer data on the ice composition towards the young stellar object Barnard 1b-S \citep{Boogert2008}, a star in the moderate density neighborhood of Barnard 1b, do not show any sign of this low s-H$_{2}$O abundance. This might be caused by several factors, such as, for example, that Spitzer observations might only show the ice composition of the envelope of these objects. It should be noted that the chemistry on grain surfaces is not very well understood yet and, in addition to the surface chemical network, depends on many parameters like the previously mentioned diffusivity on the grain surface. A more detailed study of the ice composition is outside the scope of this paper.

One additional result is the different ice composition predicted in TMC 1 and Barnard 1b. While the predicted abundance of water ice is as high as $\sim 1.3 \times 10^{-4}$ in TMC 1, its abundance is $1.7 \times 10^{-5}$ in Barnard 1b. In Barnard 1b, the oxygen is found mainly in the form of CO$_{2}$ ice, with an abundance of X(s-CO$_{2}$)$\sim 7.4\times 10^{-5}$. \citet{Vasyunin2017} proposed, based on extrapolations from Minissale's experimental results, that, in general, chemical desorption is enhanced if the ice surface is rich in CO or CH$_3$OH (instead of water). This could explain the higher abundance of H$_2$S in the high-density regions of Barnard 1b compared to TMC 1. However, this is not in agreement with Spitzer data on the ice composition towards the young stellar object Barnard 1b-S \citep{Boogert2008}, a star in the moderate density neighborhood of Barnard 1b, which do not show any sign of low s-H$_{2}$O abundance or s-CO$_2$ overabundance. This might be caused by several factors such as, for example, that Spitzer observations might show the ice composition of only the envelope of these objects instead of the dense core. It should be noted, however, that the chemistry on grain surfaces is not very well understood yet and, in addition to the surface chemical network, depends on many poorly known parameters such as the diffusivity of the grain surface. Indeed, the ratio s-CO$_{2}$/s-H$_{2}$O is very sensitive to the diffusivity of the surface and the grain temperature since the efficiency of surface processes involving O, S, CO, and SO is enhanced when surface diffusivities and ice mobilities are increased \citep{Semenov2018}. The same remains true for s-OCS abundance whose abundance is very sensitive to high mobility of heavy atoms. Further observational constraints are required to fine-tune models and give a more accurate description of the ice composition.

\begin{figure*}
	\centering
	\includegraphics[width=0.8\textwidth,keepaspectratio]{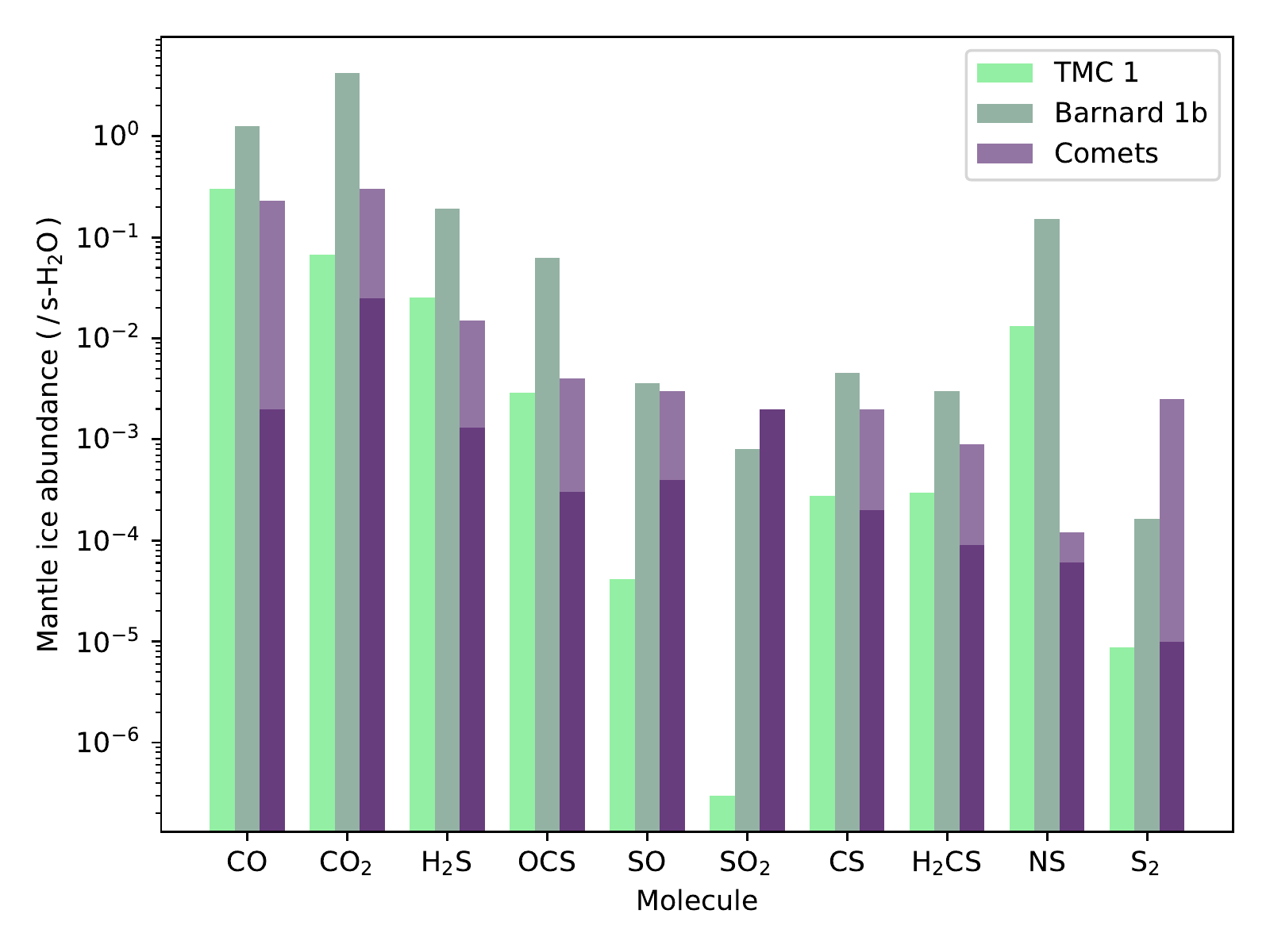}
	\caption{Predicted ice abundances in the mantle of grains in TMC 1 (light green) and Barnard 1b (dark green) of selected molecules and those observed in comets (dark and light purple) from \hyperref[tab:cometsice]{Table 5}. Light purple represents the abundance in comets given in \hyperref[fig:B1BBE]{Table 5}.}
	\label{fig:cometsice}
\end{figure*}

Under the assumption that comets are pristine tracers of the outer parts of the protosolar nebula, it is interesting to compare the composition of comets with that predicted for the ice mantles in the high extinction peaks of TMC 1 and Barnard 1b (see \hyperref[tab:cometsice]{Table 5}). There are significant variations in the observed abundances of S-bearing species among different comets. To account for this variation, we compare our predictions with a range of values derived from the compilation by \citet{BockeleMorvan2017}. In addition, we have added a specific column to compare with the most complete dataset on comet 67P/Churyumov-Gerasimenko \citep{Calmonte2016}. The ice abundances predicted for TMC 1 and Barnard 1b at the end of the simulation are, in general, close to the values observed in comets (see \hyperref[tab:cometsice]{Table 5} and \hyperref[fig:cometsice]{Fig. 13}). These values, however, need to be taken with caution. The abundances of the species in comets are usually calculated relative to water. Our calculations show that the abundance of water in solid phase varies of about one order of magnitude from TMC 1 and Barnard 1b, making the comparison with comets very dependent on the abundance of solid water itself.

\begin{table*}
	\centering
		\caption{Most important reactions to H$_{2}$S in the model}
			\begin{tabular}{ccc}
				\toprule
				\multirow{2}{*}{Visual Extinction} & \multicolumn{2}{c}{\bf TMC 1} \\ \cmidrule{2-3}
				  & Minissale bare & Minissale ice \\ \midrule
				 & \multicolumn{2}{c}{H$_{2}$S Production} \\ \midrule
				\multirow{2}{*}{$A_{\rm v}\sim 5$ mag} & s-H + s-HS $\rightarrow$ H$_{2}$S (91.0\%) & s-H + s-HS $\rightarrow$ H$_{2}$S (90.3\%)  \\ 
				& H$_{3}$CS$^{+}$ + e$^{-}\rightarrow$ H$_{2}$S + CH (8.2\%) & H$_{3}$CS$^{+}$ + e$^{-}\rightarrow$ H$_{2}$S + CH (9.2\%) \\ \midrule
				\multirow{2}{*}{$A_{\rm v}\sim 20$ mag} & s-H + s-HS $\rightarrow$ H$_{2}$S (55.1\%) &  s-H + s-HS $\rightarrow$ H$_{2}$S (43.6\%)  \\ 
				& H$_{3}$S$^{+}$ + e$^{-}\rightarrow$ H + H$_{2}$S (31.9\%) & H$_{3}$S$^{+}$ + e$^{-}\rightarrow$ H + H$_{2}$S (43.3\%) \\
				\midrule
				& \multicolumn{2}{c}{H$_{2}$S Destruction} \\ \midrule
				\multirow{2}{*}{$A_{\rm v}\sim 5$ mag} & H$_{2}$S + S$^{+}\rightarrow$ H$_{2}$ + S$_{2}^{+}$ (40.1\%) & H$_{2}$S + S$^{+}\rightarrow$ H$_{2}$ + S$_{2}^{+}$ (39.1\%)  \\ 
				& C + H$_{2}$S $\rightarrow$ HCS + H (16.7\%) & C + H$_{2}$S $\rightarrow$ HCS + H (18.0\%) \\ \midrule
				\multirow{2}{*}{$A_{\rm v}\sim 20$ mag} & H$_{2}$S + H$_{3}^{+}\rightarrow$ H$_{2}$ + H$_{3}$S$^{+\ \dagger}$ (51.7\%) & H$_{2}$S + H$_{3}^{+}\rightarrow$ H$_{2}$ + H$_{3}$S$^{+}$ (72.9\%)  \\ 
				 & H$_{2}$S + S$^{+}\rightarrow$ H$_{2}$ + S$_{2}^{+}$ (11.2\%) & H$_{2}$S + H$^{+}\rightarrow$ H + H$_{2}$S$^{+}$ (6.4\%) \\ \midrule
				\multirow{2}{*}{Visual Extinction} & \multicolumn{2}{c}{\bf Barnard 1b} \\ \cmidrule{2-3}
				 & Minissale bare & Minissale ice \\ \midrule
				 & \multicolumn{2}{c}{H$_{2}$S Production} \\ \midrule
				\multirow{2}{*}{$A_{\rm v}\sim 5$ mag} & s-H + s-HS $\rightarrow$ H$_{2}$S (99.2\%) & s-H + s-HS $\rightarrow$ H$_{2}$S (85.4\%)  \\ 
				& S + H$_{2}$S$^{+}\rightarrow$ H$_{2}$S + S$^{+}$ (0.3\%) & s-H$_{2}$S $\rightarrow$ H$_{2}$S (0.3\%) \\ \midrule
				\multirow{2}{*}{$A_{\rm v}\sim 70$ mag} & H$_{3}$S$^{+}$ + e$^{-}\rightarrow$ H + H$_{2}$S (48.4\%) & H$_{3}$S$^{+}$ + e$^{-}\rightarrow$ H + H$_{2}$S (47.0\%)  \\ 
				& s-H + s-HS $\rightarrow$ H$_{2}$S (42.8\%) & s-H + s-HS $\rightarrow$ H$_{2}$S (46.8\%) \\
				\midrule
				& \multicolumn{2}{c}{H$_{2}$S Destruction} \\ \midrule
				\multirow{2}{*}{$A_{\rm v}\sim 5$ mag} & H$_{2}$S + S$^{+}\rightarrow$ H$_{2}$ + S$_{2}^{+}$ (39.2\%) & H$_{2}$S + S$^{+}\rightarrow$ H$_{2}$ + S$_{2}^{+}$ (38.0\%) \\ 
				& C + H$_{2}$S $\rightarrow$ HCS + H (32.1\%) & C + H$_{2}$S $\rightarrow$ HCS + H (32.4\%) \\ \midrule
				\multirow{2}{*}{$A_{\rm v}\sim 70$ mag} & H$_{2}$S + H$_{3}^{+}\rightarrow$ H$_{2}$ + H$_{3}$S$^{+}$ (73.8\%) & H$_{2}$S + H$_{3}^{+}\rightarrow$ H$_{2}$ + H$_{3}$S$^{+}$ (69.8\%)  \\ 
				& H$_{2}$S + H$^{+}$ $\rightarrow$ H + H$_{2}$S$^{+}$ (10.9\%) & H$_{2}$S + H$^{+}\rightarrow$ H + H$_{2}$S$^{+}$ (20.6\%) \\ \bottomrule
			\end{tabular}
			\flushleft
			{\small
			\ \ \ \ \ \ \ \ \ \ \ \ \ \ \ \ \ \ \ \ \ \ \ \ \ \ \ \ \ \ \ \ $^{\dagger}$ This reaction recycles H$_{2}$S since the products are reactants in the production reactions.\\
			}
			\label{tab:reactions}
\end{table*}

\begin{table*}
	\centering
		\caption{Abundances of selected molecules in ice mantles}
			\begin{tabular}{ccccccc}
				\toprule
				\multirow{2}{*}{Molecule} &  \multicolumn{2}{c}{Predicted abundances (\,/\,H\,) } &  \multicolumn{2}{c}{Predicted abundances (\,/\,s-H$_{2}$O\,)} & \multirow{2}{*}{\shortstack[c]{Abundance in \\ comets (\,/\,s-H$_{2}$O\,)}\footnote{}} & \multirow{2}{*}{\shortstack[c]{67P/Churyumov-\\Gerasimenko\footnote{}}} \\ \cmidrule{2-3}\cmidrule{4-5}
				 & TMC 1 & Barnard 1b & TMC 1 & Barnard 1b \\ \midrule
					CO & $4.0\times 10^{-5}$ & $2.2\times 10^{-5}$ & $3.0\times 10^{-1}$ & $1.3$ & $0.002-0.23$ & $-$\\
					
					CO$_{2}$ & $9.0\times 10^{-6}$ & $7.4\times 10^{-5}$ & $6.8\times 10^{-2}$ & $4.2$ & $0.025 - 0.30$ & $-$\\
					
					H$_{2}$S & $3.4\times 10^{-6}$ & $3.3\times 10^{-6}$ & $2.5\times 10^{-2}$ & $1.9\times 10^{-1}$ & $0.0013-0.015$ & $0.0067-0.0175$\\
					
					OCS & $3.8\times 10^{-7}$ & $1.1\times 10^{-6}$ & $2.9\times 10^{-3}$ & $6.2\times 10^{-2}$ & $(0.3-4)\times 10^{-3}$ & $0.00017-0.00098$ \\
					
					SO & $5.5\times 10^{-9}$ & $6.3\times 10^{-8}$ & $4.1\times 10^{-5}$ & $3.6\times 10^{-3}$ &  $(0.4-3)\times 10^{-3}$ & $0.00004-0.000014$ \\
					
					SO$_{2}$ & $4.0\times 10^{-11}$ & $1.4\times 10^{-8}$ & $3.0\times 10^{-7}$ & $8.0\times 10^{-4}$ & $2\times 10^{-3}$ & $0.00011-0.00041$\\
					
					CS & $3.7\times 10^{-8}$ & $7.9\times 10^{-8}$ & $2.8\times 10^{-4}$ & $4.6\times 10^{-3}$ & $(0.2-2)\times 10^{-3}$ & $-$\\
					
					H$_{2}$CS & $4.0\times 10^{-8}$ & $5.2\times 10^{-8}$ & $3.0\times 10^{-4}$ & $3.0\times 10^{-3}$ & $(0.9-9)\times 10^{-4}$ & $-$ \\
					
					NS & $1.8\times 10^{-6}$ & $2.7\times 10^{-6}$ & $1.3\times 10^{-2}$ & $1.5\times 10^{-1}$ & $(0.6-1.2)\times 10^{-4}$ & $-$ \\
					
					S$_{2}$ & $1.2\times 10^{-9}$ & $2.8\times 10^{-9}$ & $8.7\times 10^{-6}$ & $1.6\times 10^{-4}$ & $(0.1-25)\times 10^{-4}$ & $0.000004-0.000013$\\
					
					H$_{2}$O & $1.3\times 10^{-4}$ & $1.7\times 10^{-5}$ & 1 & 1 & $-$ & $-$\\
    			\bottomrule
			\end{tabular}
			\flushleft
			{\small
			$^{2}$ Data taken from \citet{BockeleMorvan2017}\\
			$^{3}$ Taken from \citet{Calmonte2016}
			}
			\label{tab:cometsice}
\end{table*}

\subsection*{Sulphur abundance}
A large theoretical and observational effort has been done in the last five years to understand the sulphur chemistry that is progressively leading to a new paradigm  \citep{Vidal2017, Fuente2016, Fuente2019, LeGal2019, Laas2019}. Gas-grain models provide a new vision of the interstellar medium in which both gaseous and ice species are predicted. \citet{Vidal2017} updated the sulphur chemistry network and used this new chemical network to interpret previous observations towards the prototypical dark core TMC 1-CP. They found that the best fit to the observations was obtained when adopting the cosmic abundance of S as initial condition, and an age of ~1 Myr.  At this age,  more than 70 \% of the sulphur would remain in gas phase and only ~30\% would be in ice. Using the same chemical network but with 1D modeling, \citet{Vastel2018} tried to fit the abundances of twenty-one S-bearing species towards the starless core L1544. The authors found that all the species cannot be fitted with the same sulphur abundance, with variations of a factor of 100, being models with S/H $ = 8.0 \times 10^{-8}$ at an age of 1 - 3 Myr, those that best fitted the abundances of all the twenty-one species. A new gas-grain astrochemical network was proposed by \citet{Laas2019} to account for the complex sulphur chemistry observed in comets.

In this paper, we attempt to carry out a comprehensive study of the sulphur chemistry in TMC 1 and Barnard 1b by modeling the CS, SO, and H$_2$S observations in three cold cores using a fully updated gas-grain chemical network. Our chemical model is unable to fit the observations of the three species with the same parameters. By assuming a cosmic abundance of sulphur, we have a reasonable  agreement between H$_2$S and SO observations with model predictions. However, we overestimate the CS abundance by more than one order of magnitude. Adopting a different approach, \citet{Fuente2019} obtained S/H $\sim(0.4-2.2)\times 10^{-6}$, one order of magnitude lower than the cosmic sulphur abundance, in the translucent part of the the TMC 1 filament. Given the uncertainty in the sulphur chemistry,  we can conclude that  the value of S/H that best fits our data  $\sim$ a few 10$^{-6}$ to 1.5 $\times$ 10$^{-5}$.

\section{Summary and Conclusions}
We performed a cloud depth dependent observational and modeling study, determining the physical structure and chemistry of three prototypical dense cores, two of them located in Taurus (TMC 1-C, TMC 1-CP)  and the third one in Perseus (Barnard 1b). 

\begin{itemize}

\item We used the dust temperature to carry out a rough estimate of the incident UV field which is $\chi=3-10$ Draine field in TMC 1 and $\chi=24$ in Barnard 1b. Moreover, we modeled the physical conditions of TMC 1- C, TMC 1- CP and Barnard 1b assuming a BE sphere. TMC 1- C and TMC 1- CP are well-fitted with the same parameters while Barnard 1b requires a higher central density consistent with a more evolved collapse.

\item Using the full gas-grain model \textsc{Nautilus} and the physical structure derived from our observations, we investigated the chemistry of H$_2$S in these cores. Chemical desorption reveals as the most efficient release path for H$_2$S in cold cores. Our results show that the abundance of H$_2$S is well-fitted assuming high values of chemical desorption (bare grains) for densities $n_{\rm H} < 2 \times 10^4$ cm$^{-3}$.  For higher densities, our model overestimates the H$_2$S abundance suggesting that chemical desorption becomes less efficient. We propose that this critical density is related with a change in the chemical composition of the surface of the grains. 

\item To further investigate the relationship between the H$_2$S and grains properties, we examined the chemical composition of the icy mantles along the cores, as predicted by our model. Interestingly, the decrease of the H$_2$S chemical desorption occurs when the abundance of s-H$_2$O and s-CO achieves the maximum value in both molecular clouds.

\item One additional result is that our model predicts different ice compositions in TMC 1 and Barnard 1b. While the abundance of s-H$_2$O is as high as $\sim$ 10$^{-4}$ in TMC 1,  its abundance is $\sim$10$^{-5}$ in Barnard 1b.  The abundances of s-H$_2$S are however very similar in the two clouds, about one fifth of the sulphur cosmic abundance.

\item In addition to H$_2$S, we compared the abundances of CS and SO with model predictions. Our chemical model is unable to fit the observations of the three species simultaneously. Given the uncertainty in the sulphur chemistry,  we can only conclude that  the value of S/H that best fits our data is the cosmic value within a factor of 10.

\end{itemize}

This paper presents an exhaustive study of the sulphur chemistry in TMC 1 and Barnard 1b by modeling the CS, SO, and H$_2$S observations. Our chemical model is unable to fit the observations of the three species at a time, but it does manage to fit, with reasonable accuracy, our H$_2$S and SO observations. More theoretical and experimental work needs to be done in this area, especially in improving the constraints on the chemistry of CS. Given the uncertainty in the sulphur chemistry,  we can only conclude that the value of S/H that best fit our data is the cosmic value within a factor of 10.

\begin{acknowledgements}
We thank the Spanish MINECO for funding support from AYA2016-75066-C2-1/2-P, AYA2017-85111P, and ERC under ERC-2013-SyG, G. A. 610256 NANOCOSMOS. JM acknowledges the support of ERC-2015-STG No. 679852 RADFEEDBACK. SPTM and JK acknowledges to the European Union's Horizon 2020 research and innovation program for funding support given under grant agreement No~639459 (PROMISE). 
\end{acknowledgements}

\bibliography{h2s}
\bibliographystyle{aa}

\begin{appendix}

\begin{landscape}
\thispagestyle{empty}
	\vspace*{\fill}
	\section{Physical conditions and chemical abundances in TMC 1}
	\begin{threeparttable}[h!]
		{\tiny
		\caption{TMC 1 physical conditions and chemical abundances}
		\centering
		\begin{tabular}{lcccccccc}
		\toprule
		Source name\  & \ ${\rm T}_{\rm dust}$ (K) & $A_{\rm v}$ (mag) &  ${\rm T}_{\rm gas}$(K) & n$_{H}$ (cm$^{-3}$) & ${\rm N}(^{13}$CS$)$ (cm$^{-2}$)$^{\ a}$ & N$(^{13}$CS) / N(H$_{\rm T}$)$^{\ c}$ & ${\rm N(ortho-H}_{2}{\rm S})({\rm cm}^{-2})$ & N$({\rm H}_{2}{\rm S})$ / N(H$_{\rm T}$) \\ \midrule
		
		TMC 1-CP+0 & 11.92 & 18.20 & $9.7\pm 0.8$	& $(3.0\pm 0.8)\cdot 10^{4}$ & $(3.9 \pm 1.0)\cdot 10^{12}$ & $(1.1\pm 0.3)\cdot 10^{-10}$ & $(3.1\pm 0.9)\cdot 10^{13}$ & $(1.1\pm 0.3)\cdot 10^{-9}$ \\ \midrule 
		
		TMC 1-CP+30 & 12.00 & 16.71 & $10.2\pm 0.2$	 & $(4.6\pm 0.4)\cdot 10^{4}$ & $(1.6 \pm 0.2)\cdot 10^{12}$  & $(4.7\pm 0.6)\cdot 10^{-11}$ & $(5.1\pm 1.4)\cdot 10^{13}$ & $(2.0\pm 0.6)\cdot 10^{-9}$ \\ \midrule 
		
		TMC 1-CP+60 & 12.24 & 13.74 & $11.3\pm 2.2$	 & $(7.6\pm 3.8)\cdot 10^{4}$ & $(1.2 \pm 0.4)\cdot 10^{11}$  & $(4.3\pm 1.4)\cdot 10^{-11}$ & $(1.2\pm 0.6)\cdot 10^{13}$ & $(5.7\pm 0.3)\cdot 10^{-10}$ \\ \midrule
		
		TMC 1-CP+120 & 13.16 & 7.27 & $12.5\pm 1.3$ & $(3.0\pm 0.8)\cdot 10^{3}$ & $(2.8\pm 1.1)\cdot 10^{12}$  & $(1.9\pm 0.8)\cdot 10^{-10}$ & $(5.5\pm 1.6)\cdot 10^{13}$ & $(5.1\pm 1.4)\cdot 10^{-9}$ \\ \midrule
		
		TMC 1-CP+180 & 13.86 & 4.77 & $16.0\pm 2.6$ & $(5.4\pm 1.6)\cdot 10^{3}$ & $(5.4\pm 1.5)\cdot 10^{12}$ & $(5.7\pm 1.6)\cdot 10^{-11}$ & $(4.3\pm 1.3)\cdot 10^{13}$ & $(5.9\pm 1.8)\cdot 10^{-9}$ \\ \midrule
		
		TMC 1-CP+240 & 14.39 & 3.25 & $14.7\pm 1.1$ & $(3.2\pm 2.0)\cdot 10^{3}$ & $(5.2\pm 3.2)\cdot 10^{11}$ & $(8.0\pm 5.0)\cdot 10^{-12}$ & $(2.1\pm 1.3)\cdot 10^{13}$ & $(4.2\pm 2.1)\cdot 10^{-9}\ ^{b}$ \\ \midrule
		
		TMC 1-C+0 & 11.26 & 19.85 & $8.5\pm 2.0$ & $(9.2\pm 6.8)\cdot 10^{4}$ & $(1.1 \pm 0.5)\cdot 10^{12}$  & $(2.8\pm 1.2)\cdot 10^{-11}$ & $(2.0\pm 1.5)\cdot 10^{13}$ & $(6.0\pm 4.9)\cdot 10^{-10}$ \\ \midrule
		
		TMC 1-C+30 & 11.32 & 18.47 & $10.3\pm 2.0$ & $(8.8\pm 4.6)\cdot 10^{4}$ & $(7.0 \pm 3.9)\cdot 10^{11}$ & $(1.9\pm 1.1)\cdot 10^{-11}$ & $(2.5\pm 1.3)\cdot 10^{13}$ & $(8.6\pm 5.0)\cdot 10^{-10}$ \\ \midrule
		
		TMC 1-C+60 & 11.67 & 13.34 & $11.6\pm 2.2$	 & $(2.4\pm 1.0)\cdot 10^{4}$ & $(9.2\pm 2.3)\cdot 10^{11}$ & $(3.5\pm 0.9)\cdot 10^{-11}$ & $(5.4\pm 2.0)\cdot 10^{13}$ & $(2.6\pm 1.0)\cdot 10^{-9}$ \\ \midrule
		
		TMC 1-C+120 & 13.13 & 4.79 & $11.1\pm 1.9$ & $(1.1\pm 0.5)\cdot 10^{4}$ & $(6.2\pm 1.8)\cdot 10^{11}$  & $(6.5\pm 1.9)\cdot 10^{-11}$ & $(1.0\pm 0.5)\cdot 10^{13}$ & $(1.4\pm 0.6)\cdot 10^{-8}$ \\ \midrule
			
		TMC 1-C+180 & 14.08 & 2.20 & $13.5\pm 1.1$ & $(1.1\pm 2.8)\cdot 10^{4}$ & $(1.8\pm 0.9)\cdot 10^{11}$  & $(4.1\pm 2.1)\cdot 10^{-11}$ & $(6.5\pm 3.3)\cdot 10^{12}\ ^{b}$ & $(1.9\pm 1.0)\cdot 10^{-9}\ ^{b}$ \\ \midrule
		
		TMC 1-C+240 & 14.53 & 1.63 & $13.5\pm 2.7$ & $(5.2\pm 1.8)\cdot 10^{3} $ & $(1.6\pm 1.0)\cdot 10^{11}$ & $(4.9\pm 3.0)\cdot 10^{-11}$ & $(1.5\pm 1.0)\cdot 10^{13}\ ^{b}$ & $(6.0\pm 4.2)\cdot 10^{-9}\ ^{b}$ \\ \bottomrule
		\end{tabular}
		}
		
		\begin{tablenotes}{
      		\small
      		\item {\bf Notes:}
      		\small 
      		\item $^{a}$ When $^{13}$CS or C$^{34}$S isotopologues are not detected, $^{13}$CS column densities are determined from that of C$^{34}$S or CS, applying the ratios CS/$^{13}$CS $\approx$ 60 and C$^{34}$S/$^{13}$CS $\approx$ 8/3}.
      		\small
      		\item $^{b}$ Upper bound values.
      		\item $^{c}$ N(H$_{\rm T}$) stands for the total hydrogen column density: N(H$_{\rm T}$) = N(H) + 2 N(H$_{2}$). 
    	\end{tablenotes}
	\end{threeparttable}
	\label{table:TMC1conditions}
	\vspace*{\fill}
\end{landscape}

\newpage
\onecolumn
\thispagestyle{empty}
\subsection{TMC 1-C spectra}
\begin{figure*}[hbt]
	\centering
  	\includegraphics[width=0.99\textwidth]{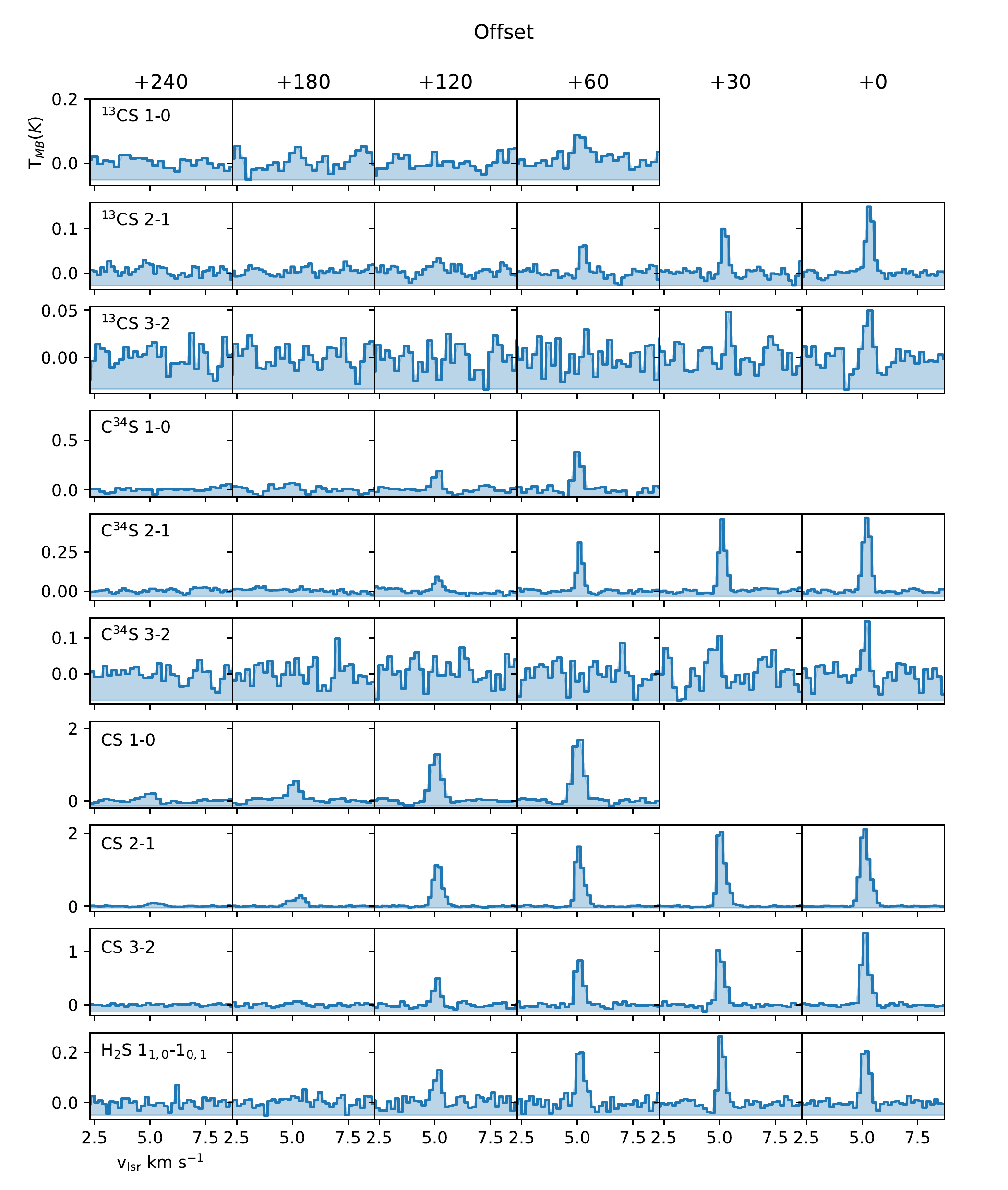}
  	\caption{Single dish spectra of $^{13}$CS $1\rightarrow 0$, $^{13}$CS $2\rightarrow 1$, $^{13}$CS $3\rightarrow 2$, C$^{34}$S $1\rightarrow 0$, C$^{34}$S $2\rightarrow 1$, C$^{34}$S $3\rightarrow 2$, CS $1\rightarrow 0$, CS $2\rightarrow 1$, C$^{34}$S $3\rightarrow 2$ transitions towards TMC 1-C positions with offsets $(+0",0")$, $(+30",0")$, $(+60",0")$, $(+120",0")$, $(+180",0")$, $(+240",0")$. The systemic velocity is $v_{\rm lsr} = 6.5$ km s$^{-1}$.}
  	\label{fig:TMC1_C_spectra}
\end{figure*}

\newpage
\onecolumn
\thispagestyle{empty}
\subsection{TMC 1-CP spectra}
\begin{center}
	\begin{figure*}[hbt]
		\centering
  		\includegraphics[width=0.99\textwidth]{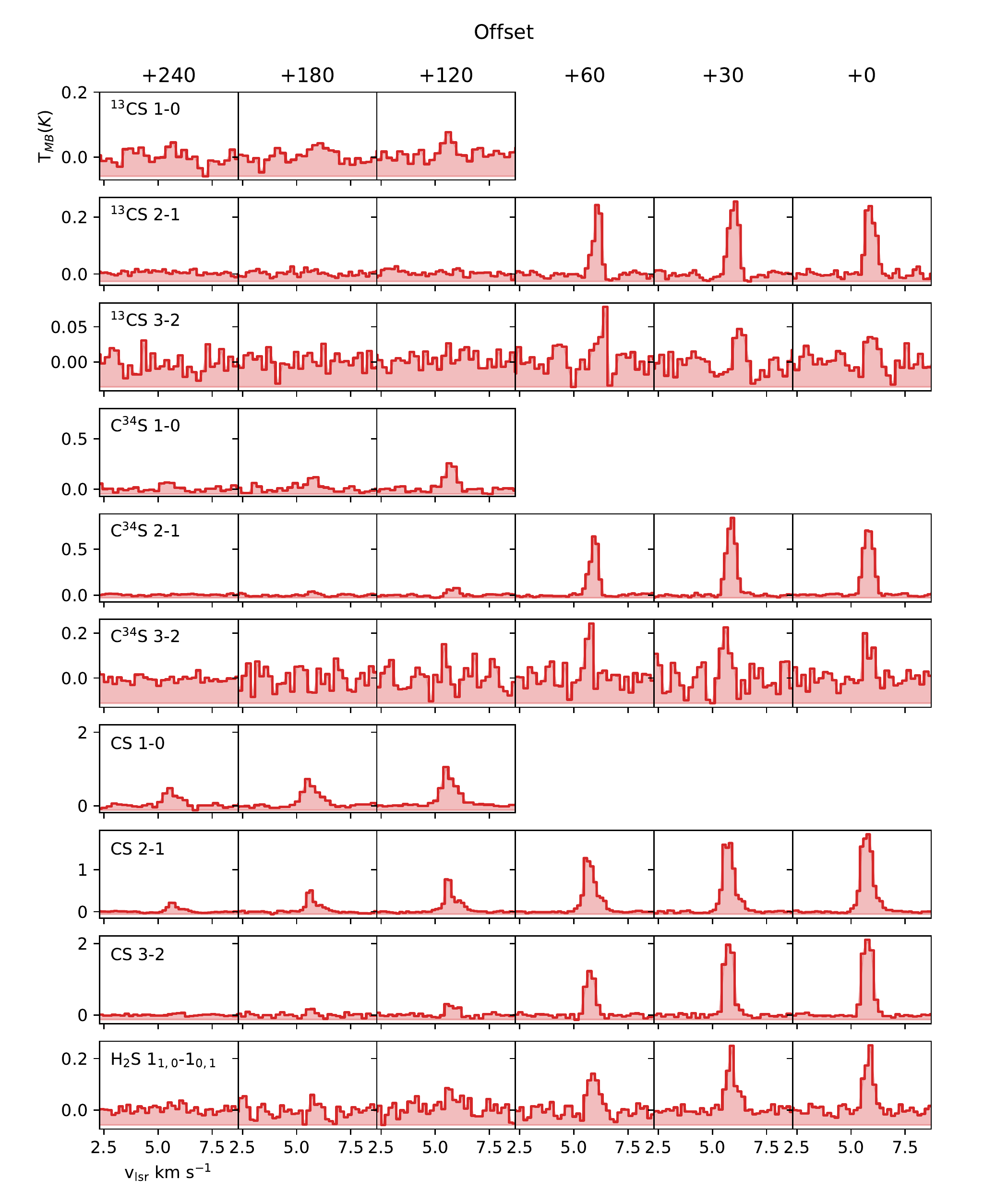}
  		\caption{Single dish spectra of $^{13}$CS $1\rightarrow 0$, $^{13}$CS $2\rightarrow 1$, $^{13}$CS $3\rightarrow 2$, C$^{34}$S $1\rightarrow 0$, C$^{34}$S $2\rightarrow 1$, C$^{34}$S $3\rightarrow 2$, CS $1\rightarrow 0$, CS $2\rightarrow 1$, C$^{34}$S $3\rightarrow 2$ transitions towards TMC 1-CP positions with offsets $(+0",0")$, $(+30",0")$, $(+60",0")$, $(+120",0")$, $(+180",0")$, $(+240",0")$. The systemic velocity is $v_{\rm lsr} = 6.5$ km s$^{-1}$.}
  		\label{fig:TMC1_CP_spectra}
	\end{figure*}
\end{center}

\newpage

\begin{landscape}
\thispagestyle{empty}
%\newgeometry{left=0cm, top=3cm}
	\vspace*{\fill}
	\section{Physical conditions and chemical abundances in Barnard 1b}
	\centering
	\begin{threeparttable}[h!]
		{\small
		\caption{Barnard 1b physical conditions. $^{13}$CS and H$_{2}$S abundances.}
		\centering
		\begin{tabular}{lccccccccc}
		\toprule
		Source name\  & \ ${\rm T}_{\rm dust}$ (K) & $A_{\rm v}$ (mag) & ${\rm T}_{\rm gas}$(K) & n$_{H}$ (cm$^{-3}$) & ${\rm N}(^{13}$CS$)$ (cm$^{-2}$)$^{\ a}$  & N($^{13}$CS) / N(H$_{\rm T}$) & ${\rm N(ortho-H}_{2}{\rm S})({\rm cm}^{-2})$ & N(${\rm H}_{2}{\rm S}$) / N(H$_{\rm T}$) \\ \midrule
		
		B1B-cal-0\_0 & 11.90 & 76.00 & $9.8\pm 1.4$ & $(6.3\pm 3.0)\cdot 10^{5}$ & $(2.0 \pm 1.6)\cdot 10^{12}$  & $(2.6\pm 2.1)\cdot 10^{-11}$ & $(5.2\pm 2.5)\cdot 10^{13}$ & $(4.5\pm 2.1)\cdot 10^{-10}$ \\ \midrule 
		
		B1B-cal-10\_0 & 11.72 & 59.80 & $10.1\pm 1.5$ & $(4.2\pm 2.8)\cdot 10^{5}$ & $(1.3 \pm 0.9)\cdot 10^{12}$  & $(2.2\pm 1.5)\cdot 10^{-11}$ & $(1.8\pm 1.2)\cdot 10^{14}$  $^{c}$ & $(1.9\pm 1.2)\cdot 10^{-9}$ $^{c}$ \\ \midrule
		
		B1B-cal-20\_0 & 11.72 & 45.80 & $10.9\pm 1.7$ & $(1.6\pm 0.9)\cdot 10^{5}$ & $(1.9 \pm 1.2)\cdot 10^{12}$  & $(4.0\pm 2.6)\cdot 10^{-11}$ & $(3.2\pm 2.1)\cdot 10^{14}$  $^{c}$ & $(4.5\pm 2.9)\cdot 10^{-9}$ $^{c}$ \\ \midrule
		
		B1B-cal-30\_0 & 11.54 & 38.70 & $11.9\pm 1.0$ & $(1.0\pm 0.4)\cdot 10^{5}$ & $(1.1\pm 0.5)\cdot 10^{12}$ & $(2.8\pm 1.2)\cdot 10^{-11}$ & $(3.1\pm 1.4)\cdot 10^{14}$ & $(5.2\pm 2.3)\cdot 10^{-9}$ \\ \midrule
		
		B1B-cal-40\_0 & 11.54 & 28.39 & $12.1\pm 1.1$ & $(1.0\pm 0.4)\cdot 10^{5}$ & $(9.5\pm 4.4)\cdot 10^{11}$  & $(3.3\pm 1.6)\cdot 10^{-11}$ & $(2.0\pm 0.9)\cdot 10^{13}$ & $(4.5\pm 2.0)\cdot 10^{-9}$ \\ \midrule
		
		B1B-cal-50\_0 & 12.39 & 20.00 & $13.2\pm 1.3$ & $(4.7\pm 2.2)\cdot 10^{4}$ & $(1.7\pm 0.8)\cdot 10^{11}$  & $(8.7\pm 4.2)\cdot 10^{-11}$ & $(1.9\pm 0.9)\cdot 10^{13}$ & $(6.0\pm 2.9)\cdot 10^{-9}$ \\ \midrule
		
		B1B-cal-60\_0 & 12.67 & 20.00 & $12.3\pm 0.9$ & $(3.1\pm 1.6)\cdot 10^{4}$ & $(2.2\pm 1.1)\cdot 10^{12}$  & $(1.1\pm 0.5)\cdot 10^{-10}$ & $(3.4\pm 1.7)\cdot 10^{14}$ & $(1.1\pm 0.6)\cdot 10^{-8}$ \\ \midrule
		
		B1B-cal-80\_0 & 13.24 & 17.05 & $13.2\pm 1.8$ & $(5.5\pm 2.4)\cdot 10^{4}$ & $(1.0\pm 0.3)\cdot 10^{12}$  & $(6.1\pm 1.8)\cdot 10^{-11}$ & $(1.0\pm 0.4)\cdot 10^{14}$ & $(3.9\pm 1.7)\cdot 10^{-9}$ \\ \midrule
		
		B1B-cal-110\_0 & 13.98 & 14.46 & $14.4\pm 1.9$ & $(5.2\pm 2.1)\cdot 10^{4}$ & $(9.8\pm 2.7)\cdot 10^{11}$  & $(6.8\pm 1.9)\cdot 10^{-11}$ & $(8.2\pm 3.3)\cdot 10^{13}$ & $(3.7\pm 1.5)\cdot 10^{-9}$ \\ \midrule
		
		B1B-cal-140\_0 & 14.53 & 11.87 & $14.2\pm 1.0$ & $(9.5\pm 2.5)\cdot 10^{3}$ & $(5.4\pm 2.7)\cdot 10^{12}$  & $(4.5\pm 2.3)\cdot 10^{-10}$ & $(8.3\pm 2.2)\cdot 10^{14}$ & $(4.5\pm 1.2)\cdot 10^{-8}$ \\ \midrule
		
		B1B-cal-180\_0 & 16.21 & 8.57 & $15.3\pm 1.2$ & $(3.6\pm 1.7)\cdot 10^{3}$ & $(6.2\pm 3.3)\cdot 10^{12}$  & $(7.3\pm 3.9)\cdot 10^{-10}$ & $(4.2\pm 2.0)\cdot 10^{14}$ & $(3.2\pm 1.5)\cdot 10^{-8}$ \\ \midrule
		
		B1B-cal-240\_0 & 16.70 & 6.16 & $16.4\pm 1.0$ & $(3.8\pm 1.8)\cdot 10^{3}$ & $(1.6\pm 0.8)\cdot 10^{12}$ & $(2.6\pm 1.4)\cdot 10^{-10}$ & $(1.5\pm 0.7)\cdot 10^{14}$ & $(1.6\pm 0.8)\cdot 10^{-8}$ \\ \midrule
		
		B1B-cal-500\_0 & 18.23 & 3.44 & $18.0\pm 5.4$ & $(9.6\pm 2.2)\cdot 10^{2}$ & $(5.5\pm 3.0)\cdot 10^{11}$ & $(1.6\pm 0.9)\cdot 10^{-10}$ & $(1.8\pm 0.4)\cdot 10^{13}\ ^{b}$ & $(3.4\pm 0.8)\cdot 10^{-9}\ ^{b}$ \\
		
		\bottomrule
		\end{tabular}
		}
		
		\begin{tablenotes}{
      		\small
      		\item {\bf Notes} 
      		\item $^{a}$ When $^{13}$CS or C$^{34}$S isotopologues are not detected, $^{13}$CS column densities are determined from that of C$^{34}$S or CS, applying the isotopic ratios CS/$^{13}$CS $\approx$ 60 and C$^{34}$S/$^{13}$CS $\approx$ 8/3}.
      		\small
      		\item $^{b}$ Upper bound values.
      		\small
      		\item $^{c}$ Column densities are obtained from that of the isotopologue H$_{2}$$^{34}$S, using H$_{2}$S/H$_{2}$$^{34}$S $\approx$ 22.5. 
    	\end{tablenotes}
	\end{threeparttable}
	\label{table:B1Bconditions}
	\vspace*{\fill}
\end{landscape}

\newpage

\begin{landscape}
\thispagestyle{empty}
%\newgeometry{left=1cm, top=10.7cm}
	\vspace*{\fill}
	\centering
	\begin{threeparttable}
		{\small
		\caption{Barnard 1b physical conditions and SO abundances.}
		\centering
		\begin{tabular}{lcccccc}
		\toprule
		Source name\  & \ ${\rm T}_{\rm dust}$ (K) & $A_{\rm v}$ (mag) & ${\rm T}_{\rm gas}$(K) & n$_{H}$ (cm$^{-3}$) & ${\rm N}($SO$)$ (cm$^{-2}$)  & N(SO) / N(H$_{\rm T}$) \\ \midrule
		
		B1B-cal-0\_0 & 11.90 & 76.00 & $9.8\pm 1.4$ & $(6.3\pm 3.0)\cdot 10^{5}$ & $(2.4\pm 0.6)\cdot 10^{14}$ & $(1.5\pm 0.4)\cdot 10^{-9}$ \\ \midrule
		
		B1B-cal-10\_0 & 11.72 & 59.80 & $10.1\pm 1.5$	 & $(4.2\pm 2.8)\cdot 10^{5}$ & $(2.5\pm 0.9)\cdot 10^{14}$ & $(2.1\pm 0.8)\cdot 10^{-9}$ \\ \midrule
		
		B1B-cal-20\_0 & 11.72 & 45.80 & $10.9\pm 1.7$ & $(1.6\pm 0.9)\cdot 10^{5}$ & $(2.7\pm 1.0)\cdot 10^{14}$ & $(2.9\pm 1.1)\cdot 10^{-9}$ \\ \midrule
		
		B1B-cal-30\_0 & 11.54 & 38.70 & $11.9\pm 1.0$ & $(1.0\pm 0.4)\cdot 10^{5}$ & $(2.4\pm 1.6)\cdot 10^{14}$ & $(3.2\pm 2.1)\cdot 10^{-9}$ \\ \midrule
		
		B1B-cal-40\_0 & 11.54 & 28.39 & $12.1\pm 1.0$ & $(1.0\pm 0.4)\cdot 10^{5}$ & $(9.9\pm 4.5)\cdot 10^{13}$ & $(1.7\pm 0.8)\cdot 10^{-9}$ \\ \midrule
		
		B1B-cal-50\_0 & 12.39 & 20.00 & $13.2\pm 1.0$ & $(4.7\pm 2.2)\cdot 10^{4}$ & $(5.2\pm 1.7)\cdot 10^{13}$ & $(1.3\pm 0.4)\cdot 10^{-9}$ \\ \midrule
		
		B1B-cal-60\_0 & 12.67 & 20.00 & $12.3\pm 1.0$ & $(3.1\pm 1.6)\cdot 10^{4}$ & $(4.8\pm 1.6)\cdot 10^{13}$ & $(1.2\pm 0.4)\cdot 10^{-9}$ \\ \midrule
		
		B1B-cal-80\_0 & 13.24 & 17.05 & $13.2\pm 1.8$ & $(5.5\pm 2.4)\cdot 10^{4}$ & $(2.2\pm 0.8)\cdot 10^{13}$ & $(6.5\pm 2.3)\cdot 10^{-10}$ \\ \midrule
		
		B1B-cal-110\_0 & 13.98 & 14.46 & $14.4\pm 1.9$ & $(5.2\pm 2.1)\cdot 10^{4}$ & $(3.4\pm 1.9)\cdot 10^{13}$ & $(1.2\pm 0.7)\cdot 10^{-9}$ \\ \midrule
		
		B1B-cal-140\_0 & 14.53 & 11.87 & $14.2\pm 1.0$ & $(9.5\pm 2.5)\cdot 10^{3}$ & $(4.2\pm 2.2)\cdot 10^{13}$ & $(1.8\pm 0.9)\cdot 10^{-9}$ \\ \midrule
		
		B1B-cal-180\_0 & 16.21 & 8.57 & $15.3\pm 1.2$ & $(3.6\pm 1.7)\cdot 10^{3}$ & $(4.9\pm 2.4)\cdot 10^{13}$ & $(2.9\pm 1.4)\cdot 10^{-9}$ \\ \midrule
		
		B1B-cal-240\_0 & 16.70 & 6.16 & $16.4\pm 1.0$ & $(3.8\pm 1.8)\cdot 10^{3}$ & $(4.9\pm 2.6)\cdot 10^{13}$ & $(4.0\pm 2.1)\cdot 10^{-9}$ \\ \midrule
		
		B1B-cal-500\_0 & 18.23 & 3.44 & $18.0\pm 5.4$ & $(9.6\pm 2.2)\cdot 10^{2}$ & $(1.5\pm 0.3)\cdot 10^{12}\ ^{a}$ & $(2.2\pm 0.5)\cdot 10^{-10}\ ^{a}$ \\
		
		\bottomrule
		\end{tabular}
		}
		
		\begin{tablenotes}{
      		\small
      		\item {\bf Notes} 
      		\item $^{a}$ Upper bound values} 
    	\end{tablenotes}
	\end{threeparttable}
	\label{table:B1BconditionsSO}
	\vspace*{\fill}
\end{landscape}

\newpage
\onecolumn
\thispagestyle{empty}
\subsection{Barnard 1b spectra}
\begin{figure*}[hbt]
	\centering
  	\includegraphics[width=0.99\textwidth]{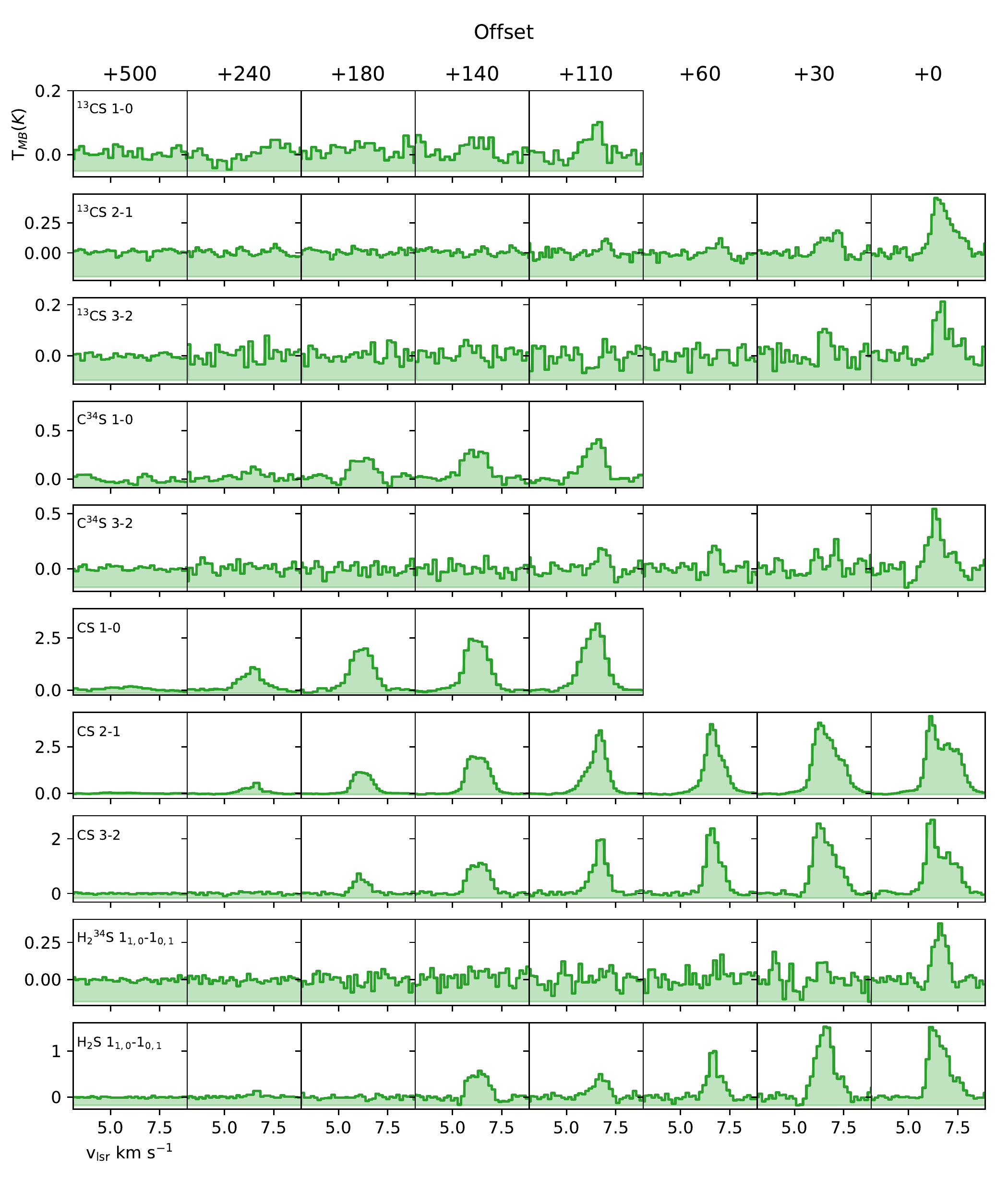}
  	\caption{Single dish spectra of $^{13}$CS $1\rightarrow 0$, $^{13}$CS $2\rightarrow 1$, $^{13}$CS $3\rightarrow 2$, C$^{34}$S $1\rightarrow 0$, C$^{34}$S $2\rightarrow 1$, C$^{34}$S $3\rightarrow 2$, CS $1\rightarrow 0$, CS $2\rightarrow 1$, C$^{34}$S $3\rightarrow 2$ transitions towards positions with offsets $(+0",0")$, $(+30",0")$, $(+60",0")$, $(+120",0")$, $(+180",0")$, $(+240",0")$ in the Barnard 1b filament. The systemic velocity is $v_{\rm lsr} = 6.5$ km s$^{-1}$.}
  	\label{fig:B1B_spectra}
\end{figure*}

%\clearpage
%\subsection{Barnard 1b spectra}
%\begin{figure*}[hbt]
%	\centering
%  	\includegraphics[width=0.82\textwidth]{SO_SpectraB1B.pdf}
%  	\caption{Single dish spectra of SO $2_{2}\rightarrow 1_{1}$, SO $2_{3}\rightarrow 1_{2}$, SO $3_{2}\rightarrow 2_{1}$, SO $3_{4}\rightarrow 2_{3}$, SO $4_{4}\rightarrow 3_{3}$, $^{34}$SO $2_{3}\rightarrow 1_{2}$, $^{34}$SO $3_{2}\rightarrow 2_{1}$, $^{34}$SO $4_{4}\rightarrow 3_{3}$ transitions towards positions with offsets $(+0",0")$, $(+30",0")$, $(+60",0")$, $(+120",0")$, $(+180",0")$, $(+240",0")$ in the Barnard 1b filament. The systemic velocity is $v_{\rm lsr} = 6.5$ km s$^{-1}$.}
%\end{figure*}

\end{appendix}

\end{document}